\documentclass[11pt,a4paper,twoside]{iopjournal}

\usepackage{ragged2e}

\usepackage{amsmath,times}

\usepackage{amssymb}

\usepackage{amsfonts}  
\usepackage{mathrsfs}

\usepackage[utf8]{inputenc}
\DeclareUnicodeCharacter{03BC}{\mbox{$\mu$}}
\DeclareUnicodeCharacter{03B4}{$\delta$}
\DeclareUnicodeCharacter{2009}{-}

\usepackage{placeins,float}

\usepackage[british]{babel}

\makeatletter
\def\bbl@set@language#1{%
	\edef\languagename{%
		\ifnum\escapechar=\expandafter`\string#1\@empty
		\else\string#1\@empty\fi}%
	\@ifundefined{babel@language@alias@\languagename}{}{%
		\edef\languagename{\@nameuse{babel@language@alias@\languagename}}%
	}%
	\select@language{\languagename}%
	\expandafter\ifx\csname date\languagename\endcsname\relax\else
	\if@filesw
	\protected@write\@auxout{}{\string\select@language{\languagename}}%
	\bbl@for\bbl@tempa\BabelContentsFiles{%
		\addtocontents{\bbl@tempa}{\xstring\select@language{\languagename}}}%
	\bbl@usehooks{write}{}%
	\fi
	\fi}
\newcommand{\DeclareLanguageAlias}[2]{%
	\global\@namedef{babel@language@alias@#1}{#2}%
}
\makeatother

\DeclareLanguageAlias{en}{english}

\def\G {\scriptscriptstyle{{G}}}
\def\B {\scriptscriptstyle{B}}
\def\Z {\scriptscriptstyle{{(Z)}}}
\def\F {\scriptscriptstyle{{F}}}
\def\B {\scriptscriptstyle{{B}}}
\def\P {\scriptscriptstyle{(P)}}

\usepackage{makecell}

\def\L{\mathsf L}

\numberwithin{equation}{section}

\begin{document}

\articletype{Topical Review}

\title{\textbf{The Lieb--Liniger model}}

\author{Zoran Ristivojevic\orcid{0000-0003-1827-4547}}

\affil{Universit\'{e} de Toulouse, CNRS, Laboratoire de Physique Th\'{e}orique, Toulouse, France}

%\email{zoran.ristivojevic@cnrs.fr}
\vspace{1cm}
\keywords{bosons in one dimension, $\delta$-interaction, Lieb--Liniger model,  Bethe ansatz, exact results, Yang--Yang thermodynamics, circular capacitor}

\begin{abstract}
\justifying
The Lieb--Liniger model describes one-dimensional bosons with contact interactions. This many-body system admits an exact solution in terms of the Bethe ansatz. Some of the exact and perturbative results for this model are reviewed. Particular attention is devoted to the explicit evaluation, in terms of the interaction parameter, of physical quantities that can be formally exactly extracted from the Bethe ansatz solution. Another goal of this review is to stress exact relations between various quantities. The technical developments are explained in detail. The most relevant experimental realisations of the studied problems are eventually discussed. This review also contains several new results such as the study of convergence of the ground-state energy series at strong interactions, the excitation spectrum at high energies, and the evaluation of the boundary energy.
\end{abstract}

%\newpage
\tableofcontents
\justifying

\section{Introduction}

The system of one-dimensional bosons with contact interactions is a well-known example of correlated quantum systems that has arrived at the stage of experimental realisations. From the theoretical perspective, this system is special as it can be exactly solved by the Bethe ansatz. It is known under the name Lieb--Liniger model. Over the years many exact results have been obtained for this model. Without limitations on parameters, such findings are of paramount importance. They are a firm ground for theoretical studies, benchmarks for computer simulations, and challenges for experimental probes.

The history of the Lieb--Liniger model started more than sixty years ago. It can be split into the period before the year 2000 and after that, where the delimiting age corresponds to the years when experimental realisations started to appear. Chronologically, the most important discoveries that shaped our knowledge about this model today are as follows. It all started in 1960 when Girardeau found the similarity between the systems of infinitely repulsive bosons and free fermions \cite{girardeau_relationship_1960}. Shortly after that, in 1963, Lieb and Liniger solved the model exactly by the Bethe ansatz technique and studied the ground-state properties and the excitation spectrum  \cite{lieb_exact_1963,lieb_exact_1963b}. In 1969, Yang and Yang found the exact description of the thermodynamics of the model at arbitrary temperatures \cite{yang_thermodynamics_1969}. The development of quantum inverse scattering method and algebraic Bethe ansatz at the end of the 1970s resulted in various results for the matrix elements and the correlation functions, reviewed in the book \cite{korepin}. Additional insights arrived at the beginning of the 1980s using the Luttinger liquid approach and bosonisation \cite{haldane_effective_1981}. The development of conformal field theory \cite{affleck_universal_1986,blote_conformal_1986} is another milestone that stimulated activity at the end of 1980s \cite{korepin}. Experimental realisations in a wide interaction range finally happened at the beginning of 2000s \cite{paredes_tonksgirardeau_2004,kinoshita_observation_2004}. This moment is special and marked a new era in several respects. First, it caused a surge of both experimental and theoretical activities. The latter have become more inclined to experimentally relevant questions and thus the Lieb--Liniger model has begun to be a research topic outside the field of mathematical physics. Second, around that time the use of computers became really necessary. The theorists started practically evaluating certain correlation functions \cite{caux_dynamical_2006} that were previously addressed only very formally. Around the same time the theoretical question of how the Lieb--Liniger gas and other integrable models relax from a highly nonequilibrium initial state emerged from the remarkable experiment on quantum Newton's cradle \cite{kinoshita_quantum_2006}. The most important theoretical results that have led towards better understanding of nonequilibrium dynamics are the notion of generalised Gibbs ensemble  \cite{rigol_relaxation_2007} and related hydrodynamic theory that accounts for infinitely many conservation laws, called generalised hydrodynamics  \cite{castro-alvaredo_emergent_2016,bertini_transport_2016}.

% due to several reasons. First, experiments provide only approximate realisations of the model. On the other hand, they probe the physics inherent of the model. The data analysis necessarily requires various adaptations to the experimental situation and solutions of nonintegrable models that describe experiments, yet very close to the original Lieb--Liniger model. Such adaptations include the remnants of three-dimensionality and the realistic effects: transverse motion in the trapping potentials, longitudinal trapping potential, finite number of weakly-interacting tubes, each being an approximate realization of the model, etc. 

%They present an extremely rich phenomenology and, since they allow us to find exact relations be- tween physical quantities, they represent a nice way to test principles and paradigms of the physical theories. 

This review is based on the scientific publications of the author over the last decade \cite{ristivojevic_excitation_2014,ristivojevic_decay_2016,petkovic_spectrum_2018,ristivojevic_conjectures_2019,reichert_exact_2019,reichert_fluctuation-induced_2019,reichert_analytical_2020,ristivojevic_exact_2021,ristivojevic_dispersion_2022,ristivojevic_method_2022,ristivojevic_exact_2023,petkovic_density_2023,panfil_local_2025}. It addresses the standard topics related to the Lieb--Liniger model such as the ground-state energy, the spectrum of elementary excitations, the conserved charges, the local correlation functions, the boundary energy, the low-temperature thermodynamics, and the circular capacitor. In addition, the polaron energy spectrum in the Yang--Gaudin Bose gas is studied. All these topics are well-known to a specialist. The exact results for many of the relevant quantities are derived and studied elsewhere starting from the initial papers of Lieb and Liniger \cite{lieb_exact_1963,lieb_exact_1963b} and Yang and Yang \cite{yang_thermodynamics_1969}. Despite its long history, several relevant quantities lacked closed-form results (or had incorrect ones); these are addressed in this review. The fact that many obtained exact results had not been explicitly evaluated in terms of the microscopic parameters, see for example the book of Korepin \textit{et~al.~}\cite{korepin}, probably originates from the lack of interest as the Lieb--Liniger model used to be a topic of mathematical physics. Since nowadays the model describes the systems realised in experiments, it is beyond its initial perimeter and explicitly evaluated physical quantities might be important. This is achieved in the present review, which provides the evaluated form for almost all addressed quantities, which usually take the form of power series. The author has always tried to produce systematic procedures that can be further explored by interested readers. Obtaining the systematic expansion in terms of the interaction strength parameter for, e.g., the ground-state energy might appear useless to someone. However, a good counterexample for this is the local $N$-body correlation function. In order to obtain analytically its leading-order term at strong interactions from the exact closed-form expression, one needs to know $N(N-1)$ subleading terms in the ground-state energy. Already for the case of $N=3$, which is experimentally probed \cite{tolra_observation_2004,haller_three-body_2011}, there is a need for an unusually large number of terms in the series of the ground-state energy.

It is widely known that the symmetries of the system impose constraints on certain physical parameters. For example, the relation $mvK=\pi\hbar n$ follows from Galilean invariance \cite{haldane_effective_1981}. Here $m$ is the mass of bosons, $v$ is the sound velocity, $K$ is the Luttinger liquid parameter, $n$ is the particle density, and $\hbar$ is the reduced Planck constant. In the review we have shown that this relation is in fact only the tip of the iceberg and can be seen as the initial member of a family of relations that connect the ground-state values of the consecutive conserved charges. Unlike the above-written relation that is generically correct in all Galilean-invariant one-dimensional models, the other relations of the family  apply in cases the model is, in addition, integrable. Similar relations that connect the derivatives of the rapidity distribution at the Fermi level also exist and will be discussed. These connections simplify the analysis of the excitation spectrum and the low-temperature thermodynamics, for example.

There are several textbooks and monographs that treat the Lieb--Liniger model. The classic ones are written by Gaudin \cite{gaudin}, Korepin \textit{et al.~}\cite{korepin}, and Takahashi \cite{takahashi}. More recent ones were written by \v{S}amaj and Bajnok \cite{samaj}, Franchini \cite{franchini},  and Eckle \cite{eckle}. In addition we must mention the textbook by Sutherland \cite{sutherland} that treats integrable models from a more general perspective. There are also several review articles that address particular aspects of the Lieb--Liniger model. Some of the recent ones are the reviews written by Bouchoule and Dubail \cite{bouchoule_generalized_2022} where the model is reviewed from the point of view of generalised hydrodynamics, by Zwerger \cite{zwerger_liebliniger_2022} where general physical aspects of confinement of higher-dimensional systems that effectively lead to the experimental realisation of the Lieb--Liniger model are addressed, by Cazalilla \textit{et al.} \cite{cazalilla_one_2011} as well as by Yiang, Chen, and Guan \cite{jiang_understanding_2015} where various results of the model obtained over several decades are reviewed. The present review is mostly complementary to these works as it provides new physical results obtained during the last years, new technical advances, and new points of view.

A note about physical units is in order. In the original paper of Lieb and Liniger \cite{lieb_exact_1963} as well as in the books \cite{gaudin,korepin,takahashi}, for example, the units where $\frac{\hbar^2}{2m}=1$ were used. In this review we explicitly keep $\hbar$ and $m$ in all expressions. On the other hand, we use the units where the Boltzmann constant is set to unity, $k_{\B}=1$. Unlike recovering $\hbar$ and $m$ in final results that may require some effort, this is not the case with the Boltzmann constant. An interested reader can trivially recover $k_{\B}$ using a simple replacement of the temperature $T$ by $k_{\B} T$.

In the remainder of this review, in section \ref{section1}, the Lieb--Liniger model is introduced and a brief summary of its Bethe ansatz solution is given. The quasimomentum distribution as a central object is introduced. It satisfies Lieb's integral equation that does not admit a closed-form solution. The zero-temperature thermodynamics is discussed. In section \ref{section:numericalexperiment}, the ground-state energy is addressed in the regime of weak interactions using a method based on experimental mathematics. It enabled exact analytical evaluation of the series expansion to a large number of terms, which is difficult to achieve using more conventional methods. The complementary regime of strong interactions is studied in section \ref{section:simplesolution}. The developed analytical method is used to study the radius of convergence of the obtained series expansion of the ground-state energy in the inverse interaction parameter. In section \ref{section:method} a new method is developed that results in difference-differential equations. In particular, the moments of the quasimomentum distribution are addressed and evaluated, as well as the Fermi quasimomentum and the local correlation functions. In section \ref{section:derivativesrho}, a partial differential equation for the quasimomentum distribution is used to derive expressions for its derivatives at the Fermi quasimomentum. They are used to define new dimensionless parameters beyond the standard Luttinger liquid parameter. They control the coefficients of the expansion of the spectrum of elementary excitations at low energies studied in section \ref{section:elementaryexcitations}. In the same section, the spectrum at high energies is also studied. In section \ref{section:thermodynamics}, the thermodynamics of the Lieb--Liniger model is addressed. The method to treat the Yang--Yang equation at low temperatures is developed, which enables the evaluation of the Helmholtz free energy beyond the well-known result obtained earlier using conformal field theory. In section \ref{section:boundaryenergy}, the boundary energy of the model is calculated. It represents the subleading term in the energy of the system subject to zero boundary conditions. In section \ref{section:capacitance}, a well-known problem of capacitance of the circular capacitor is solved analytically using its mathematical analogy with the Lieb--Liniger model. In section \ref{section:Yang-Gaudin gas}, the polaron problem is addressed using the Bethe-ansatz solvable Yang--Gaudin model and its energy spectrum is calculated. Experimental achievements relevant for the subjects of this review are reviewed in section \ref{section:experiments}. Various technical details that were not suitable for the main text are given in several appendices.

\section{The Lieb--Liniger model}\label{section1}

The Lieb--Liniger model is one of the simplest models of interacting quantum particles in the continuum \cite{lieb_exact_1963}. It describes one-dimensional bosons with contact interactions. Despite being introduced in 1963, the model continues to fascinate the scientific community sixty years later. It is remarkable in many respects. For example, the case of attractive interactions has subtle connections with classical two-dimensional systems, in particular with a surface growth described by the Kardar–Parisi–Zhang equation. Another very relevant aspect is that this model describes the physics of realistic systems as it could be nowadays directly realised in experiments with cold gases. Finally, the Lieb--Liniger model is integrable and admits analytically exact solution in terms of the Bethe ansatz. This is the most important feature for our purposes and will be exploited here. The model thus serves as a benchmark for different effective theories that unavoidably contain various levels of approximations \cite{imambekov_one-dimensional_2012,Giamarchi}. The known results for the Lieb--Liniger model form a cornerstone of the one-dimensional quantum physics of many-body systems, enabling better understanding of the correlation effects \cite{cazalilla_one_2011,guan_fermi_2013}.

\subsection{Model}

The Lieb--Liniger model describes $N$ nonrelativistic bosons in one dimension with short-range two-body interactions. It is defined by the Hamiltonian
\begin{align}\label{eq:Horiginal}
	H=-\frac{\hbar^2}{2m}\sum_{j=1}^N \frac{\partial^2}{\partial x_j^2}+\frac{\hbar^2 c}{2m}\sum_{j,l=1\atop j\neq l}^{N} \delta(x_j-x_l).
\end{align}
Here $m$ is the mass of particles that are in the continuum space at positions $x_1,x_2,\ldots,x_N$. We consider the system of the length $L$. The interaction potential between the particles has the form of $\delta$-function parametrised by the coupling constant $c$, which can be of an arbitrary sign. However, the case of attractive interactions, $c<0$, is peculiar as the ground-state energy per particle scales with $N$ as $c^2N^2$ \cite{takahashi}. Thus the thermodynamic limit of the system is ill defined in the conventional sense, i.e., fixed $c$ and $N,L\to\infty$ with their ratio $n=N/L$ fixed. Here $n$ denotes the density of particles. Our primary goal is to study the system in the thermodynamic limit, which exists for repulsive interactions. In the following we thus consider positive coupling, $c>0$.

\subsection{The Bethe ansatz solution}

The principal problem for the Hamiltonian (\ref{eq:Horiginal}) is its diagonalisation, which was achieved by the Bethe ansatz \cite{lieb_exact_1963}. Since the system of interacting bosons is considered, our goal is to solve the Schr\"{o}dinger equation $H\psi=E\psi$ and find a completely symmetric wave function $\psi(x_1,x_2,\ldots,x_N)$ with respect to the permutations of the coordinates. A physically admissible solution must be continuous everywhere in space, but may have discontinuities with respect to the first derivative. Using the standard condition for the jump of the first derivative due to the $\delta$-function potential, such solution can be obtained. It is characterised by a set of $N$ distinct real numbers $k_1,k_2,\ldots,k_N$ and takes the form \cite{korepin,gaudin}
\begin{align}\label{eq:wave function}
	\psi=\sum_{\mathcal{P}\in\pi_N} (-1)^{[\mathcal{P}]} \exp\left(i\sum_{j=1}^{N}k_{\mathcal{P}_j} x_j \right)\prod_{j,l=1\atop{j>l}}^{N}[k_{\mathcal{P}_j}-k_{\mathcal{P}_l}-i c\, \mathrm{sgn}(x_j-x_l)].
\end{align} 
Here  the summation is over $\mathcal{P}$ that is from the set $\pi_N$ of permutations of the natural numbers $1,2,\ldots,N$ and $[\mathcal{P}]$ denotes the parity of the permutation. The eigenenergy of the system is given by
\begin{align}\label{eq:eigenenergy}
	E=\frac{\hbar^2}{2m}\sum_{j=1}^{N}k_j^2.
\end{align}
We note that $k_1,k_2,\ldots,k_N$ are known as quasimomenta or rapidities. Our task is to find them for a given value of the interaction coupling $c$ in a certain state of the system. The latter is defined at our will and can be the ground state or a specific excited state. 

To study the properties of the system, let us consider the system in a periodic box of the length $L$. We thus impose periodic boundary conditions on the wave function (\ref{eq:wave function}). We require
\begin{align}\label{eq:PBC}
	\psi(x_1,x_2,\ldots,x_j+L,\ldots,x_N)=\psi(x_1,x_2,\ldots,x_j,\ldots,x_N)
\end{align}
for each $x_j$ when all the other coordinates $x_l$, $l\neq j$, are fixed. The condition (\ref{eq:PBC}) enables the characterisation of the set of quasimomenta, leading to 
\begin{align}\label{eq:discreteBA-PBC}
	\exp(i L k_j)=\prod_{l=1\atop{l\neq j}}^{N}\frac{k_j-k_l+i c}{k_j-k_l-i c},\quad j=1,2,\ldots,N.
\end{align}  
The coupled nonlinear equations (\ref{eq:discreteBA-PBC}) are known as the discrete Bethe equations. Instead of using them, it is convenient to take their logarithm that gives
\begin{align}\label{eq:DBA}
	Lk_j=2\pi I_j+\sum_{l=1}^{N}\theta(k_j-k_l),\quad j=1,2,\ldots,N.
\end{align}
Here we have introduced the two-particle scattering phase shift
\begin{align}\label{eq:phaseshift}
	\theta(k)=-2\arctan(k/c),
\end{align}
and 
\begin{align}\label{eq:Ij}
	I_j\in \begin{cases}
		\mathbb{Z}+\frac{1}{2},\quad &N\, \textrm{even},\\
		\mathbb{Z},\quad &N\,\textrm{odd}.\\
	\end{cases}
\end{align}
The different form of $I_j$  for odd and even number of particles $N$ originates from the expression for the sum $\pi \sum_{l=1,l\neq j}^{N}\mathrm{sgn}(k_j-k_l)$ that arises after transforming the logarithm into the arctangent using
\begin{gather}
	i \ln\frac{A+i}{A-i}=2\arctan (A)-\pi\:\!\:\!\mathrm{sgn}(A),\quad A\neq 0.
\end{gather}
Note that the summation in equation~(\ref{eq:DBA}) is extended to include the $j=l$ contribution, which is zero due to $\theta(0)=0$.

The system of Bethe equations (\ref{eq:discreteBA-PBC}) has some general properties \cite{korepin}. In the case of repulsion between the bosons, $c>0$, all the quasimomenta $k_j$ are real. If $I_j>I_l$, then $k_j>k_l$. On the other hand, $I_j=I_l$ implies $k_j=k_l$. Since equation~(\ref{eq:wave function}) is an antisymmetric function of $k_j$, having equal quasimomenta would imply vanishing of the wave function. Therefore, we must use the set of distinct quantum numbers $I_j$. In essence, this property is the Pauli principle for interacting bosons in one dimension \cite{korepin}.

The Bethe ansatz equations (\ref{eq:DBA}) are central for the following considerations. Conceptually, they are rather simple. Once equations~(\ref{eq:DBA}) are solved for a specific choice of $I_j$, the wave function (\ref{eq:wave function}) and the corresponding energy (\ref{eq:eigenenergy}) are at our disposal and in principle the many-body problem is solved. However, the latter is difficult to do analytically for the quasimomenta since the equations are nonlinear at finite $c$. In order to illustrate another difficulty, let us mention that the number of summands in the wave function grows factorially with $N$ and thus it becomes challenging to straightforwardly use equation~(\ref{eq:wave function}) in the computations \cite{forrester_analytic_2006}. Instead we will study equations~(\ref{eq:DBA}) in the thermodynamic limit.

\subsection{The ground state}

The ground state of the system of $N$ particles is realised for the configuration of quasimomenta corresponding to the minimal possible energy given by equation~(\ref{eq:eigenenergy}). In the case of infinite boson repulsion, $c\to\infty$, this occurs if the quasimomenta are distributed equidistantly and symmetrically around zero. The corresponding quantum numbers are given by
\begin{align}\label{eq:quantumnumbers}
	I_j=j-\frac{N+1}{2},\quad j=1,2,\ldots,N.
\end{align}
At any finite interaction $c>0$, one expects that the same quantum numbers (\ref{eq:quantumnumbers}) determine the quasimomenta in the ground state. This follows from the previous statement that a larger quantum number (with respect to a referent one) imposes a larger quasimomentum. Moreover, on physical grounds we expect that the quasimomenta for the Lieb--Liniger model change continuously with respect to $c$. The latter argument also supports a choice for the ground-state quantum numbers (\ref{eq:quantumnumbers}) at any $c>0$, which is indeed correct.

The limiting quasimomenta $k_1$ and $k_N$ are equal in the absolute value and will be denoted by $Q$, which is the Fermi quasimomentum. All other quasimomenta are distributed between $-Q$ and $Q$. Let us introduce
\begin{align}\label{eq:Fhokj}
	\rho(k_j)=\frac{1}{L(k_{j+1}-k_j)}.
\end{align}
In the thermodynamic limit, the function~(\ref{eq:Fhokj}) becomes the density of quasimomenta. Therefore, the number of quasimomenta within the small interval between $k$ and $k+dk$ is given by $L\rho(k) dk$. This enables us to transform the sum into an integral according to
\begin{align}\label{eq:sumtointegral}
	\sum_{j=1}^{N}f(k_j)\to L\int_{-Q}^{Q} dk f(k)\rho(k),
\end{align}
where $f(k_j)$ is an arbitrary function. The thermodynamic limit of equation~(\ref{eq:DBA}) can be obtained by considering its difference for the indices $j+1$ and $j$. It leads to
\begin{align}
	L(k_{j+1}-k_j)=2\pi+\sum_{l=1}^{N}[\theta(k_{j+1}-k_l)-\theta(k_{j}-k_l)].
\end{align}
Since $k_{j+1}-k_j$ is on the order of $1/L$ and thus small \cite{korepin}, we can Taylor expand the difference of the phase shifts. Using the definition (\ref{eq:Fhokj}), after transforming the sum into an integral according to equation~(\ref{eq:sumtointegral}) we end up with
\begin{align}\label{eq:LIE}
	\rho(k,Q)+\frac{1}{2\pi}\int_{-Q}^{Q}dq \:\! \theta'(k-q)\rho(q,Q)=\frac{1}{2\pi}.
\end{align}
This is the Lieb integral equation \cite{lieb_exact_1963}. The kernel of the integral operator is given by
\begin{align}\label{eq:kernel}
	\theta'(k)=\frac{d\theta(k)}{dk}=-\frac{2c}{c^2+k^2}
\end{align}
for our case of the Lieb--Liniger model. Equation (\ref{eq:LIE}) fully determines the ground-state properties of the system in the thermodynamic limit. It should be understood as an expression that enables one to evaluate the $\rho$ dependence on $k$ for a given value of the Fermi quasimomentum $Q$. Therefore $\rho$ in fact depends on two variables and thus we use the notation $\rho(k,Q)$ instead of $\rho(k)$ that is present in equation~(\ref{eq:sumtointegral}). Let us note that the above considerations, in particular equations~(\ref{eq:DBA}) and (\ref{eq:LIE}), are more general and apply for some other integrable models provided the appropriate two-particle phase shift is used \cite{sutherland}.

Once $\rho(k,Q)$ is known, the density of particles $n=N/L$ follows directly from the prescription (\ref{eq:sumtointegral}),
\begin{align}\label{eq:n}
	n=\int_{-Q}^{Q}dk \rho(k,Q).
\end{align}
The energy (\ref{eq:eigenenergy}) is an extensive quantity. Its value per particle in the ground state is given by
\begin{align}\label{eq:eps0}
	\epsilon_0=\frac{\hbar^2}{2mn}\int_{-Q}^{Q}dk\:\! k^2 \rho(k,Q).
\end{align}
Since $\epsilon_0$ by construction depends on $Q$, we can use the density dependence $n(Q)$ of equation~(\ref{eq:n}) to express $Q$ in terms of $n$. 

The Lieb--Liniger model contains a single parameter $c$ that has the dimensions of the inverse length. Rather than using it directly, it is more convenient to introduce the Lieb parameter
\begin{align}\label{eq:gammadef}
	\gamma=\frac{c}{n},
\end{align}
which is dimensionless. Then the nontrivial dependence of the ground-state energy per particle (\ref{eq:eps0}) on the interaction can be expressed through the ground-state function $e_2(\gamma)$ as
\begin{align}\label{eq:eps0e2}
	\epsilon_0=\frac{\hbar^2n^2}{2m}e_2(\gamma),
\end{align} 
where
\begin{align}\label{eq:e2def}
	e_{2}(\gamma)=\frac{1}{n^{3}}\int_{-Q}^{Q}dk\:\!k^{2} \rho(k,Q).
\end{align}
The right-hand side of the latter expression is dimensionless, yet it depends on $Q$, which can be expressed in terms of $\gamma$ using equation~(\ref{eq:n}).

Presently there is no known closed-form solution of the integral equation (\ref{eq:LIE}) in the general case. The only exception is the case of infinite repulsion, known as the Tonks--Girardeau limit. Then the kernel nullifies leading to
\begin{align}\label{eq:Fhoinf}
	\rho(k,Q)=\frac{1}{2\pi}.
\end{align}
The Fermi quasimomentum is then given by $Q=\pi n$ and the ground-state energy is $E_0=\frac{\pi^2\hbar^2n^2 N}{6m}$, i.e.,  $e_2(\gamma\to\infty)=\pi^2/3$. The obtained expressions for $Q$ and $E_0$ are the same as the Fermi momentum and the total energy of the system of noninteracting fermions. The latter agreement is not accidental as in the limit of infinite repulsion between the bosons, the quasimomenta of the system are equal to the momenta of free fermions.\footnote{\linespread{1}\selectfont Here one should be careful stating additionally that for odd (even) $N$, the free fermions in a box should satisfy periodic (antiperiodic) boundary conditions. However, this does not have further implications in our case.} From the latter coincidence, a physical picture where the quasimomenta fill the Fermi sea till the Fermi quasimomentum emerges. The same picture actually remains at any repulsion.

\subsection{Zero-temperature thermodynamics}

The ground state energy $E_0=N\epsilon_0$ enables us to calculate the thermodynamic parameters of the system. The pressure $P$ and the chemical potential $\mu$ are given by
\begin{gather}
	P=-\left(\frac{\partial E_0}{\partial L}\right)_N,\\
	\label{eq:mu1}
	\mu=\left(\frac{\partial E_0}{\partial N}\right)_L.
\end{gather}
Here partial derivatives should be taken at constant interaction $c$ and constant parameters that are written as subscripts after the parentheses. Using equation~(\ref{eq:eps0e2}) we obtain
\begin{gather}\label{eq:PzeroT}
	P=-\frac{\hbar^2 n^3}{2m}\gamma^3 \frac{d}{d\gamma}\left(\frac{e_2(\gamma)}{\gamma^2}\right),\\
	\label{eq:mu2}
	\mu=-\frac{\hbar^2 n^2}{2m}\gamma^4 \frac{d}{d\gamma}\left(\frac{e_2(\gamma)}{\gamma^3}\right).
\end{gather}
Therefore, the only nontrivial information is contained in the ground-state function $e_2(\gamma)$. We notice that the pressure and the chemical potential satisfy the relation
\begin{align}\label{eq:GDrelation}
	E_0=-PL+\mu N.
\end{align}
The ground-state energy $E_0$ is a homogeneous function of the number of particles $N$ and the system size $L$. It satisfies $E_0(\alpha N,\alpha L)=\alpha E_0(N,L)$, where $\alpha>0$ is a real number. Differentiating the latter with respect to $\alpha$ and setting $\alpha=1$ we obtain equation~(\ref{eq:GDrelation}). Having obtained the expression for the pressure we can find the sound velocity
\begin{gather}\label{eq:velocitypressure}
	v=\sqrt{-\frac{L}{mn} \left(\frac{\partial P}{\partial L}\right)_N},
\end{gather}
which more explicitly reads
\begin{gather}\label{eq:soundvelocityexplicit}
	v=\frac{\hbar n}{\sqrt{2}m}\gamma^2 \left[\frac{d^2}{d\gamma^2}\left(\frac{e_2(\gamma)}{\gamma^2} \right)\right]^{1/2}.
\end{gather}
The expression for the compressibility $\mathcal{K}=-\frac{1}{L}\left(\frac{\partial L}{\partial P}\right)_N$ then follows immediately, $\mathcal{K}=\frac{1}{mnv^2}$.

In addition to the previously introduced thermodynamic parameters it is convenient to introduce a dimensionless parameter $K$ by the relation
\begin{align}
	K=\frac{\pi\hbar n}{mv},
\end{align}
which is known as the Luttinger liquid parameter. Its explicit form is given by
\begin{align}\label{eq:Kfullexperssion}
	K=\frac{\pi\sqrt{2}}{\gamma^2} \left[\frac{d^2}{d\gamma^2}\left(\frac{e_2(\gamma)}{\gamma^2}, \right)\right]^{-1/2}.
\end{align}
where we have used equation (\ref{eq:soundvelocityexplicit}). The parameter $K$ controls the decay of the single-particle correlation function (one-body density matrix) at large distances. Interestingly, the information about the spectrum of elementary excitations is also contained in $K$. An example is the effective mass of elementary excitations, as will be discussed later.

\section{Numerical experiment at weak interactions \label{section:numericalexperiment}}

Explicit analytical expressions for various physical quantities of interest in integrable models are often difficult to extract from the exact solution and one is typically restricted to study special cases. For the Lieb--Liniger model, the relevant information about the system's wave function (\ref{eq:wave function}) and the corresponding energy (\ref{eq:eigenenergy}) is contained in the Bethe ansatz equations (\ref{eq:DBA}), which in the thermodynamic limit becomes the Lieb integral equation (\ref{eq:LIE}). Despite a simple form of the latter, the ground-state energy is only known in the limiting cases. In particular, the limit of weak interactions is difficult to treat since the problem becomes singular~\cite{lieb_exact_1963}. In the limit $c\to 0^+$, the kernel of equation~(\ref{eq:LIE}) becomes the Lorentzian representation of the Dirac $\delta$-function, yielding $\rho(k,Q)-\rho(k,Q)=1/2\pi$. This is impossible if $\rho(k,Q)$ is bounded. This does not contradict the physics, though. Decreasing the repulsion strength $c$, the set of quasimomenta evolves according to the Bethe ansatz equations and the initial flat distribution of quasimomenta (\ref{eq:Fhoinf}) shrinks symmetrically. At weak interactions, the Fermi quasimomentum is small, $Q\sim\sqrt{n c}$, and thus the distribution becomes sharply peaked around zero quasimomentum. Recall that the integral of the distribution gives the density, which is finite. The peak marks a tendency of bosons to exhibit a Bose--Einstein condensation.

The first three terms of the asymptotic series expansion for the ground-state energy per particle (\ref{eq:eps0e2}) were known analytically at weak interactions for a long time \cite{takahashi_validity_1975,popov_theory_1977,tracy_ground_2016}. The corresponding dependence is given by
\begin{align}\label{eq:e23terms}
	e_2(\gamma)={}&\gamma-\frac{4}{3 \pi} \gamma ^{3/2} +\left(\frac{1}{6}-\frac{1}{\pi^2}\right)\gamma^2+\mathcal{O}(\gamma^{5/2}).
\end{align}
The first two terms of the expansion were obtained by Lieb and Liniger in the initial work~\cite{lieb_exact_1963} adopting the Bogoliubov prescription from three-dimensional systems. However, the remaining term of equation~(\ref{eq:e23terms}) is significantly more difficult to evaluate \cite{popov_theory_1977,tracy_ground_2016}. In this section we develop an alternative description to find the ground-state energy at weak interactions. It is based on very precise numerical results that enable one to use the integer relation algorithm \cite{ferguson_analysis_1999,borwein_applications_2000} to discover the analytical form of the coefficients in the expansion of the ground-state energy. This heuristic approach that originates from the field of experimental mathematics is very simple, yet it gives the exact results for the coefficients. The presented results rely on the published work \cite{ristivojevic_conjectures_2019}.

\subsection{A representation of the density of quasimomenta in terms of Chebyshev polynomials}

Our starting point is the dimensionless form of the integral equation (\ref{eq:LIE}) obtained by rescaling the momenta by the Fermi quasimomentum $Q$. It has the form
\begin{align}\label{f}
	\varrho(x,\lambda)-\frac{\lambda}{\pi}\int_{-1}^{1} d y \frac{\varrho(y,\lambda)}{(x-y)^2+\lambda^2}=\frac{1}{2\pi}.
\end{align}
Here $\lambda=c/Q$. The dimensionless parameters $\gamma$ and $\lambda$ are connected by the normalisation condition 
\begin{align}\label{eq:norm}
	\gamma \int_{-1}^{1}d x \varrho(x,\lambda)=\lambda,
\end{align}
which arises from equation~(\ref{eq:n}). The ground-state function (\ref{eq:e2def}) can then be expressed as
\begin{align}\label{eq:e2defdimensionless}
	e_2(\gamma)=\frac{\gamma^3}{\lambda^3}\int_{-1}^{1} d x x^2\varrho(x,\lambda),
\end{align}
where in the right hand side of the equation one should express the parameter $\lambda$ in terms of $\gamma$ using their connection via equation~(\ref{eq:norm}).

The weakly-interacting limit of the model occurs at $\gamma\to 0$ which corresponds to $\lambda\to 0$. In this limit the kernel in the integral equation (\ref{f}) becomes sharply peaked and thus difficult for a direct numerical discretization. A convenient way to solve equation~(\ref{f}) is to expand $\rho(x,\lambda)$ into a set of complete functions on $[-1,1]$ that we take to be Chebyshev polynomials of the first kind, $T_j(x)=\cos(j \arccos x)$. We thus assume the form
\begin{align}\label{eq:FhoChebyshevexpansion}
	\varrho(x,\lambda)=\sum_{j=0}^{M} c_j(\lambda) T_{2j}(x),
\end{align}
where we take only even polynomials, since $\varrho$ is an even function of $x$. The upper limit $M$ in equation~(\ref{eq:FhoChebyshevexpansion}) is infinity but will in practice be a large integer, as we discuss below. One can then analytically evaluate the integral in equation~(\ref{f}) and transform the integral equation into a set of linear algebraic equations for the coefficients $c_j(\lambda)$ that is easily solvable. 

The recurrence relations for Chebyshev polynomials \cite{abramowitz}, $T_{j+1}(x)=2x T_j(x)-T_{j-1}(x)$ for integer $j\geq 1$, greatly simplify the evaluation of the integral  in equation~(\ref{f}). Introducing
\begin{subequations}\label{eq:FG}
	\begin{gather}
		F_j(x,\lambda)=\frac{\lambda}{\pi}\int_{-1}^{1}d y \frac{T_j(y)}{(x-y)^2+\lambda^2},\\
		G_j(x,\lambda)=\frac{\lambda}{\pi}\int_{-1}^{1}d y \frac{2y\,T_j(y)}{(x-y)^2+\lambda^2},
	\end{gather}
\end{subequations}
we find the recurrence relations
\begin{subequations}\label{eq:FGrecurence}
	\begin{align}
		F_j(x,\lambda)={}&G_{j-1}(x,\lambda)-F_{j-2}(x,\lambda),\\
		G_j(x,\lambda)={}&-\frac{8\lambda}{\pi} \frac{\sin^2\left(\frac{\pi}{2}(j-2)\right)}{j(j-2)}  +4x G_{j-1}(x,\lambda)-G_{j-2}(x,\lambda)\notag\\
		&-4(x^2+\lambda^2)F_{j-1}(x,\lambda),
	\end{align}
\end{subequations}
for $j\geq 2$. At $j=2$, the term $\frac{\sin^2\left(\frac{\pi}{2}(j-2)\right)}{j(j-2)}$ should be understood as a limit $j\to 2$ and thus it is zero. We thus analytically evaluated the functions $F_j(x)$ and $ G_j(x)$. For $j=0,1$ they are to be found directly from the definition (\ref{eq:FG}), while for $j\geq2$ they should be conveniently calculated from the recurrence relations (\ref{eq:FGrecurence}). We then transform the integral equation (\ref{f}) into
\begin{align}\label{eq:clambda}
	\sum_{j=0}^M c_j(\lambda)\left[T_{2j}(x)-F_{2j}(x,\lambda)\right]=\frac{1}{2\pi}.
\end{align}
The condition (\ref{eq:norm}) now leads to the expression of the Lieb parameter
\begin{align}\label{eq:gammafinal}
	\gamma=\frac{\lambda}{\sum_{j=0}^M \frac{2c_j(\lambda)}{1-4j^2}},
\end{align}
which is then used to transform equation~(\ref{eq:e2def}) into
\begin{align}\label{eq:efinal}
	e_2(\gamma)=\frac{\sum_{j=0}^M \frac{2c_j(\lambda)(3-4j^2)}{16j^4-40j^2+9} } {\left[\sum_{j=0}^M \frac{2c_j(\lambda)}{1-4j^2}\right]^3}.
\end{align}
For $M\to\infty$, the previous three equations represent a new form of the original ones. In particular, equation~(\ref{eq:clambda}) is an exact representation of the Lieb integral equation (\ref{f}). Similarly, equations~(\ref{eq:gammafinal})-(\ref{eq:efinal}) are our new representations for the normalisation condition (\ref{eq:norm}) and the ground-state energy function (\ref{eq:e2defdimensionless}).

\subsection{The results}

For the purpose of a highly precise numerical evaluation, our starting point are equations~(\ref{eq:clambda})-(\ref{eq:efinal}). For a fixed large integer $M$ we can solve equation~(\ref{eq:clambda}) at $M+1$ discrete, so-called collocation points of the variable $x$. For the collocation points we select the ones where the highest Chebyshev polynomial $T_{2M}(x)$ reaches its extrema. This occurs at $x_j=\cos\left(\pi j/2M\right)$, where $j=0,1,\ldots,M$. In this way one obtains a set of $M+1$ linear equations to find the coefficients $c_j(\lambda)$. The functions $F_{2j}(x_k,\lambda)$ are obtained efficiently from the recurrence relations (\ref{eq:FGrecurence}). The approximate solution of the integral equation is then given by substituting them into equation~(\ref{eq:FhoChebyshevexpansion}), while $\gamma$ and $e_2(\gamma)$ are  obtained from equations~(\ref{eq:gammafinal}) and (\ref{eq:efinal}). 

To give an example of the efficiency of the method, for $\lambda=1/10$ and $M=100$ one obtains $\gamma$ and $e_2(\gamma)$ with a relative error of the order of $10^{-36}$ in less than a second of time on a personal computer. By increasing the value of $M$ one obtains progressively more precise results, as detailed in table \ref{table0}. The latter feature of the Chebyshev representation is very important since one can always slightly increase $M$ to verify the precision of the results obtained at smaller values of $M$. 

\begin{table}
\caption{Illustration of the efficiency of our algorithm for two values $\lambda=1/10$ (corresponding to $\gamma=3.403\times 10^{-2}$), $\lambda=1/100$ ($\gamma=3.906\times 10^{-4}$), and for several values of $M$. We give the relative error in the evaluated value of $\gamma$, which is of the same order as for $e_2(\gamma)$. The evaluation times for the highest $M$ were on the order of a minute on a personal computer.\label{table0}}
\centering
\begin{tabular}{c|c c}
\hline
$M$ & relative error for $\lambda=1/10$ &  relative error for $\lambda=\frac{1}{100}$ \\ \hline
			$100$ & $10^{-36\phantom{0}}$ & $10^{-16}$ \\
			$200$ & $10^{-65\phantom{0}}$  & $10^{-26}$  \\
			$300$& $10^{-93\phantom{0}}$  & $10^{-35}$  \\
			$400$ & $10^{-121}$ & $10^{-45}$   \\
			$600$ & $10^{-177}$  & $10^{-62}$  \\
			$800$ & $10^{-233}$ & $10^{-80}$ \\
			\hline
\end{tabular}
\end{table}

For the purpose of obtaining the series expansion of $e_2(\gamma)$ that includes several subleading terms one needs $\gamma$ and $e_2(\gamma)$ at very high precision corresponding to, e.g., $M=600$. We calculated the dependence $e_2(\gamma)$ at small $\gamma$ by evaluating the system of equations for $50$ different values of $\lambda$ from the interval $(1/60,1/10]$.  Instead of $\gamma$ and $e_2(\gamma)$ we find it convenient to study the related quantities
\begin{align}\label{eq:alpha}
	\alpha=\frac{\sqrt{\gamma}}{2\pi},\quad \epsilon(\alpha)=\frac{e_2(\gamma)}{\gamma}.
\end{align}
We then fitted the numerical data with the function $\epsilon(\alpha)=\sum_{j=0}^{49} a_j \alpha^j$, obtaining $a_0=1.0000\ldots$ that satisfies $|a_0-1|\sim 10^{-59}$. We therefore identified the exact value $a_0=1$. Subtracting the latter unity from the numerically evaluated $\epsilon(\alpha)$ we then fitted the obtained data with the function $\sum_{j=1}^{49} a_j \alpha^j$ which yields $a_1=-2.6666\ldots$. It satisfies $|a_1+8/3|\sim 10^{-56}$, enabling us to identify the exact value $a_1=-8/3$. We continued such a procedure and found the remaining seven coefficients numerically and then found their presumed analytical form. For the coefficient in front of $\alpha^8$ we obtained the numerical value $a_8=-0.3604\ldots$ that differs from the exact value of the coefficient $a_8$ in absolute value by $10^{-47}$. Such highly precise fitting coefficients $a_0,\ldots,a_8$ which had at least $46$ correct digits were sufficient to use the integer relation algorithm \cite{ferguson_analysis_1999,borwein_applications_2000} that recognises the approximate number as a certain linear combination with rational coefficients of basis vectors that we take to be $1$, $\zeta$ functions \cite{prolhac_ground_2017,lang_correlations_2018}, their powers, and combinations of $\zeta$ functions, contrary to the conjecture of reference~\cite{prolhac_ground_2017}. We have tested the validity of presumably exact value for $a_8$ by solving the set of equations for $M=800$ and 60 values of $\lambda$ from the interval $(1/40,1/100]$. After fitting the polynomial of the order 59 we have obtained the numerical value that differs in the absolute value from the analytical form for $a_8$ by the order of $10^{-57}$. We have therefore no serious doubts that all the coefficients $a_0,\ldots,a_8$ obtained from experimental mathematics are exact. Their values are
\begin{gather}
	a_0=1,\quad a_1=-\frac{8}{3},\quad a_2=\frac{2\pi^2}{3}-4,\quad a_3= -4+3\zeta(3),\quad a_4=-\frac{8}{3}+2\zeta(3),\notag\\
	a_5=-1+\frac{15}{8}\zeta(3)-\frac{45}{32}\zeta(5),\quad a_6=-\frac{45}{32}\zeta(5)+\frac{3}{8}\zeta(3)+\frac{9}{16}\zeta(3)^2, \notag\\
	a_7=- \frac{2835}{2048}\zeta(7)-\frac{91}{128}\zeta(3)+\frac{105}{256}\zeta(5) +\frac{63}{64}\zeta(3)^2+\frac{1}{6},\notag\\
	a_8=-\frac{2457}{512}\zeta(7)-\frac{9}{32}\zeta(3)+\frac{135}{64}\zeta(5)[1+\zeta(3)].
\end{gather}
Using the relations (\ref{eq:alpha}) we then obtain
\begin{align}\label{eq:e2nineterms}
	e_2(\gamma)={}&\gamma-\frac{4}{3 \pi} \gamma ^{3/2} +\frac{\pi ^2-6}{6 \pi ^2} \gamma^2 - \frac{4-3 \zeta (3)}{8 \pi ^3}\gamma^{5/2} - \frac{4-3 \zeta (3)}{24 \pi^4} \gamma^3\notag\\ & - \frac{45 \zeta(5)-60 \zeta (3)+32}{1024 \pi ^5} \gamma^{7/2} - \frac{3 \left[15 \zeta(5)-4 \zeta (3)-6\zeta (3)^2\right]}{2048 \pi ^6}\gamma^4 \notag\\ &-\frac{8505\zeta(7)-2520\zeta(5)+4368\zeta(3)-6048\zeta(3)^2 -1024}{786432 \pi^7}\gamma^{9/2}\notag\\
	&-\frac{9[273\zeta(7)-120\zeta(5)+16\zeta(3)-120\zeta(3)\zeta(5)]} {131072 \pi ^8}\gamma^5 +\mathcal{O}(\gamma^{11/2}).
\end{align}
Equation (\ref{eq:e2nineterms}) is the asymptotic series for the ground-state function (\ref{eq:e2def}) in the regime of weak interactions, $\gamma\ll 1$. Although obtained heuristically, the series is exact. It can be alternatively obtained using recently developed analytical method to represent the Bethe ansatz equations in the form of a Riemann--Hilbert problem~\cite{volin_quantum_2011}, which is applied to the Lieb--Liniger model in reference~\cite{marino_exact_2019}, see also section \ref{section:capacitance}. We eventually note that prior efforts to identify numerically evaluated coefficients of the series (\ref{eq:e2nineterms}) were based on the double extrapolation of the numerical solution of the discrete Bethe ansatz equations \cite{prolhac_ground_2017,lang_correlations_2018}. The obtained results were limited only to three coefficients beyond equation~(\ref{eq:e23terms}) due to imprecision. For example, the coefficient in front of $\gamma$ was numerically evaluated with the relative error less than $10^{-24}$, while the others had progressively bigger errors. Using the present method, relative errors less than $10^{-100}$ are easily achievable.

The successful use of the integer relation algorithm to recognise a numerical constant requires its high precision, which grows with increasing the number of basis vectors \cite{borwein_applications_2000}. This is a serious limiting factor in practice. In the present problem we were able to avoid the numerical integration, which is always a source of numerical errors, in the integral equation (\ref{f}) by making use of the derived recurrence relations (\ref{eq:FGrecurence}) and thus easily produce very precise numerical data. Moreover, the precision can be further increased, when necessary, by simply increasing the number of Chebyshev polynomials $M$, as illustrated in table \ref{table0}.

\section{Simple solution at strong interactions \label{section:simplesolution}}

The Tonks--Girardeau limit of the Lieb--Liniger model, $c\to\infty$, is easily solvable as the Bethe equations (\ref{eq:DBA}) become trivial. The density of quasimomenta $\rho(k,Q)$ has a constant value given by equation~(\ref{eq:Fhoinf}). This is in striking contrast to the other limit $c\to0^{+}$ that is singular and thus it is nontrivial to develop a perturbation theory. Consider the kernel of the integral operator in the Lieb equation (\ref{eq:LIE}). At $c\gg Q$, the denominator changes very little since the quasimomenta are between $-Q$ and $Q$ and thus one can develop a perturbation theory. In this section we present a systematic way to do this, which can be used to obtain the analytical results in terms of a power series of an arbitrary order with respect to the small parameter $1/\gamma$. The presented method is developed in the published work \cite{ristivojevic_excitation_2014}.

\subsection{The expansion in Legendre polynomials}

At strong interactions, equation~(\ref{f}) can be solved by the power-series method. We assume the solution in the form
\begin{align}\label{fassumption}
	\varrho(x,\lambda)=\sum_{j=0}^{2M} a_j(\lambda) P_{j}(x),
\end{align}
where $P_{j}(x)$ are Legendre polynomials and the coefficients $a_j(\lambda)$ are to be determined. We assume $-1\le x\le 1$. The upper limit $M$ in the sum is infinity but in practice it will be an integer that determines the order of expansion in the final results. At $\lambda>|x-y|$, we expand the kernel of equation~(\ref{f}) into power series. Let us introduce $F_j^l=\int_{-1}^{1} dx\,x^l P_j(x)$. This expression is nonzero only at $l\ge j$ provided $l+j$ is an even integer, and then it becomes
\begin{align}
	F_j^l=\frac{2^{j+1}l\:\! !\left(\frac{l+j}{2}\right)!} {(l+j+1)!\left(\frac{l-j}{2}\right)!}.
\end{align}
Using the orthogonality of Legendre polynomials, we obtain the relations between the coefficients of $\varrho(x,\lambda)$,
\begin{align}\label{set}
	\frac{2a_j}{2j+1}-\sum_{m=0}^{M} \sum_{l=0}^{2m}\sum_{r=0}^{l} \frac{(-1)^{m+l}(2m)!}{\pi\:\!  l\:\!!(2m-l)!} \frac{a_r}{\lambda^{2m+1}} F_r^l F_j^{2m-l}=\frac{1}{\pi}\delta_{j,0},\quad j=0,1,\ldots,2M.
\end{align}
This is a set of linear equations that determines $a_j$ in equation~(\ref{fassumption}). Due to the spectral properties of the integral equation (\ref{f}), equation~(\ref{set}) with zero on the right-hand side would only have the trivial solution $a_j=0$. However, the right-hand side is nonzero and thus the equations involve only  $a_j$ with even $j$. This is expected, as $\varrho(x,\lambda)$ is an even function of $x$. Note that we have not restricted the sum in equation~(\ref{fassumption}) to even polynomials for later convenience. Since $r\le 2m$ in the summation in equation~(\ref{set}), all the coefficients $a_j$ for $j\ge 2$ scale at least as fast as  $\lambda^{-j-1}$ at large $\lambda$. This enables us to systematically solve equation~(\ref{set}) at finite $M$, which makes finite set of equations for the coefficients $a_{2j}$, where $j=0,1,\ldots,M$.

At $M=0$, equation~(\ref{set}) leads to a simple rescaling of $\varrho(x,\lambda\to\infty)=1/2\pi$. Already $M=1$ is sufficient to obtain a nontrivial $x$-dependence of the distribution, which starts at order $\mathcal{O}(1/\lambda^3)$. For the solution of equation~(\ref{f}), we obtain
\begin{align}\label{eq:Fho0large}
	\varrho(x,\lambda)=\frac{1}{2\pi}+ \frac{1}{\pi^2\lambda} + \frac{2}{\pi^3\lambda^2} + \frac{12-\pi^2}{3\pi^4\lambda^3} +\frac{8-2\pi^2}{\pi^5\lambda^4}-\left(\frac{1}{\pi^2\lambda^3} +\frac{2}{\pi^3\lambda^4}\right){x^2} +\mathcal{O}(\lambda^{-5}).
\end{align}
Using equation~(\ref{eq:norm}) we now find the inverse series $\lambda=(\gamma+2)/\pi -4\pi/3\gamma^2+16\pi/3\gamma^3+\mathcal{O}(\gamma^{-4})$, while $Q=\gamma n/\lambda$ expressed as a function of $\gamma$ follows directly. This enables us to find the ground-state function (\ref{eq:e2def}). Substituting equation~(\ref{eq:Fho0large}) into equation~(\ref{eq:e2defdimensionless}) yields
\begin{align}\label{eq:e2largefinal}
	e_2(\gamma)=\frac{\pi^2}{3}-\frac{4\pi^2}{3\gamma} +\frac{4\pi^2}{\gamma^2} -\frac{32\pi^2 (15-\pi^2)}{45\gamma^3}
	-\frac{16\pi^2(4\pi^2-15)}{9\gamma^4} +\mathcal{O}(\gamma^{-5}).
\end{align}
The first four terms in $e_2(\gamma)$ are in agreement with the result of reference~\cite{guan_polylogs_2011}. However, our procedure is systematic and can be easily extended to values $M>1$ by solving the system (\ref{set}) in order to obtain further terms in the series (\ref{eq:e2largefinal}). Each increase of $M$ by one produces two new terms in the power series for $e_2(\gamma)$.

\subsection{Does $\gamma=5$ belong to the regime of weak or strong interactions?}\label{section:radius}

The Lieb--Liniger model is characterised by the single dimensionless parameter $\gamma$ defined by equation~(\ref{eq:gammadef}). Many important physical quantities, for example the ground-state energy, the sound velocity, the Luttinger liquid parameter, etc., are fully determined via a function that only depends on $\gamma$. For all three examples the relevant object is the ground-state function $e_2(\gamma)$, see equation~(\ref{eq:e2def}). We have calculated it in the regimes of weak and strong interactions, obtaining two power series given by equations~(\ref{eq:e2nineterms}) and (\ref{eq:e2largefinal}). Since we are able to calculate many terms in the expansions, it is natural to understand the domains of their practical validity.  

From the mathematical perspective, the series  for the regime of weak interactions (\ref{eq:e2nineterms}) is expressed in integer powers of the parameter $\sqrt\gamma$. Since the series cannot be continued to the negative values of $\gamma$ as it becomes complex and thus nonphysical, we can conclude that the series (\ref{eq:e2nineterms}) is asymptotic. Therefore, for a given small value of $\gamma$, one needs to sum a finite number of terms, which depends on $\gamma$, in order to obtain the result closest to the exact numerical value.

On the other hand, the series for the regime of strong interactions (\ref{eq:e2largefinal}) does not have a limitation with respect to the change of positive to negative $1/\gamma$ and thus there is no obvious argument telling us that  the series is asymptotic. In fact, the result (\ref{eq:e2largefinal}) is obtained using the Taylor series expansion of the kernel of the integral operator in equation~(\ref{f}). For $-1\le x,y\le 1$, the expansion can be performed as long as $\lambda>2$, which according to equation~(\ref{eq:norm}) corresponds to $\gamma>4.53$. Since the series (\ref{eq:e2largefinal}) relies on the latter, it is naively expected that it converges for all $\gamma$ greater than $4.53$. This is only partly correct statement. A more careful analysis presented below shows that the series (\ref{eq:e2largefinal}) is indeed convergent for $\gamma>\gamma_R$, where our estimate is 
\begin{align}\label{eq:gammaradius}
	\gamma_R\approx6.95.
\end{align}

Now we describe how equation~(\ref{eq:gammaradius}) is obtained. First we substitute the form (\ref{fassumption}) into equation~(\ref{eq:norm}) yielding
\begin{align}\label{eq:gammalambdaseries}
	\gamma=\frac{\lambda}{2a_0(\lambda)}.
\end{align}
Here the coefficient $a_0(\lambda)$ has a power-series form in $1/\lambda$ and it is obtained by solving equation~(\ref{set}). Its leading-order term is $a_0(\lambda\to\infty)=1/2\pi$.  The series (\ref{eq:gammalambdaseries}) is convergent for $\lambda>2$ and can be understood as the function $\gamma(\lambda)$. In order to evaluate the right-hand side of equation~(\ref{eq:e2defdimensionless}), we need the function $\lambda(\gamma)$ that can be obtained from the inversion of the series (\ref{eq:gammalambdaseries}). Since there is no theorem that determines the radius of convergence of inverse series, we have to study our particular case. Assuming the inverse series is of the form
\begin{align}\label{eq:lamdagammaseries}
	\lambda=\frac{\gamma}{\pi}+\sum_{j=0}^\infty b_j \gamma^{-j},
\end{align}
we can find the coefficients $b_j$ from equation~(\ref{eq:gammalambdaseries}), i.e., from the dependence $a_0(\lambda)$. Once sufficiently many coefficients $b_j$ are found, its behaviour at large $j$ determines the domain of convergence of the series. From the Cauchy--Hadamard theorem, $\gamma_R$ corresponds to the limit of $|b_j|^{1/j}$ as $j\to\infty$. The convergence is however slow for $256$ coefficients that we calculated. Instead, we used the Mercer--Roberts procedure \cite{mercer_centre_1990}. There we plot $(b_{j+1}b_{j-1}-b_j^2)/(b_{j}b_{j-2}-b_{j-1}^2)$ as a function of $1/j$ and find the vertical intercept as $j\to\infty$, which corresponds to $\gamma_R$. We obtained almost perfectly linear dependence that is easily extrapolated towards $1/j\to 0$ limit, giving equation~(\ref{eq:gammaradius}).

\begin{figure}
	\centering
	\includegraphics[width=0.6\textwidth]{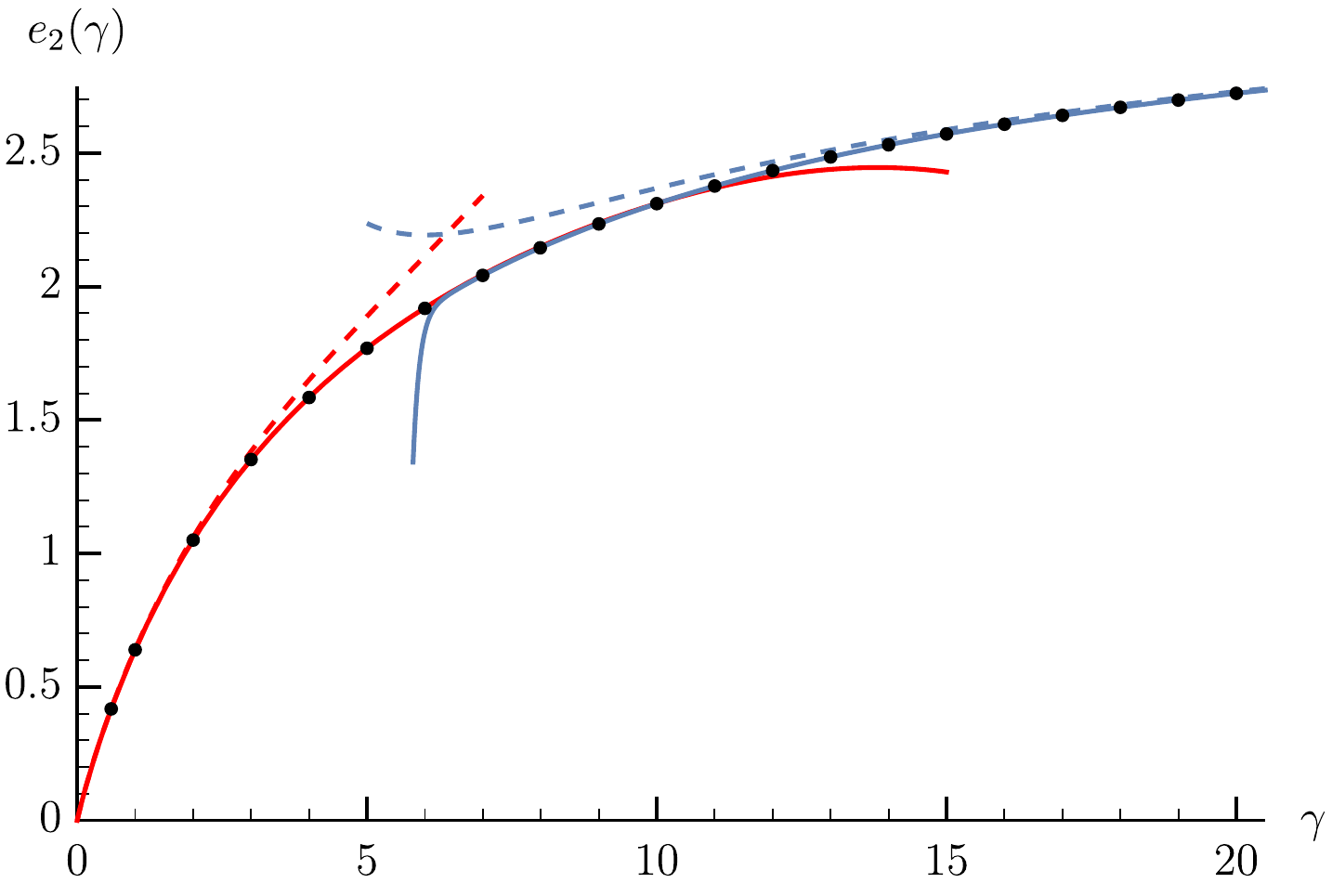}
	\caption{\justifying Plot of the ground-state function $e_2(\gamma)$. The dots represent numerically exact values, while the curves are obtained from the analytical results. In the regime of small $\gamma$, the dashed line is drawn using the first three terms and the solid line is drawn using all nine terms of the series (\ref{eq:e2nineterms}). In the regime of large $\gamma$, the dashed line is plotted using the first three terms of the series (\ref{eq:e2largefinal}) and the solid line is plotted using fifty one term. For $\gamma=7$, the relative error between the exact result and the low-$\gamma$ series is on the order of $10^{-3}$ and progressively decreases for smaller $\gamma$. For $\gamma=7$, the relative error between the exact result and the high-$\gamma$ series is on the order of $10^{-5}$ and progressively decreases for larger $\gamma$.}\label{fig:e2}
\end{figure}

It remains to evaluate $e_2(\gamma)$. From equation~(\ref{eq:e2defdimensionless}) and the assumption (\ref{fassumption}) we obtain
\begin{align}
	e_2(\gamma)=\left[\frac{2}{3}a_0(\lambda)+\frac{4}{15}a_2(\lambda)\right] \frac{\gamma^3}{\lambda^3}.
\end{align}
The series for $a_0(\lambda)$ and $a_2(\lambda)$ are convergent for $\lambda>2$. However, after substituting $\lambda$ as in equation~(\ref{eq:lamdagammaseries}), the obtained series in powers of $1/\gamma$ are convergent for $\gamma>\gamma_R$. The resulting final expression for $e_2(\gamma)$, which for the initial terms has the ones of equation~(\ref{eq:e2largefinal}), is convergent under the same condition. In figure~\ref{fig:e2}, the ground-state function $e_2(\gamma)$ is shown together with its analytical forms. We can see that its large-$\gamma$ approximation, which starts with the terms given by equation~(\ref{eq:e2largefinal}), can be well applied only for $\gamma\gtrsim 7$, which is fully supported by the previous analysis of the convergence properties of the series.

\begin{figure}
	\centering
	\includegraphics[width=0.6\textwidth]{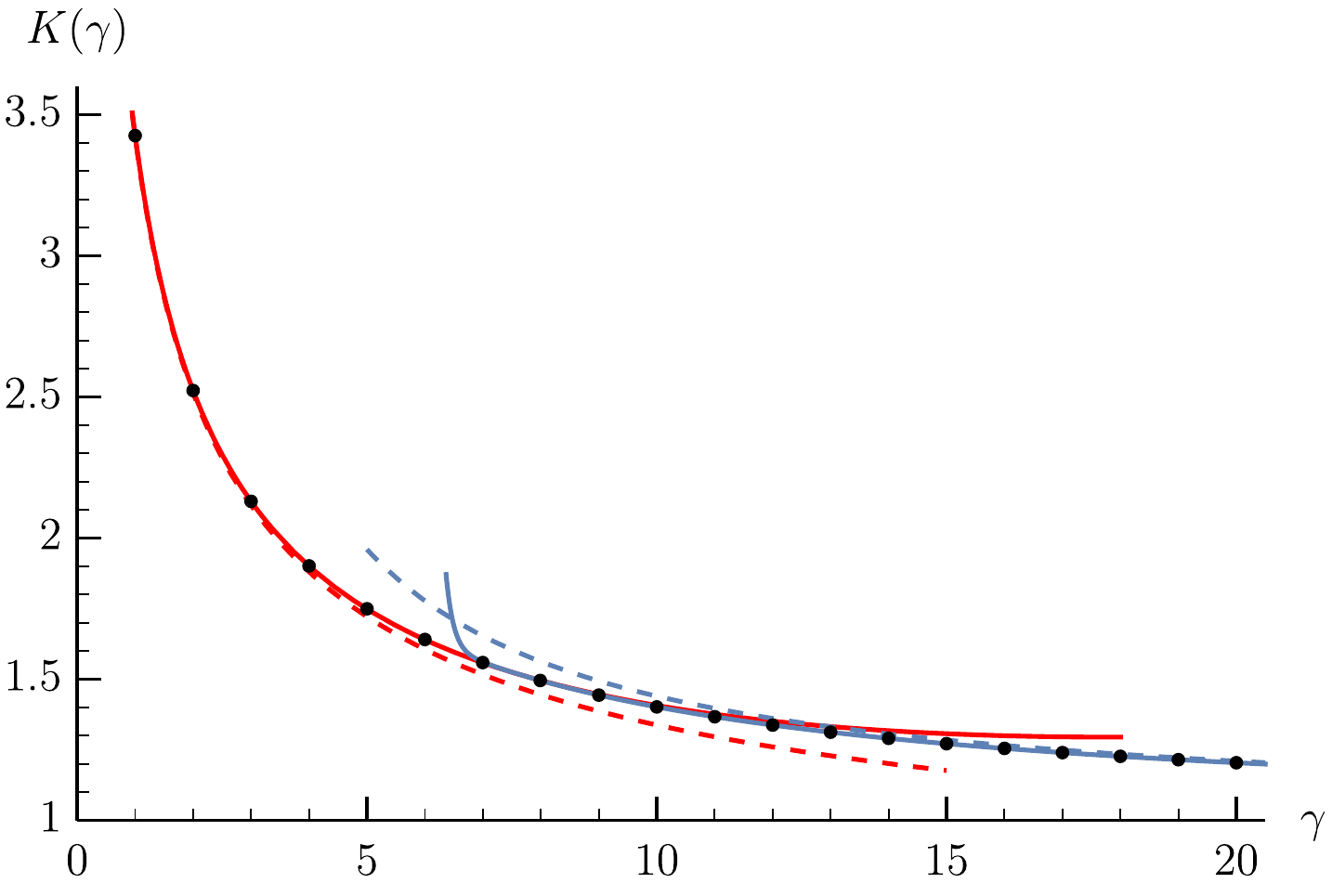}
	\caption{\justifying Plot of the Luttinger liquid parameter $K$ as a function of $\gamma$ that closely parallels the one of figure~\ref{fig:e2}. The dots represent numerically exact values, while the curves are obtained from the analytical results. In the regime of small $\gamma$, the dashed line is drawn using the first three terms and the solid line is drawn using nine terms of the corresponding series. In the regime of large $\gamma$, the dashed line is plotted using the first three terms of the series and the solid line is plotted using fifty one term. For $\gamma=7$, the relative error between the exact result and the low-$\gamma$ series is on the order of $10^{-4}$. For the high-$\gamma$ series, the relative error is on the order of $10^{-3}$.}\label{fig:K}
\end{figure}

We are now in a position to comment on the question from the beginning whether $\gamma=5$ belongs to the regime weak or strong interactions. Looking at figure~\ref{fig:e2}, we see that the ground-state energy at $\gamma=5$ cannot be well described by the series obtained from the expansion at large $\gamma$. On the other hand, accounting for several subleading terms from the expansion at small $\gamma$ it is possible to well describe $e_2(5)$. Therefore, the system at $\gamma=5$ belongs much more to the regime of weak than to the regime of strong interactions. In a related vein, $\gamma=1$ is very well described by the old result (\ref{eq:e23terms}) (relative error on the order of $10^{-3}$), and there is no way to extrapolate there the series for large $\gamma$. We note that $\gamma=1$ is nowadays customary used in the literature to denote the border between the regimes of weak and strong interactions. Based on the previous considerations it would make more sense for the border to use $\gamma=7$ or perhaps $\gamma=\pi^2$, which are around an order of magnitude larger. This is also illustrated in figure~\ref{fig:K}, where the Luttinger liquid parameter (\ref{eq:Kfullexperssion}) is shown.

\section{Method of difference-differential equations \label{section:method}}

In the analysis of the Lieb--Liniger model, the Lieb integral equation (\ref{eq:LIE}) plays the central role. It can be classified as a Fredholm integral equation of the second kind with a difference kernel on a finite interval. This equation does not generally admit a closed-form solution and hence its analysis is quite complicated. In this section we study a family of such equations concentrating on their moments. We find exact relations between the moments in the form of difference-differential equations. The latter results significantly advance the analysis, enabling one to practically determine all the moments from the explicit knowledge of the lowest one. As applications, several examples are considered. First, we study the moments of the quasimomentum distribution in the Lieb--Liniger  model, which represent the ground-state expectation values of the conserved charges and find explicit analytical results. The latter moments determine several basic quantities, e.g., the $N$-body local correlation functions. We prove the equivalence between different expressions found in the literature for the three-body local correlation functions and find an exact result for the four-body local correlation function in terms of the moments of the quasimomentum distributions. We eventually find the analytical results for the three- and four-body correlation functions in the form of power series in the regimes of weak and strong interactions. This section is based on publications \cite{ristivojevic_method_2022,ristivojevic_exact_2023}.

\subsection{General results \label{sec:generalch4}}

Instead of using the full form of the integral operator of equation~(\ref{eq:LIE}), we will use a simplified notation. Let $\mathcal{F}$ be a linear integral operator that acts on a real function $\rho(k,Q)$ of real variables as 
\begin{align}\label{eq:F}
	\mathcal{F}[\rho(k,Q)]=\rho(k,Q)+\frac{1}{2\pi}\int_{-Q}^{Q} dq\;\!\theta'(k-q)\rho(q,Q).
\end{align}
Consider a finite integration limit $Q$ and a kernel $\theta'(k)$ that is an even real function. We want to study the properties of a class of equations
\begin{align}\label{eq:Feq}
	\mathcal{F}[\rho_j(k,Q)]=\frac{k^j}{j!},
\end{align}
where $j\ge 0$ is an integer. Equation (\ref{eq:Feq}) can be classified as a Fredholm integral equation of the second kind with a difference kernel on a finite interval. Without going into the mathematical rigor, we consider continuous  $\theta'(k)$ and assume that equation~(\ref{eq:Feq}) admits a unique non-trivial solution that is a differentiable function. The solution $\rho_j(k,Q)$ of equation~(\ref{eq:Feq}) is an even function of the first argument for even $j$ and odd for odd $j$. It thus satisfies
\begin{align}\label{eq:parity}
	\rho_j(k,Q)=(-1)^j\rho_j(-k,Q).
\end{align}

Solutions of equation~(\ref{eq:Feq}) for different $j$ are not independent. Let us derive some relations among them by applying the derivatives to the operator (\ref{eq:F}). Differentiating equation~(\ref{eq:Feq}) with respect to $k$ and performing the partial integration one obtains
\begin{align}\label{eq:Feqderk}
	\mathcal{F}\left[\frac{\partial\rho_j}{\partial k}\right]={}&\frac{\varrho_j}{2\pi}\left[\theta'(k-Q)-(-1)^{j}\theta'(k+Q)\right] +\frac{ k^{j-1}}{(j-1)!}.
\end{align}
Here, we have employed the parity property (\ref{eq:parity}), introduced the abbreviation 
\begin{align}\label{eq:varrho}
	\varrho_j(Q)=\rho_j(Q,Q),	
\end{align}
and omitted the explicit dependence on the coordinates  where unambiguous. Note that equation~(\ref{eq:Feqderk}) also applies for $j=0$. In this case the last term on the right-hand side is zero, which is also formally correct since $1/(-1!)=0$. Similarly, differentiating equation~(\ref{eq:Feq}) with respect to $Q$ we obtain
\begin{align}\label{eq:FeqderQ}
	\mathcal{F}\left[\frac{\partial\rho_{j-1}}{\partial Q}\right]={}&-\frac{\varrho_{j-1}}{2\pi} [\theta'(k-Q)-(-1)^{j}\theta'(k+Q)]
\end{align}
for $j\ge 1$. From equation~(\ref{eq:FeqderQ}) we directly infer
\begin{align}\label{eq:Fhoder1}
	\frac{1}{\varrho_{j}(Q)}\frac{\partial\rho_{j}(k,Q)}{\partial Q}=\frac{1}{\varrho_{j+2}(Q)}\frac{\partial\rho_{j+2}(k,Q)}{\partial Q}
\end{align}	
for $j\ge 0$. On the other  hand, a linear combination of equations~(\ref{eq:Feqderk}) and (\ref{eq:FeqderQ}) together with equation~(\ref{eq:Feq}) leads to
\begin{align}\label{eq:Fhoder2}
	\frac{1}{\varrho_j(Q)}\frac{\partial\rho_{j}(k,Q)}{\partial k}+\frac{1}{\varrho_{j+1}(Q)}\frac{\partial\rho_{j+1}(k,Q)}{\partial Q}=\frac{ \rho_{j-1}(k,Q)}{\varrho_j(Q)},
\end{align}
which applies for $j\ge 0$. In the derivation of equations~(\ref{eq:Fhoder1}) and (\ref{eq:Fhoder2}) we have used the assumption that $\rho_{-1}(k,Q)=0$ is the only solution of the homogeneous equation $\mathcal{F}[\rho_{-1}(k,Q)]=0$. The latter means that an additional condition on the kernel $\theta'(k)$ might be needed in the most general case. However, it will be fulfilled automatically in our applications, as discussed in Appendix \ref{appendixA}.

\subsubsection{Moments $A_{j,l}$.} 

The central quantities of our  interest are the moments of $\rho_{j}$, which we define by
\begin{align}\label{eq:Ajl}
	A_{j,l}(Q)=\frac{1}{2\;\! l!}\int_{-Q}^{Q} dk \rho_j(k,Q) k^l,\quad j,l\ge 0.
\end{align}	 
They obey the symmetry property with respect to the exchange of indices,
\begin{align}\label{eq:AjlAlj}
	A_{j,l}=A_{l,j},
\end{align} 
which is shown in Appendix \ref{appendixb}. Due to the parity property (\ref{eq:parity}), $A_{j,l}=0$ for odd $j+l$.

The moments (\ref{eq:Ajl}) are not independent and apart from the symmetry (\ref{eq:AjlAlj}), they satisfy a number of other relations. One of them reads
\begin{subequations}\label{eq:Arelation1}
	\begin{align}
		A_{j,j}+A_{j+1,j-1}={}&\varrho_{j} \varrho_{j+1},\quad j=1,2,\ldots,\\
		\label{eq:rho0rho1}
		A_{0,0}={}&\varrho_{0} \varrho_{1}.
	\end{align}
\end{subequations}
It can be derived using the manipulations described in Appendix \ref{appendixb}. Other relations between the moments follow directly from equations~(\ref{eq:Fhoder1}) and (\ref{eq:AjlAlj}),
\begin{align}\label{eq:Arelation2}
	\frac{1}{\varrho_l}\frac{\partial A_{j,l}}{\partial Q}= \frac{1}{\varrho_{l+2}}\frac{\partial A_{j,l+2}}{\partial Q},
\end{align}
and equations~(\ref{eq:Fhoder2}) and (\ref{eq:AjlAlj}),
\begin{align}\label{eq:Arelation3}
	\frac{1}{\varrho_l}\frac{\partial A_{j,l}}{\partial Q}=\frac{A_{j,l}+A_{j-1,l+1}}{\varrho_{l+1}}.
\end{align}
We notice that the combination of equations~(\ref{eq:Arelation2}) and (\ref{eq:Arelation3}) gives a relation that does not involve the derivatives,
\begin{align}\label{eq:Arelation4}
	\frac{A_{j,l-1}+A_{j-1,l}}{\varrho_l}=\frac{A_{j-1,l+2}+A_{j,l+1}}{\varrho_{l+2}}.
\end{align}
Here a negative index should be understood as $A_{-1,l}=A_{l,-1}=0$, which is consistent with equation~(\ref{eq:Ajl}) and $\rho_{-1}=0$.

\subsubsection{Expressions for $A_{j,l}$ in terms of $\varrho_k$.}

Equations (\ref{eq:Arelation1}) and (\ref{eq:Arelation4}) enable us to express the integrals $A_{j,l}(Q)$ defined by equation~(\ref{eq:Ajl}) in terms of $\varrho_k(Q)$, see equation~(\ref{eq:varrho}). Considering the case $j=0$ we find the relation
\begin{align}\label{eq:A0l}
	A_{0,l}=\varrho_0 \varrho_{l+1},\quad l=0,2,4,\ldots.
\end{align}
Equation (\ref{eq:Arelation4}) for $j=1$ leads to
\begin{align}\label{eq:A1l}
	A_{1,l}={}&\varrho_1 \varrho_{l+1}-\varrho_0\varrho_{l+2},\quad l=1,3,5,\ldots.
\end{align}
Equations (\ref{eq:A0l}) and (\ref{eq:A1l}) enable us to reexpress equation~(\ref{eq:Arelation4}) in the form
\begin{align}\label{eq:jAlA}
	A_{j-1,l}+A_{j,l-1}=\varrho_j \varrho_l,\quad j+l \textrm{ odd}.	
\end{align}
This recurrence equation can be solved \cite{mickens2015difference}. We find
\begin{align}\label{eq:Ajlrr}
	A_{j,l}=\sum_{k=0}^{j}(-1)^{j+k}  \varrho_k\, \varrho_{j+l+1-k}	,\quad j+l \textrm{ even}.
\end{align}		
Equation (\ref{eq:Ajlrr}) contains an explicit expression for $A_{j,l}$ defined by equation~(\ref{eq:Ajl}) in terms of a sum of pairwise products of $\varrho_j$ functions, see equation~(\ref{eq:varrho}). Instead of evaluating the integral of the solution of an integral equation, for some applications it might be advantageous to solve several integral equations and evaluate the solutions at a single point according to equation~(\ref{eq:Ajlrr}).

Using equation~(\ref{eq:jAlA}), the differential equation (\ref{eq:Arelation3}) becomes
\begin{align}\label{eq:dAdQ}
	\frac{\partial A_{j,l}}{\partial Q}=\varrho_j \varrho_l,\quad j+l \textrm{ even}.
\end{align}
Instead of the derivatives with respect to $Q$, it is convenient to change the variables and consider the derivatives with respect to $n=\int_{-Q}^Q dk \rho_0(k,Q)/2\pi$, which in physical applications is proportional to the density of particles. From the definition (\ref{eq:Ajl}) it then follows $n=A_{0,0}/\pi$ and thus equation~(\ref{eq:dAdQ}) gives $\partial n/\partial Q=\varrho_0^2/\pi$. Therefore, equation~(\ref{eq:dAdQ}) eventually becomes
\begin{align}\label{eq:dAdn}
	\frac{\varrho_0^2}{\pi}\frac{\partial A_{j,l}}{\partial n}=\varrho_j \varrho_l,\quad j+l \textrm{ even}.
\end{align}

\subsubsection{Expressions for $A_{j,l}$ in terms of $A_{0,2k}$ and its derivative.}

Equations (\ref{eq:A0l}) and (\ref{eq:dAdn}) enable us to write
\begin{align}\label{eq:Fodd}
	&\varrho_{2l+1}=\frac{1}{\varrho_0}A_{0,2l},\\
	\label{eq:Fodd1}
	&\varrho_{2l}=\frac{\varrho_0}{\pi}\frac{\partial A_{0,2l}}{\partial n},\quad l=0,1,2,\ldots.
\end{align}
Substituting them into equation~(\ref{eq:Ajlrr}), we find
	\begin{subequations}\label{eq:Ajlsol}
		\begin{gather}\label{eq:Ajlsol(a)}
			A_{2j,2l}=\frac{1}{\pi} \frac{\partial A_{0,2j}}{\partial n} A_{0,2l} +\frac{1}{\pi}\sum_{k=0}^{j-1}\left(\frac{\partial A_{0,2k}}{\partial n} A_{0,2j+2l-2k}-A_{0,2k}\frac{\partial A_{0,2j+2l-2k}}{\partial n}\right),\\ \label{eq:Ajlsol(b)}
			A_{2j+1,2l+1}=\frac{1}{\pi}\sum_{k=0}^{j}\left( A_{0,2k} \frac{\partial A_{0,2j+2l+2-2k}}{\partial n} -\frac{\partial A_{0,2k}}{\partial n} A_{0,2j+2l+2-2k}\right).
		\end{gather} 
	\end{subequations}	
Remarkably, the whole class of integrals (\ref{eq:Ajl}) can be expressed only in terms of $A_{0,2k}$ and its derivative. In other words, the moments of $\rho_0(k,Q)$ determine the moments of all other functions $\rho_j(k,Q)$ defined by equation~(\ref{eq:Feq}). Our ultimate goal is therefore to study the even moments of $\rho_0(k,Q)$, i.e., $A_{0,2l}$ since $A_{0,2l+1}=0$ due to the parity.

\subsubsection{Connection between $A_{0,2l+2}$ and $A_{0,2l}$.}

Equation (\ref{eq:Ajlsol(b)}) at $j=0$ becomes
\begin{align}\label{eq:polaron}
	A_{1,2l+1}=n \frac{\partial A_{0,2l+2}}{\partial n}-A_{0,2l+2}.
\end{align}
Acting by the derivative $\partial/\partial n$ to equation~(\ref{eq:polaron}) and using equations~(\ref{eq:dAdn}) and (\ref{eq:Fodd}), one obtains 
\begin{align}\label{eq:main}
	\frac{\partial^2 A_{0,2l+2}}{\partial n^2}=\pi^2\frac{A_{0,2l}}{\varrho_0^4}.
\end{align} 
Equation (\ref{eq:main}) is another remarkable result. It shows that different moments of $\rho_0$ are actually not independent, but obey the difference-differential equation. In the special case $l=0$, equation~(\ref{eq:main}) leads to
\begin{align}\label{eq:aux}
	\frac{\partial^2 A_{0,2}}{\partial n^2}=\frac{\pi^3 n}{\varrho_0^4},
\end{align}
which is a connection between $\varrho_0$  and the second derivative of $A_{0,2}$. One can eventually eliminate $\varrho_0$ from equation~(\ref{eq:main}) using equation~(\ref{eq:aux}), getting an expression that only involves the moments.

\subsection{The moments of the quasimomentum distribution \label{sec:LL}}

The results of Sec.~\ref{sec:generalch4} are general and go beyond any physical application. They have been derived under the minimal assumptions on the kernel in the integral operator (\ref{eq:F}). Let us apply them to the Lieb--Liniger model. In this case the kernel is given by equation~(\ref{eq:kernel}) and the quasimomentum distribution satisfies equation~(\ref{eq:LIE}), which can be expressed as
\begin{align}\label{eq:LIEF}
	\mathcal{F}[\rho(k,Q)]=\frac{1}{2\pi}.
\end{align}
In Appendix \ref{appendixA} we show that the integral equation (\ref{eq:LIEF}) for the kernel (\ref{eq:kernel}) obeys necessary conditions in order to apply the formalism of Sec.~\ref{sec:generalch4}.

Consider the family of dimensionless functions
\begin{align}\label{eq:e2l}
	e_{2l}(\gamma)=\frac{1}{n^{2l+1}} \int_{-Q}^{Q} dk k^{2l}\rho(k,Q),
\end{align}
where $l$ is a nonnegative integer. The right-hand side of equation~(\ref{eq:e2l}) is formally a function of the Fermi quasimomentum $Q$, while on the left-hand side we wrote the dependence on $\gamma$. This is possible since $Q/n$ can be expressed only in terms of $\gamma$. The functions $e_{2l}(\gamma)$ are proportional to the even moments of the quasimomentum distribution. We notice that the odd moments vanish. Omitting the trivial proportionality factor, $e_{2l}(\gamma)$ will be loosely called the moments in the following. Up to the proportionality factor, they are the ground-state expectation values of the conserved charges of the model \cite{davies_higher_1990}. Previously we studied the ground-state function $e_2(\gamma)$ that determines the ground-state energy.

The family of functions (\ref{eq:e2l}) is directly related to the quantities defined by equation~(\ref{eq:Ajl}) by
\begin{align}\label{eq:e2ldefinition}
	A_{0,2l}=\frac{\pi}{(2l)!}{n^{2l+1}}{e_{2l}(\gamma)}.
\end{align}
Here we used the connection $\rho_0=2\pi \rho$ that arises from the comparison of equations~(\ref{eq:Feq}) and (\ref{eq:LIEF}). Therefore, various identities that we previously derived for $A_{0,2l}$ translate into a new set of identities among $e_{2l}$ functions. Consider equation~(\ref{eq:main}). Let us first transform the differentiation with respect to $n$ into the one with respect to $\gamma$, where we should use the rules
\begin{align}\label{eq:derivatives}
	\frac{\partial}{\partial n}=-\frac{\gamma}{n}\frac{\partial}{\partial \gamma}, \quad \frac{\partial^2}{\partial n^2}=\frac{2\gamma}{n^2} \frac{\partial}{\partial \gamma}+\frac{\gamma^2}{n^2}\frac{\partial^2}{\partial \gamma^2}.
\end{align}
From equation~(\ref{eq:main}) we then directly obtain \cite{ristivojevic_method_2022,ristivojevic_exact_2023}
\begin{align}\label{eq:diff-diffeq}
	\gamma^{2l+4}	\frac{d^2}{d\gamma^2} \left(\frac{e_{2l+2}(\gamma)}{\gamma^{2l+2}}\right)= \frac{(2l+1)(2l+2)}{16\pi^2}\frac{e_{2l}(\gamma)}{\rho(Q,Q)^4}.
\end{align}
Equation~(\ref{eq:diff-diffeq})  is an exact relation between the moments (\ref{eq:e2l}). It has a form of the difference-differential equation.

In the special case $l=0$, equation~(\ref{eq:diff-diffeq}) becomes
\begin{align}\label{eq:defK}
	\gamma^{4}	\frac{d^2}{d\gamma^2} \left(\frac{e_{2}(\gamma)}{\gamma^{2}}\right)= \frac{1}{8\pi^2\rho(Q,Q)^4},	
\end{align}
where we have used $e_0=1$ obtained from the definition (\ref{eq:e2l}). Using the expression for the sound velocity given by equation~(\ref{eq:soundvelocityexplicit}), equation~(\ref{eq:defK}) becomes equivalent to
\begin{align}\label{eq:mvK1}
	\frac{mv}{\hbar n}=\frac{1}{4\pi\rho(Q,Q)^2}.
\end{align} 
At this point we need the information from the microscopic theory that the Luttinger liquid parameter $K$ is defined by the relation \cite{haldane_effective_1981,korepin}
\begin{align}\label{eq:Kll}
	2\pi\rho(Q,Q)=\sqrt{K}.
\end{align}
This translates equation~(\ref{eq:mvK1}) into
\begin{align}\label{eq:mvK}
	mvK=\pi\hbar n.
\end{align} 
We have thus derived the relation between the sound velocity $v$ and $K$ in the Lieb--Liniger model. However, the relation (\ref{eq:mvK}) is more general and applies to all Galilean-invariant models \cite{haldane_effective_1981}. Note that in section \ref{section1} we defined $K$ by the relation (\ref{eq:mvK}). If we adopt this reasoning, then equation~(\ref{eq:mvK1}) leads to the connection (\ref{eq:Kll}).

Eliminating $\rho(Q,Q)$ from equation~(\ref{eq:diff-diffeq}) using equation~(\ref{eq:defK}), we obtain
\begin{align}\label{eq:e2l1}
	\frac{d^2}{d\gamma^2} \left(\frac{e_{2l+2}(\gamma)}{\gamma^{2l+2}}\right)= (l+1)(2l+1) \frac{d^2}{d\gamma^2} \left(\frac{e_{2}(\gamma)}{\gamma^{2}}\right) \frac{e_{2l}(\gamma)}{\gamma^{2l}}.
\end{align}
For $l=0$, equation~(\ref{eq:e2l1}) reduces to an identity, while for $l>0$ it gives the connections between the consecutive terms of the family (\ref{eq:e2l}). Equation~(\ref{eq:e2l1}) is our starting point for the evaluation of  $e_{2l}(\gamma)$ for $l>1$ using the knowledge of $e_2(\gamma)$, which serves as an initial value of the family $e_{2l}(\gamma)$ that generates $l>1$ terms. Since we calculated $e_2(\gamma)$ analytically in terms of the power series in the regimes of weak and strong interactions, see equations~(\ref{eq:e2nineterms}) and (\ref{eq:e2largefinal}), we will be able to evaluate $e_{2l}(\gamma)$ in the two regimes.

Haldane \cite{haldane_effective_1981} noticed that the Luttinger liquid parameter $K$ in the Lieb--Liniger model can be expressed only in terms of the single point value of the density of quasimomenta $\rho(Q,Q)=\varrho_{0}(Q)/2\pi$, see equation~(\ref{eq:Kll}). From our analysis performed in Sec.~\ref{sec:generalch4}, it follows that all the moments of the quasimomentum distribution and their derivatives can be expressed in terms of the related quantities $\varrho_j(Q)$ defined by equation~(\ref{eq:varrho}). Indeed, using equations~(\ref{eq:Fodd}), (\ref{eq:Fodd1}), and (\ref{eq:e2ldefinition}) we obtain
\begin{align}
	e_{2l}(\gamma)={}&\frac{(2l)!\sqrt{K}}{\pi}\frac{\varrho_{2l+1}(Q)}{n^{2l+1}},\\
	\gamma \frac{d 	e_{2l}(\gamma)}{d\gamma}={}&(2l+1)e_{2l}(\gamma)- \frac{(2l)!}{\sqrt{K}}\frac{\varrho_{2l}(Q)}{n^{2l}}.
\end{align}
Here in the right-hand sides one should eventually express $Q$ in terms of $\gamma$ [see equation~(\ref{eq:Qfin}) below], which will cancel the powers of $n$ in the denominators.

\subsubsection{Weak interactions.}

In the regime of weak interactions, $\gamma\ll 1$, the leading-order solution of equation~(\ref{eq:LIE}) is $\rho(k,Q)=\sqrt{Q^2-k^2}/2\pi c$ \cite{lieb_exact_1963}. This yields the order of magnitude estimate for the leading-order term in equation~(\ref{eq:e2l}),
\begin{align}
	e_{2l}(\gamma)\sim \frac{1}{\gamma}\left(\frac{Q}{n}\right)^{2l+2}.
\end{align} 
Using $e_0=1$, we find $Q\sim n\sqrt{\gamma}$ and thus $e_{2l}(\gamma)\sim \gamma^l$. Since the subsequent terms in the expansion of $e_2$ are multiplied by  $\sqrt{\gamma}$, we should assume the series
\begin{align}\label{eq:e2lass}
	e_{2l}(\gamma)=\sum_{j=0}^{\infty} a_j^{(2l)}\gamma^{l+j/2},
\end{align}
where the values of the coefficients $a_j^{(2l)}$ for $l>1$ will be calculated using the known values of $a_j^{(2)}$. Substitution of the form (\ref{eq:e2lass}) into equation~(\ref{eq:e2l1}) yields the connection between the coefficients $a_{k}^{(2l+2)}$ from the left-hand side of equation~(\ref{eq:e2l1}) and the ones from the right-hand side,
\begin{align}\label{eq:akoff}
	&\left(2l+2-k\right) \left(2l+4-k\right)	a_{k}^{(2l+2)} \notag\\
	&=(l+1)(2l+1) \sum_{j=0}^{k} (j-2)(j-4) a_{j}^{(2)} a_{k-j}^{(2l)}.
\end{align}
For $l=0$, equation~(\ref{eq:akoff}) becomes trivial since $a_{k-j}^{(0)}=\delta_{k,j}$, while for $l>1$ it enables us to evaluate the coefficients in the series (\ref{eq:e2lass}) for $e_{2l}$ using the ones of $e_2$.

\begin{table}
	\caption{Values of the coefficients in the series (\ref{eq:e2lass}) evaluated from equations~(\ref{eq:a0a1a2a3}) using the known values of $a_k^{(2)}$. \label{table}}
\centering
\begin{tabular}{l| l l l  l}
\hline
			$a_k^{(2l)}$ &$k=0$&$k=1$ &$k=2$&$k=3$ \rule{0pt}{3.1ex}\\ \hline
			$l=1$ & $1$ & $-\frac{4}{3\pi}$& $\frac{1}{6}-\frac{1}{\pi^2}$\rule[-1.3ex]{0pt}{1ex} & $-\frac{1}{2\pi^3}+\frac{3\zeta(3)}{8\pi^3}$\rule{0pt}{2.8ex}\\
			$l=2$ & $2$ &$-\frac{88}{15\pi}$ & $1-\frac{2}{\pi^2}$ &$-\frac{4}{3\pi}+\frac{1}{\pi^3}+\frac{21\zeta(3)}{4\pi^3}$\rule[-1.3ex]{0pt}{1ex} \\
			$l=3$& $5$ & $-\frac{824}{35\pi}$ & $5+\frac{14}{3\pi^2}$ & $-\frac{44}{3\pi}+\frac{17}{\pi^3}+\frac{165\zeta(3)}{4\pi^3}$\rule[-1.3ex]{0pt}{1ex} \\
			$l=4$ & $14$ & $-\frac{29168}{315\pi}$& $\frac{70}{3}+\frac{3452}{45\pi^2}$ & $-\frac{1648}{15\pi}+\frac{1438}{15\pi^3}+\frac{525\zeta(3)}{2\pi^3}$
\rule[-1.3ex]{0pt}{1ex}\\
\hline
		\end{tabular}
\end{table}

For a fixed $k$, equation~(\ref{eq:akoff}) can be explicitly solved since it is equivalent to a first-order linear difference equation \cite{mickens2015difference}. Using the known values of $a_j^{(2)}$ listed in table \ref{table} that originate from equation~(\ref{eq:e2nineterms}), we obtain
\begin{subequations}\label{eq:a0a1a2a3}
	\begin{align}
		a_{0}^{(2l)}={}&\frac{(2l)!}{l\;\! !(l+1)!},\\
		a_{1}^{(2l)}={}&-\frac{16^l}{2\pi}\frac{(l\;\! !)^2 }{(2l+1)!}\sum_{w=0}^{l-1}\frac{(2w+1)!}{16^w (w\;\! !)^2}a_0^{(2w)},\\
		a_{2}^{(2l)}={}& \frac{(2l)!}{2(l-1)! l\;\! ! } \left[\frac{1}{6}-\frac{1}{\pi^2} -\frac{1}{\pi} \sum_{w=1}^{l-1} \frac{(w-1)! w\;\! ! \;\!  a_{1}^{(2w)}}{(2w)!}\right],\\
		a_{3}^{(2l)}={}& \frac{16^l(2l-1)(l!)^2}{64\pi^3 (2l)!} \sum_{w=0}^{l-1} \frac{(2w)!\left[(4-3\zeta(3))a_0^{(2w)}-32\pi^2 a_{2}^{(2w)}\right]}{16^w (2w-1)(w\;\! !)^2}.
	\end{align}
\end{subequations}
Equations (\ref{eq:a0a1a2a3}) determine the first four coefficients in equation~(\ref{eq:e2lass}) for all the moments $e_{2l}(\gamma)$ of the quasimomentum distribution (\ref{eq:e2l}) in the regime of weak interactions. This remarkable result has its roots in the integrability of the Lieb--Liniger model and is one application of the formalism previously derived in Sec.~\ref{sec:generalch4}. In table \ref{table} we give the analytical values for  $a_{k}^{(2l)}$ for $1\le l\le 4$. A motivated reader can easily generate the coefficients for higher values of $l$ using equations~(\ref{eq:a0a1a2a3}).

The cases $k=2l+2$ and $k=2l+4$ are special for equation~(\ref{eq:akoff}) since the left-hand side then nullifies. Therefore the coefficients $a_{2l+2}^{(2l+2)}$ and $a_{2l+4}^{(2l+2)}$ cannot be immediately recursively expressed though the right-hand side of equation~(\ref{eq:akoff}). However, at $k=2l+2$ the right-hand side constitutes a new relation enabling one to express the latter missing coefficient,
\begin{align}\label{eq:akoff1zzz}
	a_{2l+2}^{(2l)} = -\frac{1}{8a_0^{(2)}} \sum_{j=1}^{2l+2} (j-2)(j-4) a_j^{(2)} a_{2l+2-j}^{(2l)}.
\end{align}
For $k=2l+4$, equation~(\ref{eq:akoff}) gives another relation between the coefficients $a_j^{(2l)}$ that however does not involve $a_{2l}^{(2l)}$ and $a_{2l+2}^{(2l)}$. Nevertheless in this way one obtains a nontrivial relation among the other coefficients of the two series for $e_2$ and $e_{2l}$ that will be discussed further below. 

We have not found a way to calculate $a_{2l}^{(2l)}$  from the difference-differential equation (\ref{eq:e2l1}). On the practical side, by increasing $l$ in the series (\ref{eq:e2lass}), $a_{2l}^{(2l)}$ becomes progressively less important since it only determines the $2l$-th correction term of the series representation for $e_{2l}$. Theoretically, one can extend the developed methods for $e_2$ \cite{marino_exact_2019} to the case-by-case study of $e_4$, $e_6$, etc., in order to obtain $a_{2l}^{(2l)}$. We performed this involved work \cite{reichert_analytical_2020} (see also section \ref{section:capacitance}) and found
\begin{subequations}\label{eq:a44a66}
	\begin{align}
		a_{4}^{(4)}={}&\frac{3}{20}-\frac{1}{\pi^2}-\frac{21\zeta(3)-10}{6\pi^4},\\
		a_{6}^{(6)}={}&\frac{61}{168}-\frac{9}{4\pi^2}-\frac{5(21\zeta(3)-10)}{12\pi^4}- \frac{3}{5120\pi^6}\bigl(2048\notag\\
		&+15460\zeta(3)-43050\zeta(3)^2+122505\zeta(5)\bigr).
	\end{align}
\end{subequations}
We were able to show that once the value of $a_{4}^{(4)}$ is known, one can find the value of $a_{2}^{(2)}$. Similarly, the value of $a_{6}^{(6)}$ suffices to find $a_{4}^{(4)}$ and $a_{2}^{(2)}$.

The coefficients $a_{2l}^{(2l)}$ obviously become progressively more complicated as $l$ is increased. An interested reader can use the coefficients $a_{4}^{(4)}$ and $a_{6}^{(6)}$ and the ones of $e_2$ from equation~(\ref{eq:e2nineterms}) to easily extend the values listed in table~\ref{table} to $k\le 7$ for arbitrary $l$ by iterating equations~(\ref{eq:akoff}) and (\ref{eq:akoff1zzz}). On the other hand for $l=2$ and $l=3$ (i.e., for the evaluation of $e_4$ and $e_6$) there is no intrinsic limitation on $k$, the only one being the knowledge of $e_2$.

\subsubsection{Structure of the series for the ground-state function.}

Interestingly, the coefficients in the series for the ground-state function $e_2(\gamma)$ at weak interactions are not independent. Starting from equation~(\ref{eq:e2l1}), for the series of the form (\ref{eq:e2lass}) we have derived the relation (\ref{eq:akoff}). The latter at $k=2l+2$ leads to 
\begin{align}\label{eq:akoff1}
	\sum_{j=0}^{2l+2} (j-2)(j-4) a_j a_{2l+2-j}^{(2l)}=0.
\end{align}
Here and in the following we introduced the simplified notation by suppressing the superscript from the coefficients entering $e_2$, i.e., we use $a_j\equiv a_j^{(2)}$. For $l=1$, equation~(\ref{eq:akoff1}) reduces to	
\begin{align}\label{eq4}
	a_4+\frac{a_1a_3}{4a_0}=0.	
\end{align}	 
In the case $k=2l+4$, equation~(\ref{eq:akoff}) gives another constraint,
\begin{align}\label{eq:akoff2}
	\sum_{j=0}^{2l+4} (j-2)(j-4) a_j a_{2l+4-j}^{(2l)}=0.
\end{align}
The latter at $l=1$, leads to the second relation among the coefficients in $e_2$, 
\begin{align}\label{eq6}
	a_6+\frac{3a_1 a_5}{8 a_0}-\frac{(a_3)^2}{16 a_0}=0.
\end{align}
Both derived relations (\ref{eq4}) and (\ref{eq6}) are indeed satisfied once the numerical values of the coefficients of the series (\ref{eq:e2nineterms}) are used.

The constraints (\ref{eq:akoff1}) and (\ref{eq:akoff2}) at $l>1$ in combination with equation~(\ref{eq:akoff}) lead to infinitely many relations among the coefficients entering the series for $e_2$. Let us illustrate how to obtain the third one. Substituting $l=2$ into equation~(\ref{eq:akoff2}) gives
\begin{align}\label{eq8'}
	8 a_0 a_8^{(4)}+3a_1 a_7^{(4)}-a_3 a_5^{(4)}+3a_5 a_3^{(4)}+8 a_6 a_2^{(4)}+15 a_7 a_1^{(4)}+24 a_8a_0^{(4)}=0.
\end{align}
Equation (\ref{eq:akoff}) at $l=1$ leads to $a_k^{(4)}=\frac{6}{(4-k)(6-k)}\sum_{j=0}^{k} (j-2)(j-4)a_j a_{k-j}$. Substituting the latter into equation~(\ref{eq8'}) leads to a linear combination of terms of the form $a_{j_1}a_{j_2}a_{j_3}$, where $j_1+j_2+j_3=8$. The coefficient $a_2$ in the combination arises with the coefficient $6[16a_0a_6+6a_1a_5-(a_3)^2]$, which is zero due to  equation~(\ref{eq6}). The remainder then gives 
\begin{align}\label{eq8}
	a_8+ \frac{13  a_1 a_7}{10a_0}+\frac{7(a_1)^2 a_6}{20 a_0^2}+\frac{a_3 a_5}{2a_0} +\frac{a_1 a_3 a_4}{20 (a_0)^2}=0.
\end{align} 
One can check that equation~(\ref{eq8}) is indeed satisfied using the numerical values of the coefficients from equation~(\ref{eq:e2nineterms}).

Equations (\ref{eq4}), (\ref{eq6}), and (\ref{eq8}) are the first three relations among the coefficients of the series for $e_2$ obtained from the general considerations based on analytic properties of the integral equation (\ref{eq:LIE}) and its consequence given by equation~(\ref{eq:e2l1}). The obtained sequence of relations can be arbitrarily extended by substituting subsequently the values $l\ge 3$ in equation~(\ref{eq:akoff2}), followed by the repetitive use of equations~(\ref{eq:akoff}) and (\ref{eq:akoff1}). The obtained relations and the subsequent ones among $a_j$'s have several special features. First, the term $a_2$ does not occur in them. Second, when multiplied by a common denominator, the summands of a particular relation have a product form $a_{j_1}a_{j_2}\cdots$ with a constant sum $j_1+j_2+\cdots$. In the relations  (\ref{eq4}), (\ref{eq6}), and (\ref{eq8}), this sum is, respectively, equal to $4$, $6$, and $8$. The second feature follows directly from equation~(\ref{eq:akoff}). The third feature is the possibility to express the coefficients with an even index $a_{2j}$ in terms of the coefficients with odd indices $a_1$, $a_3$,\ldots $a_{2j-1}$ and $a_0$. This is obvious for equations~(\ref{eq4}) and (\ref{eq6}).
For the case of equation~(\ref{eq8}), expressing $a_4$ and $a_6$ obtained from equations~(\ref{eq4}) and (\ref{eq6}), we obtain
\begin{align}\label{eqlast}
	a_8=-\frac{13 a_1a_7}{10a_0}-\frac{a_3a_5}{2a_0}+\frac{21 (a_1)^3 a_5}{160(a_0)^3}-\frac{3(a_1)^2(a_3)^2}{320(a_0)^3}.
\end{align}
Equation (\ref{eqlast}) is an expression for $a_8$ in terms of the coefficients with odd indices and $a_0$, which is actually $a_0=1$. Along the same lines, one can obtain further relations corresponding to $l\ge 3$.  Speaking in mathematical terms, we have reduced the complicated problem of the series solution for $e_2$ at weak interactions to the problem of finding the coefficients of the series with odd indices.

\subsubsection{Strong interactions.}

In the regime of strong interactions, $\gamma\gg 1$, the integral in the integral operator of equation~(\ref{eq:LIE}) is subdominant. This directly leads to $\rho(k,Q)=1/2\pi$ at the leading order, and thus $e_{2l}(\gamma)\sim 1$. Since the subsequent terms in $\rho(k,Q)$ are by a factor of $1/\gamma$ smaller, the resulting series for its moments can be assumed in the form
\begin{align}\label{eq:e2gass}
	e_{2l}(\gamma)=\sum_{j=0}^{\infty} b_j^{(2l)}\gamma^{-j}.
\end{align}
Substituting equation~(\ref{eq:e2gass}) into the expression (\ref{eq:e2l1}) we find an equation
\begin{align}\label{eq:diffg}
	b_k^{(2l+2)}={}&\frac{(l+1)(2l+1)}{(2l+2+k)(2l+3+k)}\sum_{j=0}^{k}(2+j)(3+j)b_j^{(2)} b_{k-j}^{(2l)}
\end{align}
that relates the coefficients of equation~(\ref{eq:e2gass}). Expression (\ref{eq:diffg}) is a difference equation that has a similar structure as equation~(\ref{eq:akoff}), and thus it can be solved for $l>1$. The first five terms are given by
\begin{subequations}
	\label{eeq1}
	\begin{align}
		b_0^{(2l)}={}&\frac{\pi^{2l}}{2l+1},\quad b_1^{(2l)}=-\frac{4l\;\!\pi^{2l}}{2l+1},\quad
		b_2^{(2l)}=4l\pi^{2l},\\ 
		b_3^{(2l)}={}&-\frac{16l(l+1)\pi^{2l}}{3}\biggl[1-\frac{\pi^2}{(2l+1)(2l+3)} \biggr],\\
		b_4^{(2l)}={}&\frac{8l(l+1)(2l+3)\pi^{2l}}{3}\biggl[1-\frac{4\pi^2}{(2l+1)(2l+3)}\biggr].\!
	\end{align}
\end{subequations}
Here we have used the known values of $b_j^{(2)}$ entering $e_2$, see equation~(\ref{eq:e2largefinal}). They can be recovered from equations~(\ref{eeq1}) setting $l=1$. We note that at strong interactions, the knowledge of $e_2$ suffices to find all other momenta using equation~(\ref{eq:diffg}) due to the physical reason of not having divergent moments $e_{2l}$ at $\gamma\to\infty$. This should be contrasted with the regime of weak interactions where in addition to $e_2$ one also needs the ``diagonal'' coefficients $a_{2l}^{(2l)}$ for $l>1$ in order to evaluate $a_{k}^{(2l)}$ at $k\ge 4$.

\subsection{The Fermi quasimomentum}

Let us find an expression for the Fermi quasimomentum $Q$ in terms of $\gamma$. Our starting point is
\begin{align}\label{eq:Qn}
	\frac{\partial Q}{\partial n} = \frac{1}{4\pi \rho(Q,Q)^2},
\end{align}
which is the special case of  equation~(\ref{eq:dAdQ}). By making use of equation~(\ref{eq:defK}), the Fermi quasimomentum can be expressed as
\begin{align}\label{eq:Qfin}
	Q={}&2n\sqrt\gamma\, g(\gamma),
\end{align}
where $g(\gamma)$ satisfies a differential equation
\begin{align}\label{eq:gdiff}
	\frac{d}{d\gamma}\left(\frac{g(\gamma)}{\sqrt\gamma}\right) = -\sqrt{\frac{1}{8}  \frac{d^2}{d\gamma^2} \left(\frac{e_{2}(\gamma)}{\gamma^{2}}\right)}.
\end{align}
We note that an alternative form of the right-hand side is expressed in terms of the Luttinger liquid parameter as $-\frac{\pi}{2\gamma^2 K}$. Therefore, the nontrivial dependence in $Q$ is encoded into the differential equation (\ref{eq:gdiff}), which we solve now.

In the regime of weak interactions, $\gamma\ll 1$, using the result (\ref{eq:e2nineterms}) for $e_2$ we find
\begin{align}	\label{eq:gsolutionweak}
	g(\gamma)={}&1-\frac{\sqrt\gamma}{4\pi} \left(\ln\frac{32\pi}{\sqrt\gamma}-1 \right) +\frac{\gamma}{32\pi^2}+\frac{3(\zeta(3)-1)}{256\pi^3}\gamma^{3/2} +\mathcal{O}(\gamma^2).
\end{align}
The integration constant of the first-order equation (\ref{eq:gdiff}) is the constant term proportional to $\sqrt{\gamma}$ in equation~(\ref{eq:gsolutionweak}). Its value is set using the known perturbative solution \cite{popov_theory_1977} of equation~(\ref{eq:LIE}) that is
\begin{align}\label{eq:rhoPopov}
	\rho(k,Q)=\frac{\sqrt{Q^2-k^2}}{2\pi c}+\frac{Q\left(1+\ln\frac{16\pi Q}{c}\right)-k\ln\frac{Q+k}{Q-k}}{4\pi^2\sqrt{Q^2-k^2}}+\mathcal{O}(c).
\end{align}
Equation (\ref{eq:rhoPopov}) holds in the region away from the edges, at $k$ that satisfies $Q^2-k^2\gg Qc$. Nevertheless, this limitation does not affect the determination of the first subleading term in the expression for $Q$ (i.e., the integration constant) as one can integrate $\rho(k,Q)$ from $-Q$ to $Q$ in equation~(\ref{eq:n}) and then find $Q$. We should note that the function of equation~(\ref{eq:gsolutionweak}) and thus $Q/n$ has only one logarithmic term unlike the inverse relation where the same logarithm proliferates. 

In the regime of strong interactions, $\gamma\gg 1$, the situation is simpler since the integration constant for equation~(\ref{eq:gdiff}) must be set to zero due to the physical reason of not having divergent $Q\propto \gamma$. We find
\begin{align}
	g(\gamma)=\frac{\pi}{2\sqrt\gamma}\Biggl(1-\frac{2}{\gamma} +\frac{4}{\gamma^2}+\frac{\frac{4\pi^2}{3}-8}{\gamma^3} + \mathcal{O}\left( {\gamma^{-4}}\right) \Biggr).
\end{align}
Substituting this into equation~(\ref{eq:Qfin}) we find that $Q=\pi n$ at $\gamma\to\infty$, which is identical to the Fermi momentum in a system of free fermions.

\subsection{Local correlation functions}

A local $N$-body correlation function is defined as the ground-state expectation value 
\begin{align}\label{eq:gj}
	g_N(\gamma)=\frac{1}{n^N}\Bigl\langle \Psi^\dagger(x)^N\Psi(x)^N\Bigr\rangle
\end{align}
of the  Bose field operators $\Psi^\dagger$ and $\Psi$, which satisfy the canonical commutation relation $[\Psi(x),\Psi^\dagger(y)]=\delta(x-y)$. The result for the particular case $N=2$ can be easily obtained by applying the Feynman--Hellmann theorem to the Hamiltonian of the Lieb--Liniger model, leading to \cite{gangardt_stability_2003} 
\begin{align}\label{eq:g2}
	g_2(\gamma)=\frac{de_2(\gamma)}{d\gamma}. 
\end{align}
In the case of an arbitrary integer $N$, the exact evaluation of the average value in equation~(\ref{eq:gj}) is significantly more difficult \cite{pozsgay_local_2011,bastianello_exact_2018}. The final result of reference~\cite{pozsgay_local_2011} is expressed as an integral representation
\begin{align}\label{eq:gNexact}
	g_N(\gamma)=\frac{(N!)^2}{(2\pi n)^N}\int_{-Q}^Q dq_1\ldots dq_N \prod_{1\le l<j\le N} \frac{q_j-q_l}{(q_j-q_l)^2+c^2} \prod_{j=1}^{N}(j-1)!\rho_{j-1}(q_j).
\end{align}
Here $\rho_{j}$ satisfies equation~(\ref{eq:Feq}), where the kernel in the integral operator (\ref{eq:F}) is given by equation~(\ref{eq:kernel}). In the case $N=2$, equation~(\ref{eq:gNexact}) reduces to equation~(\ref{eq:g2}). However, the treatment of equation~(\ref{eq:gNexact}) for $N>2$ is an involved task. Below we consider the cases $N=3$ and $N=4$.

\subsubsection{The three-body case.}

For $N=3$, we can split the product over the two indices in equation~(\ref{eq:gNexact}) into a sum that involves six permutations of $q_1$, $q_2$, and $q_3$, which can then be treated term by term. In this way one can obtain the final expression in the form \cite{pozsgay_local_2011}
\begin{align}\label{eq:g3exact2}
	g_3(\gamma)={}&\frac{12}{\pi n^5\gamma^2}\left(-2A_{3,1}+A_{2,2}+2A_{4,0}\right)+\frac{1}{\pi n^3}(2A_{2,0}-A_{1,1})-\frac{2}{\pi^2 n^4\gamma}A_{0,0}A_{1,1},
\end{align}
where $A_{j,l}$ is defined by equation~(\ref{eq:Ajl})\footnote{\linespread{1}\selectfont Equation (\ref{eq:g3exact2}) corresponds to equation~(7.10) of reference~\cite{pozsgay_local_2011} and to equation~(7) of reference~\cite{kormos_exact_2011} derived in a complementary way. In equation~(\ref{eq:g3exact2}) we omitted the term $A_{1,0}$ that nullifies in the ground state. We notice that the object $\{j,l\}$ used in reference~\cite{pozsgay_local_2011} is equal to $j!l! A_{l,j}/\pi$ in our notation.}. Equation (\ref{eq:g3exact2}) is expressed in terms of various moments of $\rho_j$ and thus it can be further transformed to a more convenient form that only involves the moments of $\rho_0$. Using equations~(\ref{eq:Ajlsol}) in the expression (\ref{eq:g3exact2}) we obtain
\begin{align}\label{eq:g3exact3}
	g_3(\gamma)={}&\frac{12}{\pi n^5 \gamma^2} \left(-3n\frac{\partial A_{0,4}}{\partial n}+5A_{0,4}+\frac{A_{0,2}}{\pi} \frac{\partial A_{0,2}}{\partial n}\right)+\frac{2+\gamma}{\pi\gamma n^3}\left(A_{0,2}-n \frac{\partial A_{0,2}}{\partial n}\right)+\frac{2}{\pi n^3}A_{0,2}.
\end{align}
Taking into account the definition (\ref{eq:e2ldefinition}) of $e_{2l}$ and transforming the derivative to be with respect to $\gamma$ according to equation~(\ref{eq:derivatives}), equation~(\ref{eq:g3exact3}) becomes 
\begin{align}\label{eq:g3exact1}
	g_3(\gamma)=\frac{3e'_4}{2\gamma}-\frac{5e_4}{\gamma^2} +\left(1+\frac{\gamma}{2}\right)e_2'-\frac{2e_2}{\gamma}-\frac{3e_2e_2'}{\gamma}+\frac{9e_2^2}{\gamma^2},
\end{align}
where $e_{2l}'={d e_{2l}(\gamma)}/{d\gamma}$. Equation (\ref{eq:g3exact1}) coincides with the expression initially found in reference~\cite{cheianov_exact_2006} using yet another approach. We have therefore proven that the exact results (\ref{eq:g3exact2}) and (\ref{eq:g3exact1}) are equivalent, which a priori was not obvious at all.

\subsubsection{The four-body case.}

The local correlation function (\ref{eq:gNexact}) in the case $N=4$ can be treated in a way similar to $N=3$. This leads to  \cite{pozsgay_local_2011}\footnote{\linespread{1}\selectfont Equation (\ref{eq:g4pozs}) corresponds to equation~(7.12) of reference~\cite{pozsgay_local_2011} where we omitted the terms proportional to $A_{j,l}$ with odd $j+l$ since they nullify in the ground state.}
\begin{align}\label{eq:g4pozs}
	g_4(\gamma)={}&\frac{8\gamma}{5\pi n^3}(2A_{2,0}-A_{1,1}) -\frac{16}{5\pi^2 n^4}A_{0,0}A_{1,1} +\frac{24}{\pi n^5\gamma}(A_{2,2}-2A_{3,1}+2A_{4,0}) \notag\\
	&+\frac{48}{5\pi^2 n^6\gamma^2}(2A_{1,1}A_{2,0}-5A_{2,0}^2 +5A_{0,0}A_{2,2}-8A_{0,0}A_{3,1}) \notag\\&
	+\frac{144}{\pi n^7\gamma^3} (2A_{6,0}-2A_{5,1}+2A_{4,2}-A_{3,3}).
\end{align}
The latter expression can be simplified. Applying equations~(\ref{eq:Ajlsol}), then using equation~(\ref{eq:e2ldefinition}), and finally  transforming the derivative to be with respect to $\gamma$ according to equation~(\ref{eq:derivatives}), we find
\begin{align}\label{eq:g4exact}
	g_4(\gamma)={}&\frac{e_6'}{\gamma^2}-\frac{28e_6}{5\gamma^3}-\left( \frac{10}{\gamma}+\frac{104}{5\gamma^2}+\frac{9e_2'}{\gamma^2}-\frac{12e_2}{\gamma^3}\right)e_4+\left(3+\frac{26}{5\gamma}+\frac{3e_2}{\gamma^2}\right)e_4' \notag\\&+\left(\frac{18}{\gamma}+\frac{168}{5\gamma^2}\right)e_2^2 +\left(\frac{8\gamma}{5}+\frac{4\gamma^2}{5}-6e_2 -\frac{84e_2}{5\gamma}\right)e_2' -\frac{16 e_2}{5}.
\end{align}
Equation (\ref{eq:g4exact}) is the exact result for the four-body local correlation function, which is defined by equation~(\ref{eq:gj}) taken at $N=4$. It is expressed in terms of the moments of the quasimomentum distributions and their first derivative and in this respect has a similar structure as equation~(\ref{eq:g3exact1}).

\subsubsection{Explicit results for the local correlation functions.}

The forms (\ref{eq:g3exact1}) and (\ref{eq:g4exact}) are particularly convenient for the analytical evaluation. Using our previously derived results for $e_{2l}$, we obtain
\begin{align}\label{eq:g3small}
	g_3(\gamma)={}&1-\frac{6\sqrt\gamma}{\pi} +\frac{3\gamma}{2}-\left( \frac{3}{\pi}-\frac{25}{4\pi^3}-\frac{69\zeta(3)}{16\pi^3}\right)\gamma^{3/2}+\mathcal{O}(\gamma^2)
\end{align}
at $\gamma\ll 1$ and
\begin{align}\label{eq:g3large}
	g_3(\gamma)=\frac{16\pi^6}{15\gamma^6}\left(1-\frac{16}{\gamma} +\frac{144-\frac{144\pi^2}{35}}{\gamma^2} -\frac{960-\frac{660\pi^2}{7}}{\gamma^3}\right) +\mathcal{O}\left(\gamma^{-10}\right)
\end{align}
at $\gamma\gg 1$. For the other case we find
\begin{align}\label{eq:g4small}
	g_4(\gamma)={}&1-\frac{12\sqrt\gamma}{\pi} +\left(4+\frac{24}{\pi^2}\right)\gamma -\left( \frac{24}{\pi}-\frac{65}{2\pi^3}-\frac{93\zeta(3)}{8\pi^3}\right)\gamma^{3/2}+\mathcal{O}(\gamma^2)
\end{align}
at $\gamma\ll 1$ and
\begin{align}\label{eq:g4large}
	g_4(\gamma)={}&\frac{1024\pi^{12}}{2625\gamma^{12}}\left(1-\frac{30}{\gamma}+\frac{480-\frac{160\pi^2}{21}}{\gamma^2} \right)+\mathcal{O}\left({\gamma^{-15}} \right)
\end{align}
at $\gamma\gg 1$. It is fascinating to note that in order to calculate the leading-order term in equation~(\ref{eq:g4large}) we need to know the twelfth subleading term in $e_2(\gamma)$. This was achieved using the systematic procedure described in section \ref{section:simplesolution}. We note that only the leading- and the subleading-order terms in $g_3$ and $g_4$ were known before \cite{gangardt_stability_2003,cheianov_exact_2006,gangardt_local_2003,nandani_higher-order_2016}. However, they were obtained using  complementary techniques that can hardly be extended to give better accuracy. On the other hand, the exact results (\ref{eq:g3exact1}) and (\ref{eq:g4exact}) together with the method described in Sec.~\ref{sec:LL} establish a way to explicitly evaluate analytically $g_3$ and $g_4$ to a large number of terms in the series, the only limitation being the knowledge of $e_2$.

\section{Derivatives of the density of quasimomenta at the Fermi quasimomentum }\label{section:derivativesrho}

The density of quasimomenta $\rho(k,Q)$ is the central function that determines many quantities in the Lieb--Liniger model. It is determined by the Lieb integral equation (\ref{eq:LIE}) that does not admit a closed-form solution. One is therefore tempted to find various exact relations that will enable understanding of the model. Two such relations are given by equations~(\ref{eq:defK}) and (\ref{eq:Kll}). They connect $\rho(Q,Q)$, i.e., the density at the Fermi quasimomentum, with the ground-state function $e_2(\gamma)$ and the Luttinger liquid parameter $K$. In this section we develop a systematic method for the calculation of partial derivatives of $\rho(Q,Q)$. The latter quantities will appear in the spectrum of elementary excitations and the low-temperature thermodynamics, which will be considered in the forthcoming sections.

\subsection{Partial differential equation for the density of quasimomenta}

The density of quasimomenta satisfies equation~(\ref{eq:LIE}). Using the methods derived in section \ref{section:method}, in particular equations~(\ref{eq:Fhoder1}) and (\ref{eq:Fhoder2}), it can be converted into a partial differential equation. An alternative more direct derivation is presented in Appendix \ref{appendixc}. The final result takes the form
\begin{align}\label{eq:PDE}
	\left(\frac{\partial^2}{\partial Q^2}-\frac{2}{\rho(Q,Q)}\frac{d\rho(Q,Q)}{dQ}\frac{\partial}{\partial Q}-\frac{\partial^2}{\partial k^2}\right)\rho(k,Q)=0.
\end{align} 
Instead of the original expression (\ref{eq:LIE}), equation~(\ref{eq:PDE}) is very convenient to study the local properties of $\rho(k,Q)$ for $k$ near the Fermi quasimomentum $Q$.

\subsection{The first derivatives}

Partial derivatives of the density of quasimomenta
\begin{align}
	\rho^{(j,0)}(Q,Q)\equiv \frac{\partial^j \rho(k,Q)}{\partial k^j}\bigg|_{k=Q}
\end{align}
are generally not known. Here it will be first shown that they satisfy certain linear differential equations and at a second step the latter equations will be solved. As a starting point we use equation~(\ref{eq:PDE}) and the expressions for the total derivative of $\rho(Q,Q)$,
\begin{align}\label{eq:totalderivative}
	\frac{d^j}{dQ^j}\rho(Q,Q)-\sum_{l=0}^{j}\binom{j}{l}\rho^{(l,j-l)}(Q,Q)=0.
\end{align}
The first term in the left-hand side of equation~(\ref{eq:totalderivative}) is assumed to be a known function. Indeed, via equation~(\ref{eq:defK}) $\rho(Q,Q)$ is directly related  to the ground-state function $e_2(\gamma)$ that is calculated in sections \ref{section:numericalexperiment} and \ref{section:simplesolution}. The derivatives with respect to $Q$ of $\rho(Q,Q)$ can then be obtained using equations~(\ref{eq:Qn}) and (\ref{eq:derivatives}).

A straightforward calculation shows that the equation 
\begin{align}\label{eq:Fho10}
	\frac{d}{dQ}\left(\frac{\rho^{(1,0)}(Q,Q)}{\rho(Q,Q)}\right)=R(Q)
\end{align}
is satisfied, where we have conveniently introduced
\begin{align}
	R(Q)=\frac{\ddot{\rho}(Q,Q)}{2\rho(Q,Q)}-\left(\frac{\dot{\rho}(Q,Q)} {\rho(Q,Q)}\right)^2.
\end{align}
Here and in the following the dots over $\rho(Q,Q)$ denote the total derivatives. For example, $\dot\rho(Q,Q)=\frac{d}{dQ}\rho(Q,Q)$, $\ddot\rho(Q,Q)=\frac{d^2}{dQ^2}\rho(Q,Q)$, etc. Equation (\ref{eq:Fho10}) has the form of a first-order linear differential equation for the variable $\frac{\rho^{(1,0)}(Q,Q)}{\rho(Q,Q)}$, since the right-hand side, $R(Q)$, is assumed to be a known function. We note that once the partial derivative $\rho^{(1,0)}(Q,Q)$ is calculated, the other $\rho^{(0,1)}(Q,Q)$ follows directly from equation~(\ref{eq:totalderivative}).

\subsection{The second and higher derivatives}

For the second partial derivative, we found that it satisfies
\begin{align}\label{eq:Fho20}
	\frac{d}{dQ}\left(\frac{\rho^{(2,0)}(Q,Q)}{\rho(Q,Q)}\right)=\frac{1}{2}\frac{dR(Q)}{dQ}+R(Q)\frac{\rho^{(1,0)}(Q,Q)}{\rho(Q,Q)}.
\end{align}
Using equation~(\ref{eq:Fho10}), it can be integrated, yielding
\begin{align}\label{eq:Fho20solution}
	\frac{\rho^{(2,0)}(Q,Q)}{\rho(Q,Q)}
	={}&\frac{1}{2} \frac{d}{dQ}\left(\frac{\rho^{(1,0)}(Q,Q)}{\rho(Q,Q)}\right)+\frac{1}{2}\left(\frac{\rho^{(1,0)}(Q,Q)}{\rho(Q,Q)}\right)^2.
\end{align}
Here the integration constant is set to zero to account for the case of very large  $\gamma$ where equation~(\ref{eq:Fho20}) can be explicitly solved. The expression (\ref{eq:Fho20solution}) is a remarkable result as it shows that the second partial derivative $\rho^{(2,0)}(Q,Q)$ is not independent, but can be expressed in terms of $\rho^{(1,0)}(Q,Q)$. The remaining second derivatives can be now straightforwardly obtained. The result for $\rho^{(0,2)}(Q,Q)$ follows from equation~(\ref{eq:PDE}), which can then be used in equation~(\ref{eq:totalderivative}) to obtain $\rho^{(1,1)}(Q,Q)$.

Let us comment how equation~(\ref{eq:Fho20}) can be derived. We form a linear combination that consists of $\frac{d}{dQ}\left(\frac{\rho^{(2,0)}(Q,Q)}{\rho(Q,Q)}\right)$ with the coefficient $1$, the left-hand sides of equation~(\ref{eq:PDE}) and its $k$- and $Q$-derivatives, all three taken at $k=Q$, as well as the left-hand sides of equation~(\ref{eq:totalderivative}) taken at $j=1,2,3$. The linear combination involves six parameters that can take arbitrary values as they multiply the expressions that are formally equal to zero. We then impose zero coefficients in front of the terms involving all the partial derivatives $\rho^{(j,l)}(Q,Q)$ but $\rho^{(1,0)}(Q,Q)$. As it turns out, the solution can be found, which once returned to the linear combination yields the right-hand side of equation~(\ref{eq:Fho20}). 

The latter procedure can be extended to find differential equations for $\frac{d}{dQ}\left(\frac{\rho^{(j,0)}(Q,Q)}{\rho(Q,Q)}\right)$ with $j>2$. Here we only note the result for the third derivative,
\begin{align}\label{eq:Fho30}
	\frac{d}{dQ}\left(\frac{\rho^{(3,0)}(Q,Q)}{\rho(Q,Q)}\right)={}&R(Q)\frac{\rho^{(2,0)}(Q,Q)}{\rho(Q,Q)}+\frac{1}{2}\frac{dR(Q)}{dQ} \frac{\rho^{(1,0)}(Q,Q)}{\rho(Q,Q)}+\frac{1}{4}\frac{d^2 R(Q)}{dQ^2}+ \frac{R(Q)^2}{2}.
\end{align}
We notice that the right-hand side of equation~(\ref{eq:Fho30}) can be expressed only in terms of $\frac{\rho^{(1,0)}(Q,Q)}{\rho(Q,Q)}$ and its derivatives. It is not obvious how to integrate equation~(\ref{eq:Fho30}) as the right-hand side contains the nonlinear term $R(Q)^2/2$.

\subsection{Explicit evaluation in terms of power series}

The obtained equations~(\ref{eq:Fho10}), (\ref{eq:Fho20}) and (\ref{eq:Fho30}) have a hierarchical structure in the sense that the solution of one of them, say $\rho^{(j,0)}(Q,Q)$ participates in the equations for $\rho^{(l,0)}(Q,Q)$ with $l>j$. They should be thus solved successively. Moreover, equations~(\ref{eq:Fho10}), (\ref{eq:Fho20}), and (\ref{eq:Fho30}) are linear differential equations. Their right-hand sides have the form of a power series with constant coefficients in our case of the Lieb--Liniger model since $\rho(Q,Q)$ is analytically known in this form. The equations can thus be solved one by one provided there is information about one integration constant for each equation. 

The integration constants can be found from the behaviour of $\rho(Q+q,Q)$ at $0<q\ll Q$. For this one should solve the integral equation (\ref{eq:LIE}) in the vicinity of the Fermi quasimomentum, which in this case reduces to the equation of  Wiener--Hopf type. The latter is solved in references~\cite{pustilnik_low-energy_2014,pustilnik_fate_2015}. At the leading order in $\gamma\ll 1$, the final result is given by
\begin{align}
	\rho(Q+q,Q)=\frac{1}{2\sqrt\pi \gamma^{1/4}} f\left(\frac{2\pi}{c} q\right),
\end{align}
where
\begin{align}\label{eq:fLLWH}
	f(y)=\frac{1}{\sqrt{2}\pi^{3/2}}\int_0^\infty dz \frac{\sin(2\pi z)\Gamma(z)e^{-z(\ln z-1+y)}}{\sqrt{z}}.
\end{align}
The expression for the derivatives of the quasimomentum distribution at the Fermi quasimomentum then follows,
\begin{subequations}\label{eq:FholeadingorderWH}
	\begin{align}\label{eq:Fholeadingorder}
		\rho^{(j,0)}(Q,Q)=\frac{1}{2\sqrt{\pi}\gamma^{1/4}}\left(\frac{2\pi}{c}\right)^j f^{(j)}(0).
	\end{align}
	Equation (\ref{eq:Fholeadingorder}) is the leading-order result for all the derivatives at $\gamma\ll 1$. The values of the function (\ref{eq:fLLWH}) and of its first three derivatives at zero are given by\footnote{\linespread{1}\selectfont Interestingly, the numbers in equation~(\ref{eq:fderivatives}) are identical to the coefficients appearing in Stirling's series for the Gamma function, $\Gamma(z)$, written in the form
		\begin{align*}
			\frac{\Gamma(-z)\sqrt{-z}e^{z(\ln(-z)-1)}}{\sqrt{2\pi}}={}&1-\frac{1}{12z}+\frac{1}{288z^2}+\frac{139}{51840z^3}+\mathcal{O}((-z)^{-4}).
		\end{align*}
	} 
	\begin{gather}\label{eq:fderivatives}
		f(0)=1,\quad f'(0)=-\frac{1}{12},\quad	f''(0)=\frac{1}{288},\quad f'''(0)=\frac{139}{51840}.
	\end{gather}	
\end{subequations}
Together with our differential equations, the information contained in equation~(\ref{eq:FholeadingorderWH}) suffices to reconstruct the whole series for all the derivatives $\rho^{(j,0)}(Q,Q)$ in the regime $\gamma\ll 1$. In the regime of strong interactions, $\gamma\gg 1$, from equation~(\ref{eq:Fho0large}) we can conclude that $\rho^{(j,0)}(Q,Q)=0$ for $j>0$ at $\gamma\to\infty$, which can be used to fix the integration constants. Let us mention that in the final step, one can also obtain all the remaining derivatives of the order $j$ if needed.

Equation (\ref{eq:defK}) enables us to express $\rho(Q,Q)$ via the ground-state function $e_2(\gamma)$ as
\begin{align}\label{eq:Fho(Q,Q)}
	\rho(Q,Q)=\frac{1}{8^{1/4}\sqrt\pi \gamma}\left[\frac{d^2}{d\gamma^2}\left(\frac{e_2(\gamma)}{\gamma^2} \right)\right]^{-1/4},
\end{align}
which can then be calculated using equations~(\ref{eq:e2nineterms}) and (\ref{eq:e2largefinal}). The derivatives with respect to $Q$ that enter equations~(\ref{eq:Fho10}), (\ref{eq:Fho20}), and (\ref{eq:Fho30}) can be transformed to be with respect to $\gamma$ using equations~(\ref{eq:Qn}) and (\ref{eq:derivatives}), where $c=\gamma\;\! n$  assumes a constant value. For the first derivative this gives
\begin{align}
	\frac{\partial}{\partial Q}={}&-\frac{\sqrt{2}}{c} \left[\frac{d^2}{d\gamma^2}\left(\frac{e_2(\gamma)}{\gamma^2} \right)\right]^{-1/2} \frac{\partial}{\partial\gamma},
\end{align}
while the higher ones can be similarly found. The last two equations enable us to solve equation~(\ref{eq:Fho10}) and obtain $\rho^{(1,0)}(Q,Q)$. Then we can obtain $\rho^{(2,0)}(Q,Q)$ either by solving equation~(\ref{eq:Fho20}) or directly from equation~(\ref{eq:Fho20solution}). The final results at $\gamma\ll 1$ are given by
\begin{align}\label{eq:rho10weak}
	\rho^{(1,0)}(Q,Q)=-\frac{\sqrt\pi}{12c\;\!\gamma^{1/4}}\left( 1-\frac{\sqrt{\gamma }}{\pi}-\frac{7 \gamma }{128 \pi^2} +\frac{3\zeta (3)+15}{256 \pi^3}\gamma ^{3/2}+\mathcal{O}(\gamma^{2})\right),
\end{align}
\begin{align}\label{eq:rho20weak}
	\rho^{(2,0)}(Q,Q)=\frac{\pi^{3/2}}{144c^2\gamma^{1/4}}\left( 1-\frac{17\sqrt{\gamma }}{8\pi}-\frac{289 \gamma }{128 \pi^2}-\frac{833-12\zeta (3)}{1024 \pi^3}\gamma ^{3/2} +\mathcal{O}(\gamma^{2})\right),
\end{align}
while for $\gamma\gg 1$ we find
\begin{align}\label{eq:rho10strong}
	\rho^{(1,0)}(Q,Q)=-\frac{2}{c\;\!\gamma^{2}}\left( 1-\frac{2}{\gamma}-\frac{4\pi^2-4}{\gamma^2} +\frac{76\pi^2-24}{3\gamma^3} +\mathcal{O}(\gamma^{-4})\right),
\end{align}
\begin{align}\label{eq:rho20strong}
	\rho^{(2,0)}(Q,Q)=-\frac{2}{\pi c^2\gamma}\left( 1-\frac{8\pi^2}{\gamma^2}+\frac{32\pi^2}{\gamma^3} +\mathcal{O}(\gamma^{-4})\right).
\end{align}
Interested reader can easily find further terms in the above series using more terms from previously found results for $e_2(\gamma)$. We do not give explicit solution for $\rho^{(3,0)}(Q,Q)$ as it will not be used here any further.

\subsection{New dimensionless parameters}

The Luttinger liquid parameter $K$ is related to the density of quasimomenta $\rho(k,Q)$ by the relation
\begin{align}
	K=4\pi^2\rho(Q,Q)^2,
\end{align}
see equation~(\ref{eq:Kll}). The parameter $K$ determines a number of properties of the system. In this section we have calculated the partial derivatives $\rho^{(1,0)}(Q,Q)$, $\rho^{(2,0)}(Q,Q)$, etc. As will be seen later, they contain the information about the excitation spectrum as well as determine the low-temperature free energy of the model. It is therefore convenient to define the corresponding dimensionless parameters. Consider a family of functions defined by
\begin{align}
	&K_0(q)=4\pi^2\rho(q,Q)^2,\\
	&K_j(q)=-\pi n\frac{d K_{j-1}(q)}{dq},\quad j=1,2,\ldots.
\end{align}
Then we define a family of parameters by
\begin{align}
	&K=K_0(Q),\\
	\label{eq:Kjdefinition}
	&K_j=K_j(Q),\quad j=1,2,\ldots.	
\end{align}	
The first member of the family is the Luttinger liquid parameter and the following two are
\begin{align}\label{eq:K1}
	&K_1=-2\pi n K \frac{\rho^{(1,0)}(Q,Q)}{\rho(Q,Q)},\\
	\label{eq:K21}
	&K_2=\frac{K_1^2}{2K}+2\pi^2 n^2 K \frac{\rho^{(2,0)}(Q,Q)}{\rho(Q,Q)}.
\end{align}
Similarly as $K$, the parameters $K_1$ and $K_2$ are dimensionless and only depend on $\gamma$. Note that due to the connection (\ref{eq:Fho20solution}), $K_2$ is not an independent parameter. It can be expressed in terms of $K_1$ as
\begin{align}\label{eq:K2}
	K_2=\frac{3K_1^2}{4K}+\frac{K K_1}{2}+\frac{\gamma}{2}(K_1' K-K_1 K').
\end{align}
Here $K_1'=\frac{dK_1}{d\gamma}$ and similarly $K'=\frac{dK}{d\gamma}$.

Let us note the series expansions at small $\gamma$,
\begin{gather}
	K_1=\frac{\pi^3}{3\gamma^{3/2}}-\frac{7\pi^2}{24\gamma}-\frac{3\pi}{64\sqrt\gamma}  +\frac{1+2\zeta(3)}{256}+\mathcal{O}(\sqrt\gamma),\\
	K_2=\frac{\pi^5}{12\gamma^{5/2}}-\frac{\pi^4}{6\gamma^2}-\frac{19\pi^3}{384\gamma^{3/2}}  -\frac{(29+3\zeta(3))\pi^2}{2304\gamma}+\mathcal{O}(1/\sqrt\gamma),
\end{gather}
and at large $\gamma$,
\begin{gather}
	K_1=\frac{8\pi^2}{\gamma^3}-\frac{32\pi^4}{\gamma^5}+\mathcal{O}(\gamma^{-6}),\\
	K_2=-\frac{8\pi^2}{\gamma^3}-\frac{16\pi^2}{\gamma^4} +\frac{64\pi^4}{\gamma^5} +\mathcal{O}(\gamma^{-6}).
\end{gather}
We see that at weak interactions, $K$, $\gamma K_1$, and $\gamma^2 K_2$ are on the same order, while $K_1$ and $K_2$ tend to zero at strong interactions.

\section{Spectrum of elementary excitations}\label{section:elementaryexcitations}

The universal low-energy theory of interacting quantum particles in one dimension is usually described by the paradigm of the Luttinger liquid. The excitations in this theory are phonons. These bosonic quasiparticles represent the waves of particle density propagating with constant velocity. The Luttinger liquid description is only the limiting theory describing the physics at longest wavelengths. Realistic models have a more complicated picture. The efforts to understand effects beyond the linear theory have yielded the picture of fermionic quasiparticles that are true low-energy excitations in one-dimensional quantum liquids \cite{rozhkov_fermionic_2005,imambekov_one-dimensional_2012}. This is an interesting result, different from the conventional picture of phonon excitations in a Luttinger liquid. In this section we study in more detail the spectrum of elementary excitations in the Lieb--Liniger model. Using the microscopic hydrodynamic theory and the Bethe ansatz solution, we find the excitation spectrum at arbitrary momenta and interaction strengths. The microscopic results fully support the phenomenological picture of fermionic quasiparticles that are lowest-energy excitations at arbitrary interaction strengths \cite{imambekov_one-dimensional_2012}. They are characterised by the quadratic in momentum subleading term in the spectrum controlled by the effective mass. Bosonic quasiparticles are characterised by the Bogoliubov form of the spectrum and do not have such terms. They exist only at weak interactions and at higher momenta. The statistics of quasiparticles thus changes as the momentum is increased. This section is based on the published papers \cite{ristivojevic_excitation_2014,petkovic_spectrum_2018}.

\subsection{The spectrum from the hydrodynamic theory}

In the regime of weak interactions, $\gamma\ll 1$, the system of interacting bosons described by the Lieb--Liniger model (\ref{eq:Horiginal}) can be described by the hydrodynamic approach \cite{popov_hydrodynamic_1972,haldane_effective_1981}. We start from the standard expression for the Hamiltonian of interacting bosons in second quantization and reexpress the bosonic single particle operator as $\Psi^\dagger(x)=\sqrt{n(x)}e^{i\theta(x)}$, where $n(x)$ and  $\theta(x)$ are the fluctuating bosonic density and the phase fields, respectively. Accounting for small density fluctuations in the standard way \cite{haldane_effective_1981}, where $n(x)=n+\nabla\varphi(x)/\pi$, after neglecting the constant terms that enter the ground-state energy, we obtain
\begin{align}\label{H}
	H={}&\frac{\hbar^2}{2m}\int dx\left[\left(n+\frac{\nabla\varphi(x)}{\pi}\right) (\nabla\theta(x))^2+\frac{(\nabla^2\varphi(x))^2}{4\pi^2n(x)}+\frac{c}{\pi^2}(\nabla\varphi(x))^2 \right].
\end{align}
The fields $\theta$ and $\varphi$ satisfy the commutation relation $[\nabla\varphi(x),\theta(y)]=-i\pi\delta(x-y)$. The Hamiltonian (\ref{H}) provides an effective description of the original one, given by equation~(\ref{eq:Horiginal}), at momenta below $\hbar n$. In this regime, the fluctuations of the field $\nabla\varphi$ are small, enabling us to use the hydrodynamic approach.

At lowest momenta, the excitation spectrum is determined by the most relevant operators of the Hamiltonian (\ref{H}). Retaining the operators of scaling dimension two, $(\nabla\varphi)^2$ and $(\nabla\theta)^2$, we obtain the Luttinger liquid Hamiltonian
\begin{align}\label{H00}
	H_0=\frac{\hbar^2 n}{2m}\int dx\left[(\nabla\theta)^2 +\frac{\gamma}{\pi^2}(\nabla\varphi)^2 \right].
\end{align}
It describes the excitations with linear spectrum $\varepsilon_p=v |p|$, where $v=\frac{\hbar n}{m}\sqrt\gamma$ is the sound velocity \cite{lieb_exact_1963b}. By $p$ is denoted the momentum. It is important to note that the Luttinger liquid Hamiltonian does not uniquely determine statistics of quasiparticle excitations. Indeed, $H_0$ exactly describes both, the excitations in a system of noninteracting bosons \cite{haldane_effective_1981} and fermions \cite{mattis_exact_1965} with linear dispersion. However, the theory (\ref{H}) has operators of higher scaling dimension that arise from amplitude fluctuations of $\Psi$. They lift the statistics degeneracy of $H_0$, and thus uniquely determine true quasiparticles. At lowest momenta we must include the leading irrelevant operator, which is the one of scaling dimension three. The resulting Hamiltonian  
\begin{align}\label{H0prime}
	H_{\F}=H_0+\frac{\hbar^2}{2\pi m}\int d x (\nabla\varphi)(\nabla\theta)^2
\end{align}
can be diagonalised by the fermionisation procedure. In this way one obtains the low-energy spectrum \cite{ristivojevic_excitation_2014}
\begin{align}\label{E_weakint_fermions}
	\varepsilon_p=v|p|+\frac{p^2}{2m^*}
\end{align}
with the quasiparticle mass $m^*=4\pi^{1/2}m/3\gamma^{1/4}$. The quadratic dispersion (\ref{E_weakint_fermions}) is in agreement with the result first obtained phenomenologically \cite{imambekov_one-dimensional_2012} and then microscopically using the Bethe ansatz \cite{pustilnik_low-energy_2014,pustilnik_fate_2015}. However, unlike in the latter study, the present microscopic theory directly identifies fermionic nature of quasiparticle excitations at lowest momenta as the Hamiltonian (\ref{H0prime}) is diagonalised using the fermionic operators.

At higher momenta, operators of higher scaling dimension could be more important than the ones of lower dimension. Keeping only the operator $(\nabla^2\varphi)^2$ of scaling dimension four, the Hamiltonian (\ref{H}) reduces to
\begin{align}\label{H0prime11}
	H_{\B}=H_0+\frac{\hbar^2}{8\pi^2 mn}\int d x (\nabla^2\varphi)^2.
\end{align}
This Hamiltonian can be diagonalised in terms of the bosonic operators. It thus describes bosonic quasiparticles that have the Bogoliubov spectrum
\begin{align}\label{E_bogoliubov}
	\varepsilon_p=v |p|\sqrt{1+\frac{p^2}{4m^2v^2}}.
\end{align}
Comparing the first subleading terms of the two spectra (\ref{E_weakint_fermions}) and (\ref{E_bogoliubov}), we infer the crossover momentum scale \cite{imambekov_one-dimensional_2012} $p^*=\hbar n\gamma^{3/4}$. At $p\ll p^*$, the operator of scaling dimension three cannot be neglected. In this regime we find the fermionic quasiparticles with the spectrum (\ref{E_weakint_fermions}). At $p\gg p^*$, the operator of scaling dimension four in equation~(\ref{H}) is the leading correction to $H_0$, yielding the spectrum (\ref{E_bogoliubov}) of bosonic quasiparticles. In this regime the neglected operator of scaling dimension three (and many others of higher scaling dimension) describe residual interactions between Bogoliubov quasiparticles, which is  responsible, e.g., for the broadening of the spectral function \cite{ristivojevic_decay_2016}.

At momenta below the crossover momentum $p_0=mv$, the bosonic dispersion (\ref{E_bogoliubov}) simplifies into
\begin{align}\label{E_Bogoliubovlowp}
	\varepsilon_p=v|p|+\frac{|p|^3}{8m^2 v},
\end{align}
and describes Bogoliubov phonons. Reexpressing the asymptote as $p_0=\hbar n\sqrt\gamma$, we observe that $p_0$ and $p^*$, when extrapolated to moderate interactions, cross each other at $\gamma_c=1$. If such extrapolation from the weakly interacting region $\gamma\ll 1$ indeed holds, this would imply limited parameter regime where phonon quasiparticles exist. Moreover, the bosonic quasiparticles with the momenta higher than $p_0$, which have the spectrum\footnote{\linespread{1}\selectfont
	The Bogoliubov spectrum (\ref {E_bogoliubov}) is derived here using the hydrodynamic theory. The same expression is actually valid at arbitrary high momenta, as one can obtain from the Bethe ansatz equations \cite{kulish_comparison_1976,pustilnik_low-energy_2014,pustilnik_fate_2015}.}
\begin{align}\label{Ehighp}
	\varepsilon_p=\frac{p^2}{2m}+\gamma\frac{\hbar^2n^2}{m},
\end{align}
are also expected to cease together with the phonons as interaction strength is increased. In the following we complement this picture by studying the spectrum of excitations of the model (\ref{eq:Horiginal}) using the Bethe ansatz. Note that in this subsection we have studied the particle-like branch of elementary excitations.

\subsection{Implicit form of the spectrum from the Bethe ansatz}

In the ground state of the system with $N$ particles, the quasimomenta $k_j$ satisfy the Bethe equations (\ref{eq:DBA}) that for the quantum numbers of the ground state read
\begin{align}\label{eq:DBAgs1}
	Lk_j=2\pi I_j+\theta(k_j-k_N)+\sum_{l=1}^{N-1}\theta(k_j-k_l), \quad j=1,2,\ldots,N,
\end{align}
where $I_j=j-\frac{N+1}{2}$. For any other choice of the quantum numbers $I_j$ the system is in an excited state. We can distinguish two types of elementary excitations called type I and type II \cite{lieb_exact_1963b}. For the former, the quasimomentum from the top of the Fermi sea is promoted above the sea, and for the latter, a quasimomentum within the sea is promoted to the Fermi quasimomentum.

Consider first the right-moving type-I excitation. There the quantum number of the largest quasimomentum has a value larger than the one in the ground state. This imposes a new configuration of quasimomenta where one quasimomentum is above $Q$, while the remaining quasimomenta are redistributed within the Fermi sea according to the Bethe ansatz equations. Here $Q$ denotes the Fermi quasimomentum, which in the thermodynamic limit corresponds to the highest quasimomentum in the ground state. The new set of quasimomenta $\{\bar k_1,\bar k_2,\ldots,\bar k_{N-1},\bar k_N\equiv\eta\}$ satisfies
\begin{align}\label{eq:DBAes1}
	L\bar k_j=2\pi I_j+\theta(\bar k_j-\eta)+\sum_{l=1}^{N-1}\theta(\bar k_j-\bar k_l), \quad j=1,2,\ldots,N-1,
\end{align}
where $\eta>Q$ can be considered as a parameter. The momentum $p$ and the excitation energy $\varepsilon$ of the state are given by
\begin{align}\label{eq:pEtypeI}
	p=\hbar\left(\eta+\sum_{j=1}^{N-1}\bar k_j\right),\quad \varepsilon=\frac{\hbar^2}{2m}\left(\eta^2+ \sum_{j=1}^{N-1}\bar k_j^2\right)-N\epsilon_0.
\end{align}
Here $N\epsilon_0$ denotes the ground-state energy of the system with $N$ particles.

In order to obtain the momentum and the energy of the excitation in the thermodynamic limit, let us subtract equation~(\ref{eq:DBAgs1}) from equation~(\ref{eq:DBAes1}). In a long system, the difference $\bar k_j-k_j=\Delta k_j=\mathcal{O}(1/L)$ is small and thus
\begin{align}\label{eq:DBAdifferenceex}
	L\Delta k_j=\theta(k_j-\eta)-\theta(k_j-Q)+\sum_{l=1}^{N-1}\theta'(k_j-k_l)(\Delta k_j-\Delta k_l) +\mathcal{O}(1/N).
\end{align} 
In equation~(\ref{eq:DBAdifferenceex}), $\bar k_j$ and $k_N$ have been replaced by $k_j$ and $k_N$, respectively, which is possible within the written accuracy. The formal expression $\rho(k)=\frac{1}{L}\sum_{j=1}^{N}\delta(k-k_j)$ substituted into equation~(\ref{eq:LIE}) gives
\begin{align}\label{eq:LIEdiscrete1}
	1-\frac{1}{L}\sum_{j=1}^N \theta'(k_j-k)=2\pi \rho(k).
\end{align}
After introducing $J(k_j)=L\rho(k_j)\Delta k_j$, equation~(\ref{eq:LIEdiscrete1}) enables us to express equation~(\ref{eq:DBAdifferenceex}) in the thermodynamic limit as an integral equation
\begin{align}\label{eq:J}
	J(k)+\frac{1}{2\pi}\int_{-Q}^{Q}dq \theta'(k-q)J(q)=\frac{\theta(k-\eta)-\theta(k-Q)}{2\pi}.
\end{align}
Here we have transformed the summation to the integral according to equation~(\ref{eq:sumtointegral}) and neglected the terms that scale as $1/L$. Substituting $\bar k_j=k_j+\Delta k_j$ in equation~(\ref{eq:pEtypeI}), the momentum and the excitation energy in the thermodynamic limit become \cite{korepin}
\begin{align}\label{eq:pEtypeIinter}
	p=\hbar\left[\eta-Q+\int_{-Q}^{Q}dk J(k)\right],\quad \varepsilon=\frac{\hbar^2}{2m}\left[\eta^2-Q^2+2\int_{-Q}^{Q}dk k J(k)\right].
\end{align}
Here we have used that the ground-state configuration has zero momentum, $\sum_{j=1}^{N}k_j=0$, and the ground-state energy $\frac{\hbar^2}{2m}\sum_{j=1}^{N} k_j^2=N\epsilon_0$. In the limiting case $\eta=Q$, we have $J(k)=0$ and thus from equation~(\ref{eq:pEtypeIinter}) it follows $p=\varepsilon=0$, as expected. 

Equation (\ref{eq:pEtypeIinter}) can be further transformed using the theorem derived in Appendix \ref{appendixb}. It directly implies
\begin{gather}
	\int_{-Q}^{Q} dk J(k)=\int_{-Q}^{Q}dk \rho(k,Q)\left[\theta(k-\eta)-\theta(k-Q)\right],\\
	\int_{-Q}^{Q} dk k J(k)=\frac{m}{2\pi\hbar^2}\int_{-Q}^{Q}dk \sigma(k,Q)\left[\theta(k-\eta)-\theta(k-Q)\right],
\end{gather}
where we have introduced $\sigma(k,Q)$ as the solution of the integral equation
\begin{align}\label{eq:sigma}
	\sigma(k,Q)+\frac{1}{2\pi}\int_{-Q}^{Q} dq\,\theta'(k-q)\sigma(q,Q) =\frac{\hbar^2 k}{m}.
\end{align}
Then the momentum and the energy of the right-moving type-I excitation are given by
\begin{align}\label{eq:pEtypeIfinal}
	p=2\pi\hbar \int_Q^\eta dk \rho(k,Q),\quad \varepsilon=\int_Q^\eta dk\;\! \sigma(k,Q).
\end{align}
The corresponding left-moving type-I excitation can be obtained by the symmetry. It has the opposite momentum and the same energy. We note that in equation~(\ref{eq:pEtypeIfinal}), the density of quasimomenta in the ground state $\rho(k,Q)$ is integrated above the Fermi quasimomentum. By the physical definition (\ref{eq:Fhokj}), $\rho(k,Q)$ exists for quasimomenta between $-Q$ and $Q$. We can nevertheless define $\rho(k,Q)$ for $|k|>Q$ by analytic continuation as
\begin{align}
	\rho(k,Q)=\frac{1}{2\pi}-\frac{1}{2\pi}\int_{-Q}^{Q}dq\:\! \theta'(k-q)\rho(q,Q),\quad |k|>Q.
\end{align}
We use analogous analytic continuation for $\sigma(k,Q)$ defined by equation~(\ref{eq:sigma}).

Let us now consider the right-moving type-II excitation. It is created when one of the quasimomenta from the ground-state configuration is promoted at the first available position above the Fermi quasimomentum $Q$. Denoting by $\eta\equiv k_{j'}$ with $1\le j'\le N$ the quasimomentum that will be eventually displaced, the Bethe ansatz equations for the ground state can be expressed as
\begin{align}\label{eq:typeIIgs}
	Lk_j=2\pi\left(j-\frac{N+1}{2}\right)+\theta(k_j-\eta)+\sum_{l=1\atop l\neq j'}^{N}\theta(k_j-k_l),\quad j=1,\ldots,j'-1,j'+1,\ldots,N.
\end{align}
In the excited state, the quasimomenta satisfy
\begin{align}\label{eq:typeIIes}
	L\bar k_j=2\pi\left(j-\frac{N+1}{2}\right)+\theta(\bar k_j-\eta')+\sum_{l=1\atop l\neq j'}^{N}\theta(\bar k_j-\bar k_l),\quad j=1,\ldots,j'-1,j'+1,\ldots,N.
\end{align}
Here $\eta'>Q$ is the quasimomentum of the first available state above $Q$. To obtain the momentum and the energy of the excited state we can proceed similarly as above for the type-I excitation, subtracting equation~(\ref{eq:typeIIgs}) from equation~(\ref{eq:typeIIes}). In the thermodynamic limit $\eta'=Q$ and we obtain similar expressions as before with $Q$ and $\eta$ interchanged. The final result takes the form
\begin{align}\label{eq:type2spectrum}
	p=2\pi\hbar \int_{\eta}^{Q}dk\rho(k,Q),\quad \varepsilon=\int_{\eta}^{Q}dk\;\!\sigma(k,Q),
\end{align}
where the parameter $\eta$ satisfies $-Q\le\eta\le Q$. In the limiting cases $\eta=-Q$ and $\eta=Q$ we have $\varepsilon=0$ and the corresponding momenta are $p=2\pi\hbar n$ and $p=0$. The type-II excitations thus exist only for a limited range of momenta.  

\subsection{The expression for the spectrum in terms of the density of quasimomenta}

The spectrum of elementary excitations is expressed in terms of the two functions $\rho(k,Q)$ and $\sigma(k,Q)$ that satisfy the integral equations (\ref{eq:LIE}) and (\ref{eq:sigma}). The two functions are not independent. It can be shown that the latter can be expressed in terms of the former by the relation
\begin{align}
	\sigma(k,Q)=-\frac{\hbar^2}{2m}\frac{d}{dk}\int_{|k|}^{Q} dq \frac{\rho(k,q)\tilde n(q)}{\rho(q,q)^2},
\end{align}
where
\begin{align}
	\tilde n(q)=\int_{-q}^{q} dk \rho(k,q).
\end{align}
This enables us to express the spectrum of elementary excitations as
\begin{align}\label{eq:pepsfinal}
	p(\eta)=2\pi\hbar \bigg{|}\int_Q^\eta dk \rho(k,Q)\bigg{|},\quad
	\varepsilon(\eta)=\frac{\hbar^2}{2m} \bigg{|} \int_{Q}^{|\eta|} dq \frac{\rho(\eta,q)\tilde n(q)}{\rho(q,q)^2}\bigg{|}.
\end{align}
Here $\eta>Q$ for type-I excitations and $|\eta|<Q$ for  type-II excitations. Thus, once the density of quasimomenta in the ground state $\rho(k,Q)$ is known, the whole spectrum of elementary excitations can be obtained since both $p$ and $\varepsilon$ in equation~(\ref{eq:pepsfinal}) are expressed in terms of $\rho(k,Q)$.

\subsection{The excitation spectrum at low momenta}

In the parametric form, the spectrum of elementary excitations is given by equation~(\ref{eq:pepsfinal}). Since the Lieb--Liniger model describes a gapless one-dimensional Galilean-invariant liquid, at low momenta it is expected that the spectrum of type-I excitations acquires an explicit form
\begin{align}\label{eq:spectrumlowp}
	\varepsilon_p=v p+\frac{p^2}{2m^*}+\frac{\chi}{6}p^3+\frac{\nu}{24}p^4+\ldots.
\end{align}
Here we consider a right-moving excitation that has a positive momentum, $p>0$. The result (\ref{eq:spectrumlowp}) can be understood as a Taylor-series expansion where $v$ denotes the velocity of excitations, $m^*$ is their effective mass, and $\chi$ and $\nu$ describe the cubic and quartic terms. The latter parameters can thus be expressed in terms of $\rho(Q,Q)$ and its derivatives. We have studied these functions in section \ref{section:derivativesrho}.

Eliminating $\eta$ from  equation~(\ref{eq:pepsfinal}) we can obtain the parameters entering equation~(\ref{eq:spectrumlowp}). The velocity of excitations $v=[\partial\varepsilon(\eta)/\partial p(\eta)]|_{\eta=Q}$ is given by 
\begin{align}\label{eq:vvv}
	v=\frac{\hbar n}{4\pi m\rho(Q,Q)^2}=\frac{\pi\hbar n}{mK},
\end{align}
where we have used $\tilde n(Q)=n$. Equation (\ref{eq:vvv}) with $v$ denoting the velocity of excitations (or Fermi velocity) is equivalent to the expression (\ref{eq:mvK1}) with $v$ being the sound velocity. Therefore, the two velocities are identical and thus denoted by the same symbol.

The effective mass of excitations is defined by the relation $1/m^*=[\partial^2\varepsilon(\eta)/\partial p(\eta)^2]|_{\eta=Q}$. It can be expressed as
\begin{align}
	\frac{m}{m^*}=\frac{1}{8\pi^2\rho(Q,Q)^2}\frac{d}{d Q}\left(\frac{n}{\rho(Q,Q)}\right).
\end{align}
Using 
\begin{align}
	\frac{dn}{dQ}=4\pi\rho(Q,Q)^2
\end{align}
we eventually obtain
\begin{align}\label{eq:m*vv}
	\frac{m}{m^*}=\left(1-\gamma\;\!\frac{d}{d\gamma}\right)\frac{1}{\sqrt{K}}.
\end{align}
Therefore, the first two parameters of the spectrum (\ref{eq:spectrumlowp}) can be expressed in terms of the Luttinger liquid parameter $K$ and its derivative with respect to $\gamma$. Equivalently, $v$ and $m^*$ only depend on the quasimomentum distribution at the Fermi quasimomentum, $\rho(Q,Q)$, and its total derivative. 

The latter statement turns out not to be true for the remaining parameters in equation~(\ref{eq:spectrumlowp}). They also depend on partial derivatives of $\rho(Q,Q)$. For the cubic coefficient $\chi=[\partial^3\varepsilon(\eta)/\partial p(\eta)^3]|_{\eta=Q}$, using equations~(\ref{eq:PDE}) and (\ref{eq:totalderivative}) we obtain
\begin{align}\label{eq:chivv}
	\chi=\frac{n(2\dot\rho^2-\rho\ddot\rho)}{32\pi^3\hbar m \rho^6}+\frac{n\dot\rho -4\pi\rho^3}{8\pi^3\hbar m \rho^6}\rho^{(1,0)}.
\end{align}
Here for easier notation we have suppressed the arguments in $\rho(Q,Q)$, $\rho^{(1,0)}(Q,Q)$, etc. An alternative form for $\chi$ is obtained if we express $\rho^{(1,0)}$ in terms of the parameter $K_1$ according to equation~(\ref{eq:K1}) and $\rho$ in terms of $K$. The final result reads
\begin{align}\label{eq:chiKj}
	\chi=\frac{1}{\pi\hbar m n}\left(\frac{K_1}{K^2} +\frac{\gamma K'K_1}{2K^3}-\frac{\gamma K'}{2K} +\frac{\gamma^2K'^2}{8K^2} -\frac{\gamma^2 K''}{4K} \right),
\end{align}
where $K'=\frac{dK}{d\gamma}$, etc. Therefore, unlike $v$ and $m^*$ that only depend on $K$, the coefficient $\chi$ also depends on the dimensionless parameter $K_1(\gamma)$.

The quartic coefficient $\nu=[\partial^4\varepsilon(\eta)/\partial p(\eta)^4]|_{\eta=Q}$ can also be expressed in terms of $\rho(Q,Q)$ and its derivatives. It is given by
\begin{align}\label{eq:nuvv}
	\nu={}&-\frac{n(\rho^2\dddot\rho -5\rho\dot\rho\ddot\rho+ 4\dot\rho^3)}{128\pi^4\hbar^2 m\rho^8}+\frac{5n (\rho\ddot\rho-2\dot\rho^2)}{64\pi^4\hbar^2 m \rho^8} \rho^{(1,0)}+\frac{3(n\dot\rho-4\pi\rho^3)}{32\pi^4\hbar^2m\rho^8} [\rho\rho^{(2,0)}-3(\rho^{(1,0)})^2].
\end{align}
The coefficient $\nu$ depends on $\rho(Q,Q)$ and its total derivatives as well as the partial derivatives $\rho^{(1,0)}(Q,Q)$ and $\rho^{(2,0)}(Q,Q)$. Note, however, that the second partial derivative is not independent as it can be expressed in terms of the first one as derived in equation~(\ref{eq:Fho20solution}). Therefore, similarly as the cubic coefficient $\chi$, the quartic one $\nu$ can be expressed only in terms of the Luttinger liquid parameter $K$ and the parameter $K_1$ as
\begin{align}\label{eq:nuKj}
	\nu={}&\frac{K^{5/2}}{\pi^2 \hbar^2 n^2m}\biggl(\frac{3\gamma K'}{4K^3}  + \frac{3\gamma^2 K''}{4K^3}+ \frac{\gamma^3 K'''}{8K^3} -\frac{3K_1}{4K^4}-\frac{7\gamma K'K_1}{8K^5} +\frac{11\gamma^2 K'^2K_1}{16K^6}\notag\\
	&-\frac{5\gamma^2  K''K_1}{8K^5} +\frac{15 K_1^2}{8 K^6}  +\frac{15\gamma K'K_1^2}{16 K^7} -\frac{3\gamma K_1'}{4K^4} -\frac{3\gamma^2 K'K_1'}{8K^5} \biggr).
\end{align}

Equations (\ref{eq:vvv}) and (\ref{eq:m*vv})--(\ref{eq:nuKj}) are the exact results for the coefficients of the spectrum (\ref{eq:spectrumlowp}) of right-moving type-I excitations. The same parameters also enter the spectrum of type-II excitations that at low momenta acquires the form
\begin{align}\label{eq:spectrumlowpII}
	\varepsilon_p=v p-\frac{p^2}{2m^*}+\frac{\chi}{6}p^3-\frac{\nu}{24}p^4+\ldots.
\end{align}
We note that the spectrum of left-moving excitations is obtained by changing the sign of $p$ in $\varepsilon_p$. More generally, one can replace $p$ by $|p|$ in $\varepsilon_p$ in order to describe excitations of both signs of momenta.

Here we have shown that the spectrum of elementary excitations can be constructed only from the knowledge of the density of quasimomenta $\rho(k,Q)$ in the ground state, see equation~(\ref{eq:pepsfinal}). At low momenta, the excitations involve the quasimomenta near the Fermi quasimomentum $Q$. It is therefore not surprising that the nontrivial dependence in the parameters of the low-momentum spectrum is controlled only by the value of $\rho(k,Q)$ at $k=Q$ as well as its partial derivatives with respect to $k$ at the same point. Since $\rho(k,Q)$ is continuously differentiable at $k=Q$, the same set of parameters describes both type-I and type-II excitations at low momenta, see equations~(\ref{eq:spectrumlowp}) and (\ref{eq:spectrumlowpII}). The above procedure can be repeated for other Galilean-invariant integrable models that do not have singular phase shifts. Therefore, $\rho(k,Q)$ is an analytic function at any $k$ and equation~(\ref{eq:pepsfinal}) can be used as a starting point. In addition to the Lieb--Liniger model that is characterised by the phase shift (\ref{eq:phaseshift}), another one that belongs to the same class is the hyperbolic Calogero--Sutherland model \cite{sutherland}. For the former model we will below give explicit expressions for the parameters at weak and strong interactions. For the latter model we, however, cannot do that as we are not aware of the analytical expressions that describe $\rho(k,Q)$ for $k$ near the Fermi quasimomentum.

\subsubsection{Explicit results in terms of the interaction parameter.}

We are now in a position to list explicit results for the parameters of the spectrum (\ref{eq:spectrumlowp}) as a function of the interaction parameter $\gamma$. In the regime of weak interactions, $\gamma\ll 1$, we use the ground-state function (\ref{eq:e2nineterms}) to obtain 
\begin{gather}
	K={}\frac{\pi}{\sqrt\gamma}+\frac{1}{4}+\frac{3\sqrt\gamma}{32\pi}+\frac{1+3\zeta(3)}{128\pi^2}\gamma+\mathcal{O}(\gamma^{3/2}),\\
	v={}\frac{\hbar n}{m}\sqrt\gamma\left(1-\frac{\sqrt\gamma}{4\pi}-\frac{\gamma}{32\pi^2} -\frac{3\zeta(3)-3}{128\pi^3}\gamma^{3/2} +\mathcal{O}(\gamma^2)\right),\\
	\frac{m}{m^*}={}\frac{3\gamma^{1/4}}{4\sqrt\pi}\left(1-\frac{\sqrt\gamma}{24\pi} +\frac{\gamma}{128\pi^2} +\frac{12\zeta(3)-9}{1024\pi^3}\gamma^{3/2} +\mathcal{O}(\gamma^2)\right).
\end{gather}
For the remaining two coefficients we also need equations~(\ref{eq:rho10weak}) and (\ref{eq:rho20weak}) leading to
\begin{gather}
	\chi={}\frac{1}{4\hbar n m \sqrt\gamma}\left(1-\frac{11\sqrt\gamma}{12\pi}+\frac{19\gamma}{192\pi^2}+\mathcal{O}(\gamma^2)\right),\\
	\nu={}\frac{5\sqrt\pi}{32\hbar^2n^2m \gamma^{5/4}}\left(1-\frac{49\sqrt\gamma}{24\pi}+\frac{203\gamma}{128\pi^2} -\frac{159+12\zeta(3)}{1024\pi^3}\gamma^{3/2}+\mathcal{O}(\gamma^2)\right).
\end{gather}
We note that $\chi$ does not have the term proportional to $\gamma$ in the expansion at small $\gamma$. This can be related to the absence of the term proportional to $\gamma^2$ in $\frac{1}{m^*\sqrt K}$ as consequence of a differential equation that relates $\frac{\partial}{\partial\gamma}(\chi m^*\sqrt K)$ to a function that involves $\gamma$ and $K$ \cite{petkovic_spectrum_2018}.

At $\gamma\ll 1$, all four corrections to the linear spectrum in equation~(\ref{eq:spectrumlowp}) are on the same order of magnitude at the momentum scale $p^*(\gamma)=\hbar n\gamma^{3/4}$. Therefore, the expansion (\ref{eq:spectrumlowp}) has a usual sense at momenta $p\ll p^*$. For momenta higher than $p^*$, the spectrum acquires the Bogoliubov form (\ref{E_bogoliubov}). The crossover function between the spectra (\ref{eq:spectrumlowp}) and (\ref{E_Bogoliubovlowp}) is calculated in reference~\cite{pustilnik_low-energy_2014}.

Closely parallel procedure can be used for the regime of strong interactions, $\gamma\gg 1$. There the central quantity is the ground-state function (\ref{eq:e2largefinal}). It yields
\begin{gather}
	K=1+\frac{4}{\gamma}+\frac{4}{\gamma^2}-\frac{16\pi^2}{3\gamma^3} +\mathcal{O}(\gamma^{-4}),\\
	v=\frac{\pi\hbar n}{m}\left(1-\frac{4}{\gamma}+\frac{12}{\gamma^2}+\frac{16(\pi^2-6)}{3\gamma^3} +\mathcal{O}(\gamma^{-4})\right),\\
	\frac{m}{m^*}=1-\frac{4}{\gamma}+\frac{12}{\gamma^2} +\frac{32(\pi^2-3)}{3\gamma^3}+\mathcal{O}(\gamma^{-4}).
\end{gather}
For the remaining coefficients we should also use equations~(\ref{eq:rho10strong}) and (\ref{eq:rho20strong}). We obtain
\begin{gather}
	\chi={}\frac{16\pi}{\hbar n m \gamma^3}\left(1-\frac{10}{\gamma}+\frac{6(10-\pi^2)}{\gamma^2} +\frac{7(43\pi^2-120)}{3\gamma^3}+\mathcal{O}(\gamma^{-4})\right),\\
	\nu={}\frac{16}{\hbar^2 n^2 m \gamma^3}\left(1-\frac{10}{\gamma}-\frac{12(\pi^2-5)}{\gamma^2} +\frac{7(89\pi^2-120)}{3\gamma^3}+\mathcal{O}(\gamma^{-4})\right).
\end{gather}
In the Tonks--Girardeau limit $\gamma\to\infty$, the spectrum becomes quadratic,
\begin{align}\label{eq:spTG}
	\varepsilon_p=\frac{\pi\hbar n}{m}p+\frac{p^2}{2m}.
\end{align}
This result is exact for any $p$. At finite $\gamma\gg 1$, on the other hand, the spectrum (\ref{eq:spTG}) acquires corrections that can be described by the additional terms present in equation~(\ref{eq:spectrumlowp}). The regime of strong interactions is treated by expanding the kernel in the integral operator, see section \ref{section:simplesolution}. It leads to the condition $p\ll p^*(\gamma)$ with $p^*(\gamma)=\hbar n\gamma$ for the applicability of the spectrum (\ref{eq:spectrumlowp}).

\subsection{The excitation spectrum at high momenta}

The type-I excitation branch extends to arbitrary high momenta. Let us find its spectrum. Instead of using equations~(\ref{eq:pEtypeIfinal}) or (\ref{eq:pepsfinal}) it will be convenient to derive their equivalent form. Integrating the Lieb integral equation (\ref{eq:LIE}) over $k$ in the interval $0< k<\infty$ we obtain the constraint
\begin{align}\label{eq:LIEconstraint1}
	\int_0^\infty dk [2\pi\rho(k,Q)-1]=\pi n.
\end{align}
The result (\ref{eq:LIEconstraint1}) implies another constraint
\begin{align}\label{eq:LIEconstraint2}
	\int_Q^\infty dk [2\pi\rho(k,Q)-1]=Q.
\end{align}
Equation (\ref{eq:LIEconstraint2}) enables us to express the momentum of the excitation from equation~(\ref{eq:pEtypeIfinal}) as
\begin{align}\label{eq:pversnionlargeeta}
	p=\hbar\eta -\hbar \int_{\eta}^{\infty} dk [2\pi\rho(k,Q)-1].
\end{align}
This can be further transformed. Using equation~(\ref{eq:LIE}) we eventually obtain
\begin{align}\label{eq:petaexact}
	p=\hbar\eta-2\hbar \int_{-Q}^{Q}dq\;\!\rho(q,Q)\arctan\left(\frac{c}{\eta-q}\right).
\end{align}
Equation (\ref{eq:petaexact}) is exact.

In order to transform the expression for the excitation energy of equation~(\ref{eq:pEtypeIfinal}), let us introduce the function $\mathcal{E}(k,Q)$ to be the solution of the integral equation 
\begin{align}\label{eq:LIEenergy}
	\mathcal{E}(k,Q)+\frac{1}{2\pi}\int_{-Q}^{Q} dq\,\theta'(k-q)\mathcal{E}(q,Q) =\frac{\hbar^2 k^2}{2m}-\mu.
\end{align}
Here the constant $\mu$ should be selected in such a way that the condition 
\begin{align}\label{eq:epsconditionzero}
	\mathcal{E}(Q,Q)=0
\end{align}
is fulfilled. Using the functions introduced by equation~(\ref{eq:Feq}) we can express $\mathcal{E}(k,Q)=\frac{\hbar^2}{m}\rho_2(k,Q)-\mu\rho_0(k,Q)$. The condition (\ref{eq:epsconditionzero}) then gives $\mu=\frac{\hbar^2\rho_2(Q,Q)}{m\rho_0(Q,Q)}$. With the help of equation~(\ref{eq:Fodd1}) we eventually obtain
\begin{align}\label{eq:muEps}
	\mu=-\frac{\hbar^2 n^2}{2m}\gamma^4 \frac{\partial}{\partial\gamma}\left(\frac{e_2(\gamma)}{\gamma^3}\right).
\end{align}
This expression coincides with equation~(\ref{eq:mu2}). Therefore, $\mu$ in equation~(\ref{eq:LIEenergy}) denotes the chemical potential of the system. Differentiating equation~(\ref{eq:LIEenergy}) with respect to $k$ we obtain the connection
\begin{align}
	\mathcal{E}^{(1,0)}(k,Q)=\sigma(k,Q),
\end{align}
where $\sigma(k,Q)$ satisfies equation~(\ref{eq:sigma}). This simplifies the expression for the excitation energy of equation~(\ref{eq:pEtypeIfinal}) to \cite{korepin}
\begin{align}\label{eq:epsexcitationenergy}
	\varepsilon=\mathcal{E}(\eta,Q).
\end{align}
We have thus found another exact form for the excitation spectrum that is given by equations~(\ref{eq:petaexact}) and (\ref{eq:epsexcitationenergy}).

Let us find the excitation spectrum at high momenta, $p\gg \hbar Q$. At such momenta, the parameter $\eta$ in equation~(\ref{eq:petaexact}) is large, while the integral is only a small contribution. We thus obtain
\begin{align}\label{eq:eta}
	\eta=\frac{p}{\hbar}+2n \arctan\left(\frac{\hbar n\gamma}{p}\right)+\ldots.
\end{align}
The terms in the ellipsis have an additional smallness with respect to the second term. These small terms have different forms that depend on the ratio of $\hbar n\gamma$ and $p$. At high values of $\eta$ that are accounted for by equation~(\ref{eq:eta}), the energy of equation~(\ref{eq:epsexcitationenergy}) is dominated by the right-hand side of equation~(\ref{eq:LIEenergy}). It yields the excitation spectrum
\begin{align}\label{eq:spectrumhighp}
	\varepsilon_p=\frac{p^2}{2m}+\frac{2\hbar n p}{m}\arctan\left(\frac{\hbar n\gamma}{p}\right)+ \frac{2\hbar^2 n^2}{m}\arctan^2\left(\frac{\hbar n\gamma}{p}\right) -\mu+\ldots.
\end{align} 
Equation (\ref{eq:spectrumhighp}) applies if $p\gg\hbar Q$ is fulfilled, but the ratio between $p$ and $\hbar n\gamma$ can be arbitrary.

Let us discuss the spectrum (\ref{eq:spectrumhighp}). At strong interactions, $\gamma\gg 1$, the Fermi quasimomentum is $Q\sim n$. At momenta that satisfy $\hbar n\ll p\ll \hbar n\gamma$, equation~(\ref{eq:spectrumhighp}) at the leading order in $\gamma$ reduces to the dispersion of the Tonks--Girardeau limit (\ref{eq:spTG}). Accounting for more terms of $1/\gamma$-expansion, one would reproduce the spectrum (\ref{eq:spectrumlowp}), which also extends to the smallest values of $p$. At highest momenta, $p\gg \hbar n\gamma$, equation~(\ref{eq:spectrumhighp}) simplifies to~\cite{lieb_exact_1963b}
\begin{align}\label{eq:LLb}
	\varepsilon_p=\frac{p^2}{2m}+2\gamma\frac{\hbar^2 n^2}{m}-\mu+\mathcal{O}\left(\frac{1}{p^2}\right).
\end{align}
Therefore, the spectrum (\ref{eq:spectrumhighp}) describes the crossover between the two limiting cases (\ref{eq:spTG}) and (\ref{eq:LLb}). In the regime of weak interactions, $\gamma\ll 1$, we have $Q\sim n\sqrt\gamma$ and thus equation~(\ref{eq:spectrumhighp}) applies at $p\gg\hbar n\sqrt\gamma$. In this case, however, the $\arctan$-functions must be expanded leading to the spectrum (\ref{eq:LLb}). Therefore, the result (\ref{eq:LLb}) applies at any $\gamma$ if $p$ is sufficiently high. Substituting  $\mu=\gamma\frac{\hbar^2 n^2}{m}$ appropriate for $\gamma\ll 1$, equation~(\ref{eq:LLb}) reproduces the result (\ref{Ehighp}) obtained by different means.

\subsection{Discussion of the obtained results}

\begin{figure}
\centering
\includegraphics[width=0.6\textwidth]{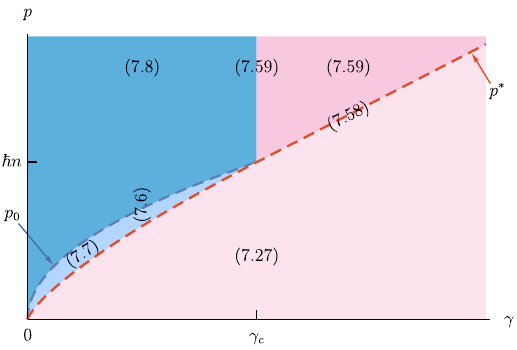}
\caption{\justifying The spectrum of type-I elementary excitations in the Lieb--Liniger model has its characteristic form in several regions of the interaction-momentum plane. The number in parentheses denotes the equation where the spectrum is given. The crossover momenta are $p_0\sim\gamma^{1/2}$ and $p^*\sim\gamma^{3/4}$ at small $\gamma$, and $p^*\sim\gamma$ at large $\gamma$. The asymptotes cross at $\gamma_c=1$, corresponding to the momentum $\hbar n$. At weak interactions, the Bogoliubov spectrum (\ref{E_bogoliubov}) describes bosonic quasiparticles that have the momenta higher than $p_*$, while at momenta lower than $p^*$, the spectrum describes fermionic quasiparticles.}\label{fig:spectrum}
\end{figure}

The spectrum of type-I excitations in the Lieb--Liniger model is nontrivial. Using the Bethe ansatz, it can be expressed parametrically in several forms that we previously derived. Such forms of the spectrum are expressed in terms of quantities that are solutions of integral equations and do not have simple forms at an arbitrary interaction. The parametric spectrum have then been analysed enabling obtainment of the explicit forms of the spectrum for certain regions of momenta. The results are summarised in figure~\ref{fig:spectrum}.

Two main features of the spectrum of elementary excitations at weak interactions that follow from the hydrodynamic approach are the existence of fermionic and bosonic quasiparticles. At lowest momenta, the initial Hamiltonian (\ref{eq:Horiginal}) is diagonalised in terms of fermionic quasiparticle operators and the leading correction to the lowest-order linear spectrum is quadratic, see equation~(\ref{E_weakint_fermions}). At slightly higher momenta, above $p^*$, the Hamiltonian (\ref{eq:Horiginal}) is diagonalised in terms of bosonic quasiparticle operators and the resulting spectrum has the standard Bogoliubov form. There, the leading correction to the spectrum is cubic, see equation~(\ref{E_Bogoliubovlowp}). 

The spectrum of elementary excitations can also be obtained from the Bethe ansatz, however, this technique does not address the statistics of quasiparticles. On the other hand, it gives the exact results. Nevertheless such exact result, for example equation~(\ref{eq:pepsfinal}), are given in the parametric form and rely on the integrals that involve the quasimomentum distribution $\rho(k,Q)$, which itself does not have a closed-form expression. Rather, $\rho(k,Q)$ is the unique solution of the integral equation (\ref{eq:LIE}). Based on the properties of the latter studied in section \ref{section:derivativesrho}, we obtained the exact form of the spectrum at small momenta, see equation~(\ref{eq:spectrumlowp}). It has the power-series form where the coefficients are expressed as closed-form expressions that depend on the quasimomentum distribution at the Fermi quasimomentum and its derivatives. 

The spectrum (\ref{eq:spectrumlowp}) is a natural extension to all interaction strengths $\gamma$ of the fermionic spectrum (\ref{E_weakint_fermions}) obtained at small $\gamma$. It is therefore expected that the lowest-energy excitations have the fermionic statistics. This observation can be further strengthened by noticing that the Lieb--Liniger model of repulsive bosons can be studied within the dual Cheon--Shigehara model of fermions with attraction \cite{cheon_fermion-boson_1999,khodas_dynamics_2007,granet_duality_2022}. There the attraction strength is inversely proportional to the repulsion. Free fermionic model has fermionic quasiparticles with the spectrum (\ref{eq:spTG}). Accounting for small attraction, the parameters of the spectrum will get slightly renormalized, but not the statistics. The obtained fermionic quasiparticles should actually be the ones that have the spectrum (\ref{eq:spectrumlowp}). Having the two quasiparticle pictures at weak and strong interactions that support fermionic quasiparticles as the ones with the lowest energies, it is very plausible that the same quasiparticles extend to any repulsion. Moreover, they should also extend to any momentum at strong interactions. In the limiting case $\gamma\to\infty$, equation~(\ref{eq:spTG}) describes quasiparticle excitations in the system of bosons. In this case the infinite repulsion prevents two bosons to share the same space position, acting effectively as Pauli principle on fermions. These considerations undoubtedly suggest that fermionic quasiparticle excitations exist at all momenta at strong interactions, and  at lowest momenta for arbitrary interactions. The existence of Bogoliubov excitations is limited only to weak interactions and sufficiently large momenta, see figure~\ref{fig:spectrum}.

\section{Low-temperature thermodynamics \label{section:thermodynamics}}

In 1969 Yang and Yang showed that the thermodynamic properties of the Lieb--Liniger model can be calculated exactly by solving so-called thermodynamic Bethe ansatz equations \cite{yang_thermodynamics_1969}.  Despite this formally exact solution, extracting analytical forms of relevant quantities valid without limitation on the interaction strength is a formidable task. A notable exception is the leading low-temperature result for the Helmholtz free energy, 
\begin{align}\label{eq:HFE}
	F=E_0-\frac{\pi L T^2}{6\hbar v}+\mathcal{O}(T^4),
\end{align}
which was obtained in the framework of conformal field theory \cite{affleck_universal_1986,blote_conformal_1986}. The leading temperature-dependent term applies at any interaction strength. It relies on the linear spectrum of low-energy excitations characterised by the velocity $v$. This approach should be contrasted to the others, based on the Bethe ansatz \cite{guan_polylogs_2011} or the effective quasiparticle picture \cite{kerr_analytic_2024,de_rosi_beyond-luttinger-liquid_2019}, that treat the nonlinear spectrum but are limited to weak or strong interactions. Here we overcome these difficulties. Using the local properties of the quasimomentum distribution derived in section \ref{section:derivativesrho} and the Yang--Yang theory, we develop a systematic way to calculate the thermodynamic quantities at low temperatures. In particular, we obtain the term proportional to $T^4$ in the free energy of equation~(\ref{eq:HFE}). It is valid at any interaction and can be understood as the leading correction to the result obtained using conformal field theory. This section is based on the work \cite{panfil_local_2025}.

\subsection{The Yang--Yang thermodynamics}

The Bethe ansatz equations (\ref{eq:DBA}) are parametrised by the quantum numbers $I_j$. Their admissible values are either integers or odd half-integers depending on the parity of the particle number $N$, see equation~(\ref{eq:Ij}). The quantum numbers that characterize the ground state of the system take the values (\ref{eq:quantumnumbers}), forming an equidistant array between $-(N-1)/2$ and $(N-1)/2$. The quantum numbers that participate in equations for quasimomenta entering the Bethe wave function (\ref{eq:wave function}) are called particle quantum numbers. Other admissible nonparticipating quantum numbers are called hole quantum numbers. In an excited state, some of the elements of the original array between $-(N-1)/2$ and $(N-1)/2$ do not participate in the description of the state. They become hole quantum numbers and, vice versa, some of the original hole quantum numbers become particle quantum numbers.

Consider extension of equation~(\ref{eq:DBA}) where $I_j$ take all admissible particle and hole quantum numbers. Let us call the extended set of $k_j$ vacancies as they denote potential positions in momentum space that quasimomenta entering the wave function (\ref{eq:wave function}) can take. Defining three densities, of vacancies $\rho_t(k)$, of particle quasimomenta $\bar\rho(k)$, and hole quasimomenta $\rho_h(k)$, similarly as in equation~(\ref{eq:Fhokj}), we trivially have $\rho_t(k)=\bar\rho(q)+\rho_h(q)$. From the extension of equation~(\ref{eq:DBA}), in the thermodynamic limit we obtain \cite{yang_thermodynamics_1969,korepin}
\begin{align}\label{eq:LIETBA}
	\rho_t(k)=\bar\rho(q)+\rho_h(q)=\frac{1}{2\pi} -\frac{1}{2\pi}\int_{-\infty}^{\infty}dq\:\!\theta'(k-q)\bar\rho(q).
\end{align}
This equation gives a connection between the densities of hole and particle quasimomenta. Note that the integral on the right-hand side only involves the density of particle quasimomenta.

At finite temperatures $T>0$, particle quasimomenta are not limited in absolute value by the Fermi quasimomentum, but spread over all momenta. Yang and Yang showed \cite{yang_thermodynamics_1969} that the state of thermal equilibrium is achieved in the case the densities of quasimomenta of particles $\bar\rho(k)$ and holes $\rho_h(k)$ obey
\begin{align}
	\frac{\rho_h(k)}{\bar\rho(k)}=e^{\bar{\mathcal{E}}(k)/T},
\end{align} 
where the function $\bar{\mathcal{E}}(k)$ satisfies the Yang--Yang equation
\begin{align}\label{eq:YYoriginal}
	\bar{\mathcal{E}}(k)=-\bar \mu+\frac{\hbar^2 k^2}{2m}+ \frac{T}{2\pi}\int_{-\infty}^{+\infty} dq\;\! \theta'(q-k)\ln\left(1+e^{-\bar{\mathcal{E}}(q)/T}\right).
\end{align}
Here $\bar\mu$ denotes the chemical potential of the system. Once $\bar{\mathcal{E}}(k)$ is found, the pressure of the gas can be expressed as
\begin{align}\label{eq:pressureYYoriginal}
	\bar P=\frac{T}{2\pi}\int_{-\infty}^{+\infty} dq \ln\left( {1+e^{-\bar{\mathcal{E}}(q)/T}}\right).
\end{align}
Equations (\ref{eq:YYoriginal}) and (\ref{eq:pressureYYoriginal}) represent exact results that can be understood as the grand-canonical description of equilibrium properties of the Lieb--Liniger model.

\subsection{The Yang--Yang equation at low temperatures}

After a partial integration, equation~(\ref{eq:YYoriginal}) can be expressed as
\begin{align}\label{eq:YY}
	\bar{\mathcal{E}}(k)=-\bar\mu+\frac{\hbar^2 k^2}{2m}+ \frac{1}{2\pi}\int_{-\infty}^{+\infty} dq\;\! \frac{\theta(q-k)\bar{\mathcal{E}}'(q)}{1+e^{\bar{\mathcal{E}}(q)/T}}.
\end{align}
The expression (\ref{eq:YY}) will serve as our starting point for the evaluation of the function $\bar{\mathcal{E}}(k)$ at a given chemical potential $\bar\mu$ and the temperature $T$. We consider $\bar\mu>0$. Without entering into mathematical rigor \cite{yang_thermodynamics_1969,kozlowski_low-t_2014}, we will assume that $\bar{\mathcal{E}}(k)$ is an even real function, which is monotonically increasing for $k>0$ with the value $\bar{\mathcal{E}}(0)<0$ for $\bar\mu>0$.\footnote{\linespread{1}\selectfont One can think about the solution obtained by iterations \cite{yang_thermodynamics_1969}.} Since $\bar{\mathcal{E}}(k)\propto k^2$ at $k\to+\infty$, the function nullifies at certain $k=\bar Q$, 
\begin{align}\label{eq:Energy-Q-condition}
	\bar{\mathcal{E}}(\bar Q)=0.
\end{align}
Our goal is to solve equation~(\ref{eq:YY}) at low temperatures and then calculate the thermodynamic parameters at an arbitrary interaction.

In the zero-temperature limit, the denominator in the integral of equation~(\ref{eq:YY}) makes the boundary of integration finite and it thus reduces to the linear equation (\ref{eq:LIEenergy}). At low temperatures, the integral in equation~(\ref{eq:YY}) admits a series expansion in even powers of temperature, as we will discuss later. This is manifested as a series
\begin{gather}\label{eq:Energy-T}
	\bar{\mathcal{E}}(k)=\mathcal{E}(k)+T^2 \mathcal{E}_2(k)+T^4\mathcal{E}_4(k)+\mathcal{O}(T^6).
\end{gather}
The temperature also affects the point $k=\bar Q$ where $\bar{\mathcal{E}}(k)$ nullifies. Therefore $\bar Q$ also admits an expansion of the form
\begin{gather}\label{eq:Q-T}
	\bar Q=Q+T^2 Q_2+T^4 Q_4+\mathcal{O}(T^6).
\end{gather}
The requirement (\ref{eq:Energy-Q-condition}) at the leading order then reduces to 
\begin{gather}\label{eq:conditionE0}
	\mathcal{E}(Q)=0,
\end{gather}
while from the higher orders one can express $Q_2$ and $Q_4$ in terms of $\mathcal{E}(Q)$, $\mathcal{E}_2(Q)$, $\mathcal{E}_4(Q)$, and the derivatives of $\mathcal{E}(k)$ at $k=Q$.

Using the previously adopted notation
\begin{align}
	\mathcal{F}[\mathcal{E}(k)]=\mathcal{E}(k)+\frac{1}{2\pi}\int_{-Q}^{Q} dq\;\!\theta'(k-q)\mathcal{E}(q),
\end{align}
the functions entering the right-hand side of equation~(\ref{eq:Energy-T}) fulfil
\begin{subequations}\label{eq:e0e2e4}
	\begin{gather}\label{eq:e0}
		\mathcal{F}[\mathcal{E}(k)]=-\bar\mu+\frac{\hbar^2 k^2}{2m},\\
		\label{eq:e2}
		\mathcal{F}[\mathcal{E}_2(k)]=b_{12}[\theta'(Q-k)+\theta'(Q+k)],\\
		\label{eq:e4}
		\mathcal{F}[\mathcal{E}_4(k)]=\sum_{i=1}^3 b_{i4}	[\theta^{(i)}(Q-k)+\theta^{(i)}(Q+k)].
	\end{gather}
\end{subequations}
Here, 
\begin{subequations} 
	\begin{gather}
		b_{12}={}\frac{\pi}{12\mathcal{E}'(Q)},\\
		b_{14}={}\frac{[\mathcal{E}_2'(Q)]^2}{4\pi \mathcal{E}'(Q)}-\frac{\pi \mathcal{E}_2'(Q)}{12[\mathcal{E}'(Q)]^2}+\frac{\pi\mathcal{E}_2(Q) \mathcal{E}''(Q)}{12[\mathcal{E}'(Q)]^3}-\frac{7\pi^3 \mathcal{E}'''(Q)}{720 [\mathcal{E}'(Q)]^4}+\frac{7\pi^3[\mathcal{E}''(Q)]^2}{240 [\mathcal{E}'(Q)]^5},\\
		b_{24}={}-\frac{\pi \mathcal{E}_2(Q)}{12 [\mathcal{E}'(Q)]^2}-\frac{7\pi^3\mathcal{E}''(Q)}{240[\mathcal{E}'(Q)]^4},\\
		b_{34}={}\frac{7\pi^3}{720[\mathcal{E}'(Q)]^3}.
	\end{gather}
\end{subequations}
The principal steps of the calculation are explained in Appendix \ref{appendix:sommerfeld}. We notice that the hierarchy of equations for each power of $T$ generated from equation~(\ref{eq:YY}) formally also contains odd powers of $T$ in equation~(\ref{eq:Energy-Q-condition}). They, however reduce to the equations $\mathcal{F}[\mathcal{E}_1(k)]=0$ and $\mathcal{F}[\mathcal{E}_3(k)]=0$ for the terms linear and cubic in $T$. Due to the property of the operator $\mathcal{F}$ that reads
\begin{align}\label{eq:propertyF}
	\mathcal{F}[g]=0\quad \textrm{implies}\quad g=0,
\end{align} 
see Appendix \ref{appendixA}, we find $\mathcal{E}_1(k)=\mathcal{E}_3(k)=0$ and thus we omitted the terms proportional to odd powers of $T$. Due to the same reason, there are no such terms in equation~(\ref{eq:Q-T}). It is worth to mention a comment about the notation. In this section we use the bar over certain symbols in order to keep the notation of previous sections unique. These symbols usually denote temperature-dependent quantities, such as $\bar Q$ and $\bar{\mathcal{E}}$. At $T=0$ these symbols reduce to $Q$ and $\mathcal{E}$, corresponding to the same quantities used in earlier sections where the case of zero temperature was implicitly assumed.

Let us consider the Lieb integral equation for the quasimomentum density,
\begin{align}\label{eq:LieQ0}
	\mathcal{F}[\rho(k,Q)]=\frac{1}{2\pi}.
\end{align}
Performing the differentiation of equation~(\ref{eq:LieQ0}) with respect to $Q$, we obtain integral equations of the form $\frac{\partial^j \mathcal{F}[\rho]}{\partial Q^j}=0$ that determine the function $\frac{\partial^j\rho(k,Q)}{\partial Q^j}\equiv\rho^{(0,j)}(k,Q)$. Here $j$ is an arbitrary positive integer. Their explicit form is given by
\begin{align}\label{eq:dQF}
	\mathcal{F}\left[\rho^{(0,j)}(k,Q) \right] +\sum_{i=1}^{j}a_{i j}[\theta^{(i)}(Q-k)+\theta^{(i)}(Q+k)]=0.
\end{align}
Here we have introduced the notation $\theta^{(i)}(k)=\frac{d^i\theta(k)}{dk^i}$. For $1\le j\le 3$, the coefficients are given by
\begin{subequations}
	\begin{gather}
		a_{11}{}=a_{22}=a_{33}=\frac{\rho(Q,Q)}{2\pi},\\
		a_{12}{}=\frac{\rho^{(0,1)}(Q,Q)+\dot{\rho}(Q,Q)}{2\pi},\\
		a_{13}{}=\frac{\rho^{(0,2)}(Q,Q)+\dot{\rho}^{(0,1)}(Q,Q) +\ddot{\rho}(Q,Q)}{2\pi},\\
		a_{23}{}=\frac{\rho^{(0,1)}(Q,Q)+2\dot{\rho}(Q,Q)}{2\pi}.
	\end{gather}
\end{subequations}
Here the dot under the symbol denotes its total derivative, $\dot{\rho}(Q,Q)=\frac{d\rho(Q,Q)}{dQ}=\rho^{(0,1)}(Q,Q) +\rho^{(1,0)}(Q,Q)$, etc.

Equation (\ref{eq:dQF}) enables us to express the derivatives of $\theta$-functions in equations~(\ref{eq:e0e2e4}) in terms of $\mathcal{F}\left[\rho^{(0,j)}(k,Q) \right]$. Using the linearity of the operator $\mathcal{F}$ and the property (\ref{eq:propertyF}), we obtain
\begin{subequations}
	\begin{gather}
		\mathcal{E}_2(k)={}c_{12} \rho^{(0,1)}(k,Q),\\
		\mathcal{E}_4(k)={} c_{14}\rho^{(0,1)}(k,Q)+c_{24}  \rho^{(0,2)}(k,Q)+c_{34}\rho^{(0,3)}(k,Q),
	\end{gather}
\end{subequations}
where
\begin{subequations}
	\begin{gather}
		c_{12}={}-\frac{b_{12}}{a_{11}},\\
		c_{14}={}-\frac{b_{14}}{a_{11}}+\frac{a_{12}b_{24}}{a_{11}a_{22}} +\frac{a_{13}b_{34}}{a_{11}a_{33}}-\frac{a_{12}a_{23} b_{34}} {a_{11}a_{22}a_{33}},\\
		c_{24}={}-\frac{b_{24}}{a_{22}}+\frac{a_{23}b_{34}}{a_{22}a_{33}},\\
		c_{34}={}-\frac{b_{34}}{a_{33}}.
	\end{gather}
\end{subequations}
We have therefore expressed the functions $\mathcal{E}_2(k)$ and $\mathcal{E}_4(k)$ that determine the low-temperature thermodynamics in terms of the partial derivatives of $\rho(k,Q)$. The advantage of using the latter functions is that they have been calculated explicitly in section \ref{section:derivativesrho}.

\subsubsection{The derivatives of the Yang--Yang energy.}

The derivatives of the Yang--Yang energy $\mathcal{E}^{(j)}(Q)$ can be calculated using the formalism of section \ref{section:method}. In terms of the functions introduced by equation~(\ref{eq:Feq}), the solution of equation~(\ref{eq:e0}) can be expressed as
\begin{align}\label{eq:eps0r0r2}
	\mathcal{E}(k)=-\bar \mu \rho_0(k,Q)+\frac{\hbar^2}{m}\rho_2(k,Q).
\end{align}
Using equations~(\ref{eq:Fhoder1}) and (\ref{eq:Fhoder2}), we obtain
\begin{align}
	\mathcal{E}'(k)=\frac{\hbar^2}{m}\rho_1(k,Q),
\end{align}
where we have accounted for the condition (\ref{eq:conditionE0}). The same equations lead to
\begin{align}
	\rho_1^{(1,0)}(k,Q)=\rho_0(k,Q)-\frac{\rho_1(Q,Q)}{\rho_0(Q,Q)} \rho_0^{(0,1)}(k,Q),
\end{align}
while equation~(\ref{eq:rho0rho1}) becomes
\begin{align}
	\rho_0(Q,Q)\rho_1(Q,Q)=\pi n.
\end{align}
The derivatives are thus given by
\begin{align}\label{eq:eps0prime}
	\mathcal{E}'(Q)=\frac{\hbar^2 n}{2m}\frac{1}{\rho(Q,Q)}
\end{align}
and
\begin{align}
	\mathcal{E}^{(j)}(Q)={}\frac{2\pi\hbar^2}{m}\rho^{(j-2,0)}(Q,Q)-\frac{\hbar^2 n}{2m\rho(Q,Q)^2}\rho^{(j-2,1)}(Q,Q),
\end{align}
for $j=2,3,4,\ldots$. Here we have used $\rho_0(k,Q)=2\pi\rho(k,Q)$.

\subsubsection{The pressure.}

The pressure of the gas (\ref{eq:pressureYYoriginal}) can be expressed as 
\begin{align}\label{eq:pressureYY}
	\bar P= \frac{1}{2\pi}\int_{-\infty}^{+\infty} dq\;\! \frac{q\;\! \bar{\mathcal{E}}'(q)}{1+e^{\bar{\mathcal{E}}(q)/T}}.
\end{align}
At low temperatures, we use the expansion detailed in Appendix \ref{appendix:sommerfeld} and obtain
\begin{align}\label{eq:Pbar}
	\bar P=P+T^2 P_2+T^4 P_4+\mathcal{O}(T^6),
\end{align}
where
\begin{subequations}
	\begin{gather}
		\label{eq:PbarP}
		P=-\frac{1}{2\pi}\int_{-Q}^{Q} dq\;\! \mathcal{E}(q),\\
		P_2=2b_{12}-\frac{1}{2\pi}\int_{-Q}^{Q} dq\;\! \mathcal{E}_2(q),\\
		P_4=2b_{14}-\frac{1}{2\pi}\int_{-Q}^{Q} dq\;\! \mathcal{E}_4(q).
	\end{gather}
\end{subequations}
The integrals of $\mathcal{E}$-functions can be calculated using the theorem discussed in Appendix \ref{appendixb} combined with the expression obtained  differentiating equation~(\ref{eq:LieQ0}) with respect to $k$. It yields 
\begin{subequations}
	\label{eq:Pbarparts}
	\begin{gather}
		\label{eq:PbarP1}
		P=\bar\mu n-\frac{E_0}{L},\\	
		P_2=4\pi b_{12}\rho(Q,Q),\\
		P_4=4\pi \left(b_{14}\rho(Q,Q)+b_{24}\rho^{(1,0)}(Q,Q)+b_{34}\rho^{(2,0)}(Q,Q)\right).
	\end{gather}
\end{subequations}
Here $n=\frac{N}{L}$ is the density of particles and $E_0$ is the ground-state energy (in the canonical ensemble where $n$ is fixed).
Equations (\ref{eq:Pbar}) and (\ref{eq:Pbarparts}) determine the pressure of the system in the grand canonical ensemble. It should be understood as an expression that depends on the chemical potential $\bar\mu$ and the temperature $T$. Interestingly, $b$-coefficients that enter equations~(\ref{eq:e0e2e4}) also enter the terms in the pressure. We note that the leading-order correction to the pressure is quadratic in $T$ with a rather simple form
\begin{align}\label{eq:P2final}
	P_2=\frac{\pi}{6\hbar v}.
\end{align}
The first subleading correction is quartic in $T$ and more complicated. It can be expressed as
\begin{align}\label{eq:P4final}
	P_4={}&\frac{224\pi^6 m^3 \rho^8}{15\hbar^6 n^5}\biggl[ 1-\frac{13n\dot\rho}{21\pi\rho^3} +\frac{19n^2\dot\rho^2}{336\pi^2\rho^6} +\frac{17n^2\ddot\rho}{672\pi^2\rho^5} +\frac{2n\rho^{(1,0)}}{7\pi\rho^3} -\frac{n^2\dot\rho\rho^{(1,0)}}{14\pi^2\rho^6} \notag\\  & +\frac{5n^2(\rho^{(1,0)})^2}{336\pi^2\rho^6} -\frac{5n^2\rho^{(2,0)}}{168\pi^2\rho^5}\biggr].
\end{align}
Here for easier notation we suppressed the arguments in $\rho(Q,Q)$, $\rho^{(1,0)}(Q,Q)$, $\dot\rho(Q,Q)$, etc., with the dot over the symbol denoting the total derivative. Using the results of section \ref{section:derivativesrho}, equation~(\ref{eq:P4final}) can be further evaluated analytically at small and large $\gamma$.

\subsubsection{The density.}

In the formalism of the grand-canonical ensemble the chemical potential $\bar\mu$ is kept fixed and the density $\bar n$ is a temperature-dependent quantity. It can be calculated from the pressure $\bar P$ via the expression
\begin{align}\label{eq:nmu}
	\bar n=\left(\frac{\partial \bar P}{\partial \bar\mu}\right)_T
	=\frac{\left(\frac{\partial \bar P}{\partial Q}\right)_T}{\left(\frac{\partial \bar\mu}{\partial Q}\right)_T}.
\end{align}
Accounting for equations~(\ref{eq:Pbar}) and (\ref{eq:PbarP1}), we obtain
\begin{align}\label{eq:barn}
	\bar n = n + \frac{T^2}{\hbar v}\frac{\partial P_2}{\partial Q} + \frac{T^4}{\hbar v}\frac{\partial P_4}{\partial Q}+\mathcal{O}(T^6),
\end{align}
where we have used 
\begin{align}\label{eq:muQ}
	\frac{\partial\bar \mu}{\partial Q}=\hbar v.
\end{align}
The latter expression can be obtained from the condition (\ref{eq:conditionE0}) leading to $\bar\mu=\frac{\hbar\rho_2(Q,Q)}{m\rho_0(Q,Q)}$. Combining with equation~(\ref{eq:Fodd1}) we then obtain $\bar\mu$ to be equal to the right-hand side of equation~(\ref{eq:mu2}). Using $\frac{\partial}{\partial Q}=-\frac{\gamma K}{\pi n} \frac{\partial}{\partial \gamma}$ and equation~(\ref{eq:defK}), the result (\ref{eq:muQ}) follows. 

\subsection{The results in the canonical ensemble}

The Yang--Yang equation naturally leads to the results in the grand-canonical ensemble. Let us now consider the canonical ensemble.

\subsubsection{The chemical potential.}

The density in the grand-canonical ensemble has the form
\begin{align}\label{eq:ngeneric}
	\bar n=n+T^2 n_2+T^4 n_4+\mathcal{O}(T^6),
\end{align}
see equation~(\ref{eq:barn}). In this ensemble the chemical potential $\bar\mu$ is fixed. If we want to obtain the results in the canonical ensemble where the density is fixed, $\bar n=n=N/L$ with $N$ being the number of particles and $L$ the system size, and the chemical potential depends on the temperature, we seek the latter in the form
\begin{align}\label{eq:mugeneric}
	\bar\mu=\mu+T^2 \mu_2+T^4\mu_4+\ldots.
\end{align}
Since $\bar\mu$, $n$, and $T$ are mutually related,\footnote{\linespread{1}\selectfont In the thermodynamic Bethe ansatz \cite{yang_thermodynamics_1969}, the density is an integral of a function that depends on $\bar\mu$ and $T$.} there is an exact relation
\begin{align}
	\left(\frac{\partial \bar\mu}{\partial T}\right)_n {\left(\frac{\partial n}{\partial \bar\mu}\right)_T}=-\left(\frac{\partial n}{\partial T}\right)_{\bar\mu}.
\end{align}
Expressing the derivative with respect to $\bar\mu$ via $Q$, we obtain
\begin{align}\label{eq:muT}
	\left(\frac{\partial \bar\mu}{\partial T}\right)_n \left(\frac{\partial n}{\partial Q}\right)_T =-\left(\frac{\partial n}{\partial T}\right)_{\bar\mu} \left(\frac{\partial\bar\mu}{\partial Q}\right)_T.
\end{align}
Substituting equations~(\ref{eq:ngeneric}) and (\ref{eq:mugeneric}) into
equation~(\ref{eq:muT}), we obtain a hierarchy of expressions with different powers of $T$ that must be satisfied. At the lowest order we obtain
\begin{align}
	\mu_2=-n_2\frac{\frac{\partial \mu}{\partial Q}}{\frac{\partial n}{\partial Q}}=-n_2\frac{\partial\mu}{\partial n},
\end{align}
which is
\begin{align}\label{eq:mu2gg}
	\mu_2=-\frac{\pi}{K}\frac{\partial P_2}{\partial Q}.
\end{align}
At the next order in small $T$, we obtain
\begin{align}
	\mu_4=-n_4 \frac{\partial\mu}{\partial n}  -\frac{\frac{\partial}{\partial Q} (n_2\mu_2)}{2\frac{\partial n}{\partial Q}},
\end{align}
leading to
\begin{align}\label{eq:mu4gg}
	\mu_4=-\frac{\pi}{K}\frac{\partial{P}_4}{\partial Q} +\frac{\pi^2}{\hbar K^2 v} \frac{\partial{P}_2}{\partial Q} \frac{\partial^2{P}_2}{\partial Q^2} - \frac{\pi}{2\hbar K v n} \left(\frac{\partial{P}_2}{\partial Q}\right)^2.
\end{align}
Instead of evaluating equations~(\ref{eq:mu2gg}) and (\ref{eq:mu4gg}) right away, we will do in a later step from the free energy.

\subsubsection{The free energy.}

Until now we have calculated the pressure $\bar P$ of the system in the grand-canonical ensemble, see equation~(\ref{eq:Pbar}). It is connected to the grand-canonical potential $\Omega$ by $\Omega=-\bar PL$, which depends on the chemical potential $\bar\mu$ and the temperature $T$. Let us find the free energy in the canonical ensemble, which depends on the density $n$ and $T$. It is defined as 
\begin{align}\label{eq:Fdefinition}
	F=\Omega+\bar\mu nL.
\end{align}
Here the chemical potential $\bar\mu$ should be understood as a function of the density $n$ and the temperature $T$. Assuming the low-temperature forms of $\bar P$ and $\bar\mu$ given by equations~(\ref{eq:Pbar}) and (\ref{eq:mugeneric}), the free energy takes the form
\begin{align}\label{eq:Fdensity}
	F=-PL+\mu nL-T^2P_2L+T^4 \left(-P_4-n_2\mu_2 -\frac{K}{2\pi\hbar v}\mu_2^2\right)L+\mathcal{O}(T^6).
\end{align}
In equation~(\ref{eq:Fdensity}) we have used the relations
\begin{gather}
	P'(\mu)=n,\quad {P}_2'(\mu)=n_2,\quad P''(\mu)=\frac{K}{\pi\hbar v},
\end{gather}
where the prime denotes the derivative with respect to $\bar\mu$ at $\bar\mu=\mu$. Substituting previously calculated expressions for $n_2$ and $\mu_2$, the final result for the free energy is given by
\begin{align}\label{eq:Free}
	F=E_0-T^2{P}_2L+T^4 \left[-{P}_4+\frac{mv^2}{2n}\left(P_2'(\mu)\right)^2\right]L+\mathcal{O}(T^6).
\end{align}
Here $E_0$ is the ground-state energy, while $P_2$ and $P_4$ are given by equations~(\ref{eq:P2final}) and (\ref{eq:P4final}). Equation (\ref{eq:Free}) applies at low temperatures, but it is exact with respect to the interaction strength. The obtained result for the quadratic correction in equation~(\ref{eq:Free}) has a universal form $-\frac{\pi T^2}{6\hbar v}L$ that was initially obtained using the conformal field theory arguments \cite{blote_conformal_1986,affleck_universal_1986}. The quartic correction is the new result. Similarly as the quadratic one that can be expressed only in terms of the Luttinger liquid parameter $K=4\pi^2\rho(Q,Q)^2$ (or the sound velocity $v=\frac{\pi\hbar n}{mK}$), the quartic correction also depends on $K$. In addition, the latter depends on another dimensionless parameter conveniently taken to be $K_1$, which is directly related to $\rho^{(1,0)}(Q,Q)$ and $\rho^{(2,0)}(Q,Q)$, see equations~(\ref{eq:K1})--(\ref{eq:K2}).

Let us introduce the dimensionless parameter $\tau=T/\epsilon$ that accounts for the temperature, where $\epsilon=\frac{\hbar^2n^2}{2m}$ is the temperature of quantum degeneracy. Then the dimensionless free energy per particle $f(\gamma,\tau)=\frac{F}{N\epsilon}$ has the form
\begin{align}\label{eq:f}
	f(\gamma,\tau)=e_2(\gamma)-\frac{K}{12}\tau^2+{C}_4 \tau^4 +\mathcal{O}(\tau^6),
\end{align}
where
\begin{align}\label{eq:C4general}
	C_4={}&-\frac{64\pi^6 \rho^8}{45}\biggl[ 1-\frac{n\dot\rho}{2\pi\rho^3} -\frac{n^2\dot\rho^2}{256\pi^2\rho^6} +\frac{17n^2\ddot\rho}{512\pi^2\rho^5}+\frac{3n\rho^{(1,0)}}{8\pi\rho^3} -\frac{3n^2\dot\rho\rho^{(1,0)}}{32\pi^2\rho^6}\notag\\
	&-\frac{5n^2\rho^{(2,0)}}{128\pi^2\rho^5}+\frac{5n^2(\rho^{(1,0)})^2}{256\pi^2\rho^6} \biggr].
\end{align}
An alternative form is given by
\begin{align}\label{eq:C4}
	C_4=\frac{29K^2K_1}{5760\pi^2}+\frac{7\gamma KK'K_1}{5760\pi^2}+\frac{\gamma K^2K_1'}{1152\pi^2}-\frac{K^4}{180\pi^2}-\frac{49\gamma K^3 K'}{5760\pi^2} -\frac{\gamma^2K^2 K'^2}{1536\pi^2} -\frac{17 \gamma^2K^3K''}{11520\pi^2}.
\end{align}
Equation (\ref{eq:C4}) is exact and expressed in terms of two dimensionless parameters $K$ and $K_1$.

\begin{figure}
\centering
\includegraphics[width=0.6\textwidth]{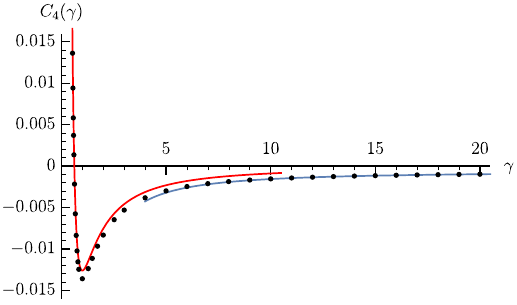}
\caption{\justifying Plot of the coefficient $C_4$. The dots represent numerically exact values and the curves are obtained from the analytical results (\ref{eq:C4small}) and (\ref{eq:C4large}).}\label{fig:C4}
\end{figure}

At the leading order in $\gamma\ll 1$, we have $K=\frac{\pi}{\sqrt\gamma}$ and $K_1=\frac{\pi^3}{3\gamma^{3/2}}$. Therefore, only the terms that involve $K_1$ in equation~(\ref{eq:C4}) give rise to the leading term given by 
$C_4=\frac{\pi^3}{960\gamma^{5/2}}$. Accounting for more terms in the expansion we find
\begin{align}\label{eq:C4small}
	C_4=\frac{\pi^3}{960\gamma^{5/2}}\left(1-\frac{35\sqrt\gamma}{12\pi} -\frac{685\gamma}{192\pi^2} -\frac{207-6\zeta(3)}{64\pi^3}\gamma^{3/2} +\mathcal{O}(\gamma^{2})\right).
\end{align}
In sharp contrast to the case $\gamma\ll 1$, at strong interactions, $\gamma\gg 1$, the summands of equation~(\ref{eq:C4}) that involve $K_1$ enter starting from the third subleading term. In practice they can thus be often neglected. The leading order term $C_4=-\frac{1}{180\pi^2}$ is obvious from the structure of equation~(\ref{eq:C4}), while further terms in the expansion are given by
\begin{align}\label{eq:C4large}
	C_4=-\frac{1}{180\pi^2}\left(1+\frac{12}{\gamma}+\frac{60}{\gamma^2} -\frac{52\pi^2-480}{3\gamma^3}+\mathcal{O}(\gamma^{-4})\right).
\end{align}
Since $C_4$ is positive at weak interactions and negative at strong ones, the sign of $C_4$ changes as the interaction strength $\gamma$ is increased. We have found that $C_4$ nullifies at $\gamma\approx 0.606$. The plot of $C_4$ is shown in figure~\ref{fig:C4}.

\begin{figure}
\centering
\includegraphics[width=0.6\textwidth]{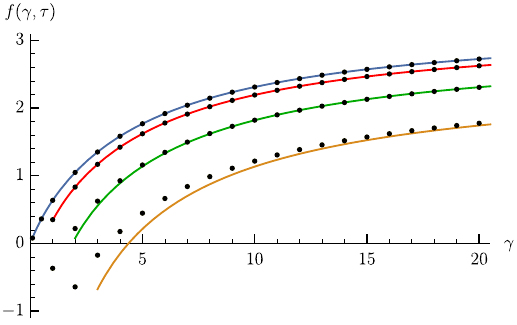}
\caption{\justifying The dimensionless free energy per particle $f(\gamma,\tau)$ obtained from the analytical expression (\ref{eq:f}) is represented by the curves and the one obtained by numerically solving the Yang--Yang equation is shown by the dots. Four distinct values of $\tau$ are considered: $0.1$, $1$, $2$, and $3$. For a fixed value of $\gamma$, the function $f(\gamma,\tau)$ decreases when $\tau$ increases. This is expected as the entropy of the system increases with the temperature. The agreement between equation~(\ref{eq:f}) and numerically exact values is perfect for $\tau$ smaller than 1 and $\gamma$ larger than $\tau$. On the resolution of the plot, the curve for $\tau=0$ would be indistinguishable from the plotted one for $\tau=0.1$.}\label{fig:freeenergy}
\end{figure}

The limiting cases of $C_4$ can be understood in terms of simple physics. At weak interactions, the free energy of the system can be calculated by studying the statistical mechanics of bosonic quasiparticles with Bogoliubov spectrum (\ref{E_bogoliubov}). It yields the result (\ref{eq:C4small}) taken at the leading order \cite{de_rosi_beyond-luttinger-liquid_2019}. Interestingly, Bogoliubov spectrum is not the correct form of the quasiparticle spectrum at smallest momenta, as it is replaced by the fermionic one of equation~(\ref{eq:spectrumlowp}). However, such picture with bosonic quasiparticles is sufficient in order to reproduce the leading-order coefficient in front of $T^4$ power in the free energy.\footnote{\linespread{1}\selectfont Recall that the ground-state energy of the Lieb--Liniger model at the leading and subleading order is well captured by the same approach with Bogoliubov quasiparticles \cite{lieb_exact_1963}.} In the limiting case of strong interactions, by calculating the free energy of the gas of noninteracting fermions we recovered the leading order term in equation~(\ref{eq:C4large}). Nevertheless, the expression (\ref{eq:C4}) applies at any interaction. In figure~\ref{fig:freeenergy} we show the dimensionless free energy per particle for different values of the temperature.

\subsubsection{Other thermodynamic parameters.}

Once the free energy is evaluated, the calculation of other thermodynamic parameters in the canonical ensemble is simple. The chemical potential is given by $\bar\mu=\frac{1}{L}\frac{\partial F}{\partial n}$, yielding
\begin{align}\label{eq:mucanonical}
	\bar \mu=\mu+\frac{K+\gamma K'}{12}\epsilon\;\!\tau^2-\left(5C_4+\gamma \frac{dC_4}{d\gamma}\right)\epsilon\;\!\tau^4+\mathcal{O}(\tau^6).
\end{align}
Here $\mu$ is given by equation~(\ref{eq:mu2}). An alternative way is to evaluate equations~(\ref{eq:mu2gg}) and (\ref{eq:mu4gg}), which leads to the same result. 

The pressure in the canonical ensemble follows from equation~(\ref{eq:Fdefinition}), $\bar P=\bar\mu n-n\epsilon f(\gamma,\tau)$. Using equations~(\ref{eq:mucanonical}) and (\ref{eq:f}) we obtain
\begin{align}
	\bar P=P+\frac{2K+\gamma K'}{12}n\epsilon\;\!\tau^2 -\left(6C_4+\gamma \frac{dC_4}{d\gamma}\right)n\epsilon\;\!\tau^4+\mathcal{O}(\tau^6).
\end{align}
Here $P$ is given by equation~(\ref{eq:PzeroT}). The entropy $S=-\frac{\partial F}{\partial T}$ and the internal energy $U=F+TS$ are given by
\begin{gather}
	S=N\left(\frac{K}{6}\tau-4C_4\tau^3+\mathcal{O}(\tau^5)\right),\\
	U=N\epsilon \left(e_2(\gamma)+\frac{K}{12}\tau^2-3C_4\tau^4+\mathcal{O}(\tau^6)\right).
\end{gather}
Finally, the specific heat $C=\frac{\partial U}{\partial T}$ reads
\begin{align}
	C=N\left(\frac{K}{6}\tau-12C_4 \tau^3+\mathcal{O}(\tau^5)\right).
\end{align}
Therefore, $C_4$ determines the leading correction in the specific heat. Changing the sign as the interaction strength is increased, see figure \ref{fig:C4}, the coefficient $C_4$ is directly responsible for the change of the specific heat behaviour at a fixed temperature from concave at $\gamma< 0.606$ to convex at $\gamma > 0.606$.

\section{Boundary energy}\label{section:boundaryenergy}

Until now we have studied the Lieb--Liniger model imposing periodic boundary conditions on the wave function. Consider now the same model in the presence of a hard-wall box potential, which causes the nullification of the wave function at the two ends. In this case the model is also exactly solvable by the Bethe ansatz as first shown by Gaudin in 1971 \cite{gaudin_boundary_1971}. The case with zero boundary conditions shows some important qualitative differences. In particular, it is characterised by the boundary energy $E_{\B}$, which represents the nonextensive part of the ground-state energy $E_0$ in the thermodynamic limit, which can be expressed as \cite{gaudin_boundary_1971,blote_conformal_1986}
\begin{align}\label{eq1}
	E_0=N\epsilon_0 + E_{\B}+\mathcal{O}(1/N).
\end{align}
Here $\epsilon_0$ is the ground-state energy per particle and $N$ is their total number. Note that the bulk energy $\epsilon_0$ is identical for the two geometries, while the boundary  energy $E_{\B}$ is a surface effect and it exists only in the case of zero boundary conditions \cite{gaudin_boundary_1971}. Although contained in the exact solution, the boundary energy in the thermodynamic limit was only approximately calculated by Gaudin, who found the leading order result at weak repulsion. Here we find the exact result for the boundary energy. This section is based on the published work \cite{reichert_exact_2019}.

\subsection{Simple solutions in the limiting cases}

The boundary energy reflects the increase in the ground-state energy due to the hard-wall potential. It creates two nodes in the many-body wave function and thus the density becomes nonuniform. As a result, there is an increase of the kinetic energy. The typical size of the density depletion near the boundary is on the order of the healing length $\xi$ and thus involves $\xi n$ particles, where $n$ is the mean boson density. This enables us to estimate the boundary energy as $E_{\B}\sim \frac{\hbar^2}{m\xi^2}(\xi n)$, where $m$ denotes the mass of bosons. Using $\xi=\hbar/\sqrt{m\mu}$, where $\mu$ is the chemical potential of the system we end up with the result
\begin{align}\label{eq:estimate}
	E_{\B}\sim \epsilon	
	\begin{cases}
		\sqrt\gamma,\quad &\gamma\ll 1\\
		1,\quad &\gamma\gg 1
	\end{cases}
\end{align}
in the two regimes of weak and strong interactions. Here $\epsilon=\frac{\hbar^2 n^2}{2m}$ is the natural unit of energy for the system.

We can easily go beyond the estimate (\ref{eq:estimate}). At weak interactions, we could use the description of the system in terms of the Gross--Pitaevskii equation \cite{pitaevskii_bose-einstein_2016}. For a semi-infinite system it is given by 
\begin{align}\label{eq:GPE}
	i\hbar\frac{\partial\psi(x,t)}{\partial t}=\left(-\frac{\hbar^2}{2m}\frac{\partial^2}{\partial x^2}+ g|\psi(x,t)|^2\right)\psi(x,t)
\end{align} 
subject to the boundary condition $\psi(0,t)=0$. Here $g=\frac{\hbar^2 c}{m}$ and $\psi(x,t)$ is the semiclassical wave function of the system. A solution that describes the ground state is given by
\begin{align}\label{eq:tanh}
	\psi(x,t)=\sqrt\frac{\mu}{g} \tanh\left(\frac{x}{\xi}\right)e^{-i \mu t/\hbar}.
\end{align}
The number of particles is connected to the wave function by
\begin{align}\label{eq:N}
	N=\int_0^L dx|\psi(x,t)|^2,
\end{align}
where $L$ denotes the length that will eventually be sent to infinity. Equation (\ref{eq:N}) enables us to express $\mu$ in terms of $N$ and then find the energy of the semi-infinite system
\begin{align}\label{eq:EGP}
	E=\mu N-\frac{g}{2}\int_{0}^{L}dx |\psi(x,t)|^4=N\epsilon\;\!\gamma +\frac{4}{3}\epsilon\sqrt\gamma +\mathcal{O}\left({1}/{N}\right).
\end{align}
The first term after the last equality is extensive and describes the ground-state energy of the system. It corresponds to equation~(\ref{eq:eps0e2}) multiplied by $N$, where  we should use $e_2(\gamma)=\gamma$ since we consider the leading order in $\gamma\ll 1$. The subleading term in equation~(\ref{eq:EGP}) describes the increase of the energy due to the nullification of the wave function at $x=0$. Since the system with zero boundary conditions has two such regions, the boundary energy is twice that term. We thus obtain
\begin{align}\label{eq:Ebweak}
	E_{\B}=\frac{8}{3}\epsilon\sqrt{\gamma}.
\end{align}
At infinite repulsion, the boundary energy follows from the expression for the ground-state energy of $N$ free fermions in a hard-wall box of the size $L$. It is given by $E_0=\frac{\pi^2\hbar^2}{2mL^2}\sum_{j=1}^N j^2$. The subleading term at large $N$ then gives the boundary energy 
\begin{align}\label{eq:Ebstrong}
	E_{\B}=\frac{\pi^2}{2}\epsilon.
\end{align}
The results (\ref{eq:Ebweak}) and (\ref{eq:Ebstrong}) are consistent with the estimate (\ref{eq:estimate}).

\subsection{The boundary energy from the Bethe ansatz}

Let us now calculate the boundary energy directly from the Bethe ansatz solution. Consider the system described by the Lieb--Liniger Hamiltonian (\ref{eq:Horiginal}) with zero boundary conditions. Its exact solution is described by the Bethe ansatz equations \cite{gaudin_boundary_1971,oelkers_bethe_2006}
\begin{align}\label{eq:BAEZBCoe}
	\exp(i 2L\bar k_j)=\prod_{l=1\atop{l\neq j}}^{N}\frac{\bar k_j-\bar k_l+ic}{\bar k_j-\bar k_l-ic}\,\frac{\bar k_j+\bar k_l+ic}{\bar k_j+\bar k_l-ic},\quad j=1,2,\ldots,N,
\end{align}
which determine the set of quasimomenta $\bar k_j$. The system (\ref{eq:BAEZBCoe}) has the property that if a certain set of quasimomenta satisfies equation~(\ref{eq:BAEZBCoe}) then the set obtained by changing the sign to an arbitrary subset of it is again the solution. Since at $c>0$ all the quasimomenta are different, it is sufficient to consider only the positive ones. Taking the logarithm of equation~(\ref{eq:BAEZBCoe}) with the help of 
\begin{align}
	\ln\frac{A+ic}{A-ic}=2i\arctan\frac{c}{A},
\end{align}
we obtain that the quasimomenta in the ground-state of the system of length $L$ with $N$ particles satisfy
\begin{align}\label{eq:GaudinBAE}
	L\bar k_j=\pi+\sum_{l=1\atop l\neq j}^{N} \left(\arctan\frac{c}{\bar k_j-\bar k_l} +\arctan\frac{c}{\bar k_j+\bar k_l}\right),\quad j=1,2,\ldots,N.
\end{align}
Using the identity $\arctan x+\arctan(1/x)=\pi\, \mathrm{sgn}(x)/2$ for $x\neq 0$, we can reexpress equation~(\ref{eq:GaudinBAE}) as
\begin{align}\label{eq:GaudinBAE1}
	L\bar k_j =\pi j+\frac{1}{2}\sum_{l=1}^N\left[\theta(\bar k_j-\bar k_l)+\theta(\bar k_j+\bar k_l) \right]-\frac{\theta(2\bar k_j)}{2},\quad j=1,2,\ldots,N,
\end{align}
where $\theta(k)=-2\arctan(k/c)$. The ground-state energy is given by $E^{\Z}(N)=\frac{\hbar^2}{2m}\sum_{j=1}^N \bar k_j^2$, where the superscript denotes zero boundary conditions.

Consider now the same system described by the Hamiltonian (\ref{eq:Horiginal}) with periodic boundary conditions. The Bethe ansatz equations for the quasimomenta in the ground-state of the system of length $2L$ with $2N$ particles follows directly from equation~(\ref{eq:DBA}),
\begin{align}\label{eq:BAEPBC}
	2Lk_{j}=2\pi \left(j-\frac{2N+1}{2}\right) +\sum_{l=1}^{2N} \theta(k_{j}-k_{l}),\quad j=1, 2,\ldots,2N.
\end{align}
The system of equations~(\ref{eq:BAEPBC}) has a unique solution with distinct quasimomenta $k_j$, where one-half of them are negative ($k_j<0$ for $1\le j\le N$), while the remaining ones are positive ($k_j>0$ for $N+1\le j\le 2N$). Moreover, the quasimomenta are positioned symmetrically around zero, i.e., $k_j=-k_{2N+1-j}$. It will be convenient to shift the indices in equation~(\ref{eq:BAEPBC}): $j\to j-N-1$ for $1\le j\le N$ and $j\to j-N$ for $N+1\le j\le 2N$, so that the property $k_j=-k_{-j}$ is satisfied. This enables us to write
\begin{align}\label{eq:BAEPBC1}
	Lk_j=\pi \left(j-\frac{1}{2}\right)+\frac{1}{2}\sum_{l=1}^{N}\left[ \theta(k_j-k_l) + \theta(k_j+k_l)\right],\quad j=1,2,\ldots,N.
\end{align}
The ground state is thus characterised by the set of $N$ positive quasimomenta obtained by solving the system~(\ref{eq:BAEPBC1}), while the negative ones are automatically obtained from them. The ground-state energy is then given by
$E^{\P}(2N)=\frac{\hbar^2}{m} \sum_{j=1}^N\ k_j^2$, where the superscript denotes periodic boundary conditions.

The boundary energy is the difference in the ground-state energy of the system with zero and periodic boundary conditions, $E_{\B}(N)=E^{\Z}(N)-E^{\P}(N)$.
For the latter case, one can show that, at the same density, the energy of the systems with $N$ and $2N$ particles are simply related as $E^{\P}(N)=E^{\P}(2N)/2+\mathcal{O}(1/N)$ \cite{gaudin_boundary_1971}. In the thermodynamic limit this yields 
\begin{align}
E_{\B}=\lim_{N\to\infty} \left[E^{\Z}(N)-E^{\P}(2N)/2\right],
\end{align}
which is
\begin{align}\label{eq:Eb}
	E_{\B}=\lim_{N\to\infty}\frac{\hbar^2}{2m}\sum_{j=1}^N (\bar k_j^2-k_j^2),
\end{align}
where the corresponding quasimomenta are the solutions of equations~(\ref{eq:GaudinBAE1}) and (\ref{eq:BAEPBC1}).

For the evaluation of the boundary energy~(\ref{eq:Eb}) we subtract equation~(\ref{eq:BAEPBC1}) from equation~(\ref{eq:GaudinBAE1}). Since in a long system the difference $\bar k_j-k_j=\Delta k_j=\mathcal{O}(1/L)$ is small, we obtain
\begin{align}\label{eq:BE1be}
	L\Delta k_j={}&\frac{\pi}{2}-\frac{\theta(2\bar k_j)}{2} + \frac{1}{2} \sum_{l=1}^N [\theta'(k_j-k_l)(\Delta k_j-\Delta k_l)+\theta'(k_j+k_l)(\Delta k_j+\Delta k_l)]\notag\\
	&+\mathcal{O}(1/N).
\end{align}
In a system of length $2L$ with periodic boundary conditions the density of quasimomenta is $\rho(k_j)=[2L(k_{j+1}-k_j)]^{-1}$, cf. Eq~(\ref{eq:Fhokj}). In the thermodynamic limit it satisfies the Lieb integral equation (\ref{eq:LIE}) with the kernel (\ref{eq:kernel}). Using the formal expression $\rho(k)=\sum_{j=1}^N [\delta(k-k_j)+\delta(k+k_j)]/2L$ and the property $\rho(k)=\rho(-k)$, we then obtain
\begin{align}
	1-\frac{1}{2L}\sum_{l=1}^N [\theta'(k-k_l)+\theta'(k+k_l)]=2\pi \rho(k).
\end{align}
The latter equation enables us to simplify equation~(\ref{eq:BE1be}). Introducing an odd function $g(k_j)=L\rho(k_j)\Delta k_j$, we obtain that it satisfies an integral equation
\begin{subequations}\label{eq:mainBE}
	\begin{align}\label{eq:integralequationforg}
		g(k)-\frac{c}{\pi}\int_{-Q}^{Q} dq\, \frac{g(q)}{c^2+(q-k)^2}=r(k),
	\end{align}
	where
	\begin{align}\label{eq:F(k)}
		r(k)=\frac{\mathrm{sgn(k)}}{4}+ \frac{\arctan\frac{2k}{c}}{2\pi}.
	\end{align}
\end{subequations}
The boundary energy can then be expressed as
\begin{align}\label{eq:Ebfinal}
	E_{\B}=\frac{\hbar^2}{m}\int_{-Q}^Q dk g(k)k.
\end{align}
Equations~(\ref{eq:mainBE}) and (\ref{eq:Ebfinal}) give the exact result for the boundary energy of the Lieb--Liniger model at an arbitrary interaction strength $c>0$. Instead of dealing directly with equation~(\ref{eq:mainBE}), we find it more convenient to reexpress the boundary energy (\ref{eq:Ebfinal}) as
\begin{align}\label{eq:Ebsigma}
	E_{\B}=\int_{-Q}^{Q}dk\sigma(k,Q) r(k),
\end{align}
where $r(k)$ is given by equation~(\ref{eq:F(k)}) and $\sigma(k,Q)$ satisfies equation~(\ref{eq:sigma}). The equivalence of equations~(\ref{eq:Ebfinal}) and (\ref{eq:Ebsigma}) follows from the result of Appendix \ref{appendixb}. We have therefore reformulated the problem of finding  the boundary energy to the problem of solving equation~(\ref{eq:sigma}) and then evaluating $E_{\B}$ using equation~(\ref{eq:Ebsigma}). The latter expression is derived in the thermodynamic limit, in which case the system size is much larger than the healing length, $L\gg\xi$. In a finite system there is an additional regime at $L\lesssim \xi$, which can occur only at very weak interactions that satisfies $\gamma\lesssim 1/N^2$. We have not studied the latter case.

\subsection{The evaluation of  the boundary energy}

The boundary energy (\ref{eq:Ebsigma}) depends on $\sigma(k,Q)$ that obeys the integral equation (\ref{eq:sigma}). At strong interactions, the latter is easily solvable using the method described in section \ref{section:simplesolution}. However, at weak interactions the situation is more involved. One possibility is to solve perturbatively equation~(\ref{eq:sigma}) using the approach of Popov \cite{popov_theory_1977} who was dealing with equation~(\ref{eq:LIE}). For the solution we obtained
\begin{align}\label{eq:sigmaPopov}
	\sigma(k,Q)=\frac{\hbar^2}{2m}\left[\frac{k\sqrt{Q^2-k^2}}{c}+\frac{\left(1+\ln\frac{16\pi Q}{c}\right)Qk+(Q^2-2k^2)\ln\frac{Q+k}{Q-k}}{2\pi \sqrt{Q^2-k^2}}\right]+ \mathcal{O}(c).
\end{align}
Equation (\ref{eq:sigmaPopov}) holds in the region away from the edges, at the values of $k$ that satisfy $Q^2-k^2\gg Qc$, similarly as equation~(\ref{eq:rhoPopov}). Using the expression for $Q$ of equation~(\ref{eq:Qfin}) that includes the first two terms of the expansion and the solution (\ref{eq:sigmaPopov}), equation~(\ref{eq:Ebsigma}) leads to
\begin{align}\label{eq:Ebsmall}
	E_{\B}=\frac{8}{3}\epsilon\sqrt{\gamma}\left[1-\frac{3}{16}\sqrt{\gamma} +\mathcal{O}(\gamma) \right].
\end{align}
At the leading order, equation~(\ref{eq:Ebsmall}) agrees with the result~(\ref{eq:Ebweak}). On the other hand, the whole result (\ref{eq:Ebsmall}) can be obtained by studying the first quantum correction to the Gross--Pitaevskii equation \cite{reichert_fluctuation-induced_2019}. 

An alternative route to evaluate $E_{\B}$ is by taking the derivative with respect to $Q$ of equation~(\ref{eq:Ebsigma}). Using the method presented in section \ref{section:method}, in particular equation~(\ref{eq:Fhoder2}), after introducing the boundary energy function $e_{\B}(\gamma)$ by 
\begin{align}
	E_{\B}=\epsilon\, e_{\B}(\gamma),
\end{align}
it can be shown that $e_{\B}(\gamma)$ satisfies the differential equation
\begin{align}\label{eq:diffeqeB}
	\frac{d}{d\gamma}\left(\frac{e_{\B}(\gamma)}{\gamma^2}\right)= -\frac{2\pi^3}{K^2\gamma^3} \left(\rho(0,Q)+\frac{2c}{\pi}\int_{-Q}^{Q}dq \frac{\rho(q,Q)}{c^2+4q^2}\right).
\end{align}
Here $K$ is the Luttinger liquid parameter and $\rho(k,Q)$ satisfies equation~(\ref{eq:LIE}).

At $c\to 0^+$, the integrand in equation~(\ref{eq:diffeqeB}) contains the representation of the Dirac $\delta$-function, leading to 
\begin{align}
	\frac{d}{d\gamma}\left(\frac{e_{\B}(\gamma)}{\gamma^2}\right)= -\frac{4\pi}{\gamma^2}\rho(0,Q).
\end{align}
Here the right-hand side is only correct at the leading order and thus $K$ is replaced by $\pi/\sqrt\gamma$. Within the same accuracy $\rho(0,Q)=1/\pi \sqrt\gamma$, see equation~(\ref{eq:rhoPopov}), yielding $e_{\B}(\gamma)=8\sqrt\gamma/3$. We can, however, evaluate $e_{\B}(\gamma)$ beyond the leading-order at $\gamma\ll 1$. To achieve that we use the result for the resolvent of equation~(\ref{eq:LIE}) that directly follows from the consideration of section \ref{section:capacitance}. It enables us to find
\begin{gather}
	\rho(0,Q)=\frac{1}{\pi\sqrt\gamma}+\frac{1}{2\pi^2}+\frac{12+\pi^2}{96\pi^3}\sqrt\gamma+\frac{\pi^2+9\zeta(3)}{192\pi^4}\gamma+\mathcal{O}(\gamma^{3/2}),\\
	\frac{2c}{\pi}\int_{-Q}^{Q}dq \frac{\rho(q,Q)}{c^2+4q^2}=\frac{1}{\pi\sqrt\gamma}-\frac{\pi-2}{4\pi^2}-\frac{\pi^2-6}{48\pi^3}\sqrt\gamma-\frac{2\pi^2-9\zeta(3)}{192\pi^4}\gamma+\mathcal{O}(\gamma^{3/2}).
\end{gather}
After the integration we obtain
\begin{align}\label{eq:eBfinalweak}
	e_{\B}(\gamma)=\frac{8}{3}\sqrt\gamma-\frac{\gamma}{2} +\frac{12\pi-24-\pi^2}{24\pi^2}\gamma^{3/2}+\mathcal{O}(\gamma^2).
\end{align}
Two comments are in order. First, the right hand-side of equation~(\ref{eq:diffeqeB}) does not contain the term proportional to $1/\gamma$ and thus there are no logarithmic terms in $e_{\B}(\gamma)$. Second, the term proportional to $\gamma^2$ in equation~(\ref{eq:eBfinalweak}) is the integration constant for the differential equation (\ref{eq:diffeqeB}) and has the value
\begin{align}\label{eq:Eb4}
	\frac{1}{48}-\frac{1}{24\pi}+\frac{1}{4\pi^2}-\frac{2}{3\pi^3}-\frac{\zeta(3)}{4\pi^3}.
\end{align}
The result (\ref{eq:Eb4}) can be obtained either by applying the formalism of section \ref{section:capacitance} or the method of  section \ref{section:numericalexperiment} to solve equation~(\ref{eq:sigma}) at $\gamma\ll 1$ and then evaluate the boundary energy (\ref{eq:Ebsigma}). The same methods, of course, also lead to the other terms of equation~(\ref{eq:eBfinalweak}).

\begin{figure}
\centering
\includegraphics[width=0.6\textwidth]{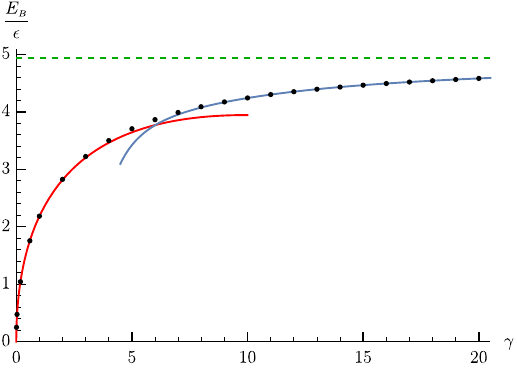}
\caption{\justifying The boundary energy $E_{\B}$ in units of $\epsilon$ as a function of the interaction strength $\gamma$. The dots represent the exact result obtained numerically. The two curves describe the series expansions at small and large $\gamma$ that are given by equations~(\ref{eq:eBfinalweak}) and (\ref{eq:Eblarge}), while the dashed line is at $\pi^2/2$ corresponding to the value at $\gamma\to\infty$. For $2\lesssim\gamma\lesssim 9$, on the plot there is a visible deviation between the exact result and the series since the latter contain only few terms.}\label{fig:boundaryenergy}
\end{figure}

At strong interactions, $\gamma\gg 	1$, it is convenient to express $\rho(0,Q)$ on the right-hand side of equation~(\ref{eq:diffeqeB}) by the integral that enters equation~(\ref{eq:LIE}). Using
\begin{align}
	\frac{c}{\pi}\int_{-Q}^Q dq \frac{\rho(q,Q)}{c^2+a q^2}=\frac{1}{\pi}\sum_{j=0}^{\infty} (-1)^j\frac{a^j e_{2j}(\gamma)}{\gamma^{2j+1}},
\end{align}
where $e_{2j}(\gamma)$ is defined by equation~(\ref{eq:e2l}), we obtain
\begin{align}\label{eq:eBdiffeqlargegamma}
	\frac{d}{d\gamma}\left(\frac{e_{\B}(\gamma)}{\gamma^2}\right)=-\frac{2\pi^2}{K^2\gamma^3} \left(\frac{1}{2}+\sum_{j=0}^{\infty}(-1)^j(1+2^{2j+1}) \frac{e_{2j}(\gamma)}{\gamma^{2j+1}}\right).
\end{align}
Since $e_{2j}(\gamma)$ approaches a constant value at large $\gamma$, the result (\ref{eq:e2largefinal}) for $e_2(\gamma)$ suffices to find the first five terms in the expansion of the boundary energy. After solving the differential equation (\ref{eq:eBdiffeqlargegamma}) we obtain
\begin{align}\label{eq:Eblarge}
	e_{\B}(\gamma)=\frac{\pi^2}{2}\left[1-\frac{4}{3\gamma}-\frac{4}{\gamma^2}+\frac{4(120+7\pi^2)}{15\gamma^3}-\frac{40(30+\pi^2)}{9\gamma^4}+	
	\mathcal{O}\left(\gamma^{-5}\right) \right].
\end{align}
We note that the integration constant is set to zero since the boundary energy cannot diverge at $\gamma\to\infty$. The result (\ref{eq:Eblarge}) is in accordance with equation~(\ref{eq:Ebstrong}). In figure~\ref{fig:boundaryenergy} we show the two series expansions together with the exact result (\ref{eq:Ebsigma}) for the boundary energy.

\subsection{A comment about the boundary energy found by Gaudin}

In reference~\cite{gaudin_boundary_1971} as well as in his book \cite{gaudin}, Gaudin studied the problem of the boundary energy. He found the integral equation of the form~(\ref{eq:integralequationforg}) but with a different right-hand side, which instead was given by $r_{\G}(k)=\mathrm{sgn(k)}/2$. Such an expression is the approximately correct right-hand side of equation~(\ref{eq:integralequationforg}) only at $c\to 0^+$, as one can see by considering equation~(\ref{eq:F(k)}) in this limit. Thus, Gaudin was able only to find the leading order expression~(\ref{eq:Ebweak}) for the boundary energy at weak interactions. Interestingly, using equation~(\ref{eq:Ebsigma}) Gaudin's formula for the boundary energy becomes 
\begin{align}\label{eq:EbG}
	E_{\B\G}=\int_0^Qdk \sigma(k).
\end{align}
This expression formally coincides with the energy of Lieb's type-II excitation in the (periodic) Lieb--Liniger model with the momentum $\pi\hbar n$, as follows from equation~(\ref{eq:type2spectrum}) for $\eta=0$. 

Equation (\ref{eq:EbG}) can be evaluated analogously as  $E_{\B}$. After introducing the dimensionless function $e_{\B\G}(\gamma)$ by 
\begin{align}
	E_{\B\G}=\epsilon\, e_{\B\G}(\gamma),
\end{align}
one can be show that it satisfies the differential equation
\begin{align}
	\frac{d}{d\gamma}\left(\frac{e_{\B\G}(\gamma)}{\gamma^2}\right)= -\frac{4\pi^3}{K^2\gamma^3} \rho(0,Q).
\end{align}
We then find
\begin{align}\label{eq:Ebgweak}
	e_{\B\G}(\gamma)=
	\frac{8}{3}\sqrt{\gamma}-\left(\frac{1}{\pi^2}-\frac{1}{12}\right)\gamma^{3/2}+ \mathcal{O}(\gamma^2)
\end{align}
at weak interactions and
\begin{align}\label{eq:Ebgstrong}
	e_{\B\G}(\gamma)=
	\pi^2-\frac{4\pi^2}{\gamma}+\frac{12\pi^2}{\gamma^2}+\mathcal{O}(\gamma^{-3})
\end{align}
at strong interactions. We note that at weak interactions $E_{\B\G}$ does not have the subleading term proportional to $\gamma$. The type-II excitation of the energy $E_{\B\G}$, which at weak interactions can be understood as a dark soliton \cite{pitaevskii_bose-einstein_2016}, at any repulsion is greater than the boundary energy. At weak interactions, there is a difference in the first beyond-mean-field correction, and at strong interactions, the boundary energy is two times smaller.

\section{Capacitance of a circular plate capacitor \label{section:capacitance}}

As initially noted by Gaudin \cite{gaudin_boundary_1971}, the classic problem of the capacitance of a circular capacitor with parallel oppositely charged plates has the problematics of the Lieb--Liniger model. While Gaudin used the information of the former to obtain the results for the latter, here we will do the opposite and present the solution of the capacitance problem in detail, starting from the initial Love integral equation. We will eventually connect the two problems and give a simple parametric expression for the capacitance that relies on the ground-state function $e_2$ of the Lieb--Liniger model calculated previously. This section is based on the published works \cite{reichert_analytical_2020} and \cite{ristivojevic_method_2022}.

\subsection{The problem and short history}

Capacitance is one of the basic concepts in electrostatics. For a capacitor, it  denotes the ratio between the charge on one of the plates and the potential difference between them. The capacitance purely depends on the geometry. In  textbooks, the standard simplification is a parallel plate capacitor in a vacuum with the characteristic plate size much larger than their separation. In this case, the capacitance acquires the familiar form 
\begin{align}\label{eq:Cideal}
	C=\epsilon_0 \frac{S}{\kappa}.
\end{align}
Here $S$ denotes the surface of the plates, $\kappa$ is their separation, and the constant $\epsilon_0$ is the vacuum permittivity. The expression (\ref{eq:Cideal}) should be understood only as a result valid in the limit of small plate separations, in which case the edge effects are neglected. 

The edge effects can be qualitatively understood through the example of a circular parallel plate capacitor. Consider a system where the coaxial thin plates have unit radius and which are at the separation $\kappa\ll 1$ (in the units of radius in the following). The capacitance of this system was initially studied by Kirchhoff \cite{kirchhoff_zur_1879} in 1877 who found 
\begin{align}\label{eq:Ckirchhoff}
	\mathcal{C}(\kappa)=\frac{1}{4\kappa}+\frac{1}{4\pi}\left(\ln\frac{16\pi}{\kappa}-1\right)+ \mathcal{O}(\kappa).
\end{align}
Here for convenience we have introduced $\mathcal{C}=C/{4\pi\epsilon_0}$, which we loosely also call the capacitance and use in the following. The first term in the expansion (\ref{eq:Ckirchhoff}) is in accordance with equation~(\ref{eq:Cideal}) and thus represents the idealised situation where the edge effects have been neglected. The remaining terms in equation~(\ref{eq:Ckirchhoff}) account for the leading correction that describes the edge effects. 

The problem of capacitance of a circular capacitor is complicated. It has attracted considerable attention from researchers working in physics and mathematics. The nonrigorous derivation of Kirchhoff was first proven by Hutson \cite{hutson_circular_1963} in 1963. Further efforts \cite{shaw_circulardisk_1970,leppington_capacity_1970,chew_microstrip_1982,tracy_ground_2016} resulted in the term proportional to $\kappa$ in the expansion (\ref{eq:Ckirchhoff}), as will be discussed below, as well as the general structure of the series in equation~(\ref{eq:Ckirchhoff}) with unevaluated coefficients \cite{soibelman_asymptotics_1996}. 

On the technical level, the problem of evaluation of the capacitance of a circular capacitor turns out to be directly related to an integral equation of Fredholm type, known as the Love equation \cite{love_electrostatic_1949} in the potential theory literature \cite{sneddon_mixed_1966}. The explicit solution of that equation is not known currently, which is reflected in a quite small number of known terms in equation~(\ref{eq:Ckirchhoff}), which require significant efforts to be obtained. Interestingly, the same integral equation was derived later by Lieb and Liniger  \cite{lieb_exact_1963}, who were studying a seemingly unrelated quantum problem of bosons in one dimension with contact repulsion. The mathematical connection between the two problems was first noted and used in reference~\cite{gaudin_boundary_1971}; see also a recent review \cite{farina_love--lieb_2022}. 

Recently, significant progress has been achieved in understanding the way to solve the Fredholm-type integral equations of the second kind with finite support, which often appear when one studies integrable models and field theories \cite{volin_mass_2010,volin_quantum_2011,marino_exact_2019,marino_resurgence_2019}. Using those achievements, here we analyse the Love equation and calculate analytically the small $\kappa$ asymptotics of the capacitance to a high order. We also systematically solve the Love equation at large $\kappa$ and then analytically calculate the capacitance in this regime.

\subsection{The Love equation}

Our goal is to find the capacitance of a circular capacitor that consists of two thin coaxial conducting discs of unit radius at the separation $\kappa$. The discs are  held at equal potentials in the absolute value, $\pm V_0/2$, which guarantees equal charges on the two surfaces, $\pm q$. By the definition, the capacitance is given by $C=q/V_0$. To find that ratio, one should solve  the Laplace equation for the potential with appropriate boundary conditions for the potential on the discs, which could be done in an elementary way \cite{carlson_circular_1994}. The central quantity that determines the capacitance is encoded into the Love integral equation \cite{love_electrostatic_1949,sneddon_mixed_1966}\footnote{\linespread{1}\selectfont Up to a simple rescaling, the Love equation (\ref{eq:Love}) is equivalent to the one of Lieb given by equation~(\ref{eq:LIE}) with the kernel (\ref{eq:kernel}).}
\begin{align}
	\label{eq:Love}
	f(x,\kappa)-\frac{\kappa}{\pi}\int_{-1}^{1} dy\, \frac{f(y,\kappa)}{\kappa^2+(y-x)^2}=1.
\end{align}
The latter equation determines the function $f(x,\kappa)$, which then enables one to express the capacitance $\mathcal{C}=C/4\pi\epsilon_0$ in the form
\begin{align}\label{eq:cap}
	\mathcal{C}(\kappa)=\frac{1}{2\pi}\int_{-1}^{1}d x f(x,\kappa).
\end{align}
From this point the problem of capacitance is reduced to the problem of evaluating the integral in equation~(\ref{eq:cap}) that involves the solution of the Love integral equation (\ref{eq:Love}). We will solve the latter in the regimes of small and large $\kappa$ and then find the series expansions (asymptotic at $\kappa\ll 1$ and with a finite radius of convergence at $\kappa\gg 1$) for $\mathcal{C}(\kappa)$ to a high order, which practically covers all the distances.

\subsection{Formal solution of the Love equation at small separations}

In order to solve the Love equation (\ref{eq:Love}), we are using a method developed by Volin \cite{volin_mass_2010,volin_quantum_2011}, which has recently been adapted to study the problem of one-dimensional gas of bosons with contact interactions \cite{marino_exact_2019}. Here we rederive the method and obtain a closed-form solution in terms of a system of linear equations. We apply the method to the problem of the circular plate capacitor to obtain its capacitance in the regime $\kappa\ll 1$ to, in principle, an arbitrary order. We also comment on the similarities of the present approach with the well-known work of Popov \cite{popov_theory_1977}.

As already noticed in reference~\cite{popov_theory_1977} (see also references~\cite{hutson_circular_1963,lieb_exact_1963}), working with a perturbative expansion in small $\kappa$ of the function $f(x,\kappa)$ defined by equation~(\ref{eq:Love}) becomes problematic at higher orders in $\kappa$ since the corrections become increasingly more divergent near the end of the support of the function, rendering the perturbative expression for $f(x,\kappa)$ nonintegrable in the expression (\ref{eq:cap}). This is a serious problem which practically hinders the calculation of corrections to Kirchhoff's result (\ref{eq:Ckirchhoff}). To overcome this issue, it is convenient to deal with the resolvent $R(z)$ of the function $f(x,\kappa)$ defined by
\begin{align}
	R(z)=\int_{-1}^{1} dx \frac{f(x)}{z-x}. \label{R}
\end{align}
Notice that for simplicity we omitted the argument $\kappa$ from $R(z)$ and $f(x)$. The resolvent is an analytic function in the complex plane except for $z\in [-1,1]$. Its discontinuity along $[-1,1]$ determines the function $f(x)$ on the same interval, since 
\begin{align}
	f(x)=\frac{i}{2\pi}[R(x+i 0)-R(x-i 0)].\label{disc}
\end{align}
Here we have used the formula
\begin{align}
	\delta(y)=\frac{i}{2\pi}\left(\frac{1}{y+i0}-\frac{1}{y-i0}\right).
\end{align}
We notice that $f(x)$ for $|x|>1$ can be found, e.g., from the integral equation (\ref{eq:Love}) by performing the integration once equation~(\ref{disc}) is used in the  integrand.

By making use of the resolvent, the integral equation (\ref{eq:Love}) is transformed into a difference equation
\begin{align}
	\left[1-\mathcal D(\kappa)\right]\widetilde R(x+i0)-\left[1-\mathcal D(-\kappa)\right]\widetilde R(x-i0)=0,\label{diff}
\end{align}
where 
\begin{align}\label{eq:FRt}
	\widetilde R(z)=R(z)-\frac{\pi z}{ \kappa}.
\end{align}
Here we introduced the shift operator $\mathcal D(\kappa)=e^{i \kappa \partial_z}$ acting as
\begin{align}
	\mathcal D(\kappa) R(z)=R(z+i \kappa).
\end{align}
Instead of solving the integral equation (\ref{eq:Love}) we should now solve the alternative equation (\ref{diff}). This will be achieved in two steps \cite{volin_mass_2010,volin_quantum_2011}, first considering the bulk regime near the origin and then in the edge regime, which is in the vicinity of $x=\pm 1$. One finally matches the two solutions for the resolvent in the overlapping regime, as discussed below. We notice that a similar procedure was also applied in the study of Popov \cite{popov_theory_1977}, who was solving the integral equation (\ref{eq:Love}) directly, rather than dealing with the resolvent.

\subsubsection{Bulk regime.}
Let us start by studying the bulk regime given by the limit
\begin{align}
	\kappa \to 0,\quad z\text{ fixed}.
\end{align}
In this case, we assume the resolvent  in the form 
\cite{volin_mass_2010,volin_quantum_2011,marino_exact_2019}
\begin{align}
	&\widetilde R_b(z)=-\frac{\pi\sqrt{z^2-1}}{\kappa} +\sum_{n,m=0}^\infty\sum_{k=0}^{n+m+1}c_{n,m,k}\kappa^{m+n} \frac{z^{\lambda_k}}{(z^2-1)^{n+1/2}}\ln^k\left(\frac{z-1}{z+1}\right), \label{Rt}
\end{align}
where $\lambda_k=[1-(-1)^k]/2$, while the subscript $b$ denotes the bulk solution.
The ansatz for the resolvent (\ref{Rt}) is proposed in reference~\cite{marino_exact_2019}. We showed that it is consistent with the ansatz for $f(x)$ that Popov \cite{popov_theory_1977} used to solve equation~(\ref{eq:Love}) (see the discussion below). In equation~(\ref{Rt}), the coefficients $c_{n,m,k}$ are unknown polynomials, as it turns out, of $\ln \kappa$. Therefore, they only weakly depend on $\kappa$. It is important to note that the ansatz (\ref{Rt}) contains many terms that diverge when $z\to \pm 1$. Moreover, each subsequent term that has higher value of $n$ is more divergent from the preceding ones. These two issues imply that the ansatz for the resolvent is justified only for $|z\pm 1|\gg \kappa$. Therefore, the ansatz (\ref{Rt}) applies in the complex plane sufficiently outside the two circles  around the centres at $\pm1$ with a small characteristic radius $\kappa$. This explains the name bulk solution, opposite to the edge solution that is derived as a series expansion near $z=1$ (see below).

The coefficients $c_{n,m,k}$ of equation~(\ref{Rt}) should be chosen in such a way that equation~(\ref{diff}) is satisfied. Substituting the ansatz (\ref{Rt}) into equation~(\ref{diff}) and using the expression for the logarithm around the branch cut of the form $(0<x<1)$
\begin{align}
	\frac{\ln^k\left(\frac{x\pm i0-1}{x\pm i0+1}\right)}{[(x\pm i0)^2-1]^{n+1/2}}=\mp i (-1)^n\frac{\left(\ln\frac{1-x}{1+x}\pm i \pi\right)^k}{(1-x^2)^{n+1/2}},\label{cont}
\end{align}
in the limit $\kappa\to 0$ and $x\to0$ one obtains the relations that $c_{n,m,k}$ should satisfy. At order $\kappa^{N+1}$ and $x^{2N^2}$, one can find all the coefficients $c_{n,m,k>0}$ for $n+m<N$. In other words, the coefficients in front of the logarithms in equation~(\ref{Rt}) can be fixed. Such a procedure parallels the one of Popov \cite{popov_theory_1977}, who was working with the ansatz for $f$. Substituting it into the integral equation (\ref{eq:Love}), Popov was able to fix the coefficients in front of the logarithms order by order at small $\kappa$. However, the determination of the remaining coefficients, in our case $c_{n,m,0}$, is not possible using the ansatz for the bulk regime. Instead one must solve the problem near the edge and match the bulk with the edge solution.

\subsubsection{Connection between the resolvent and the moments of the Love equation.}

Using the relation (\ref{disc}) and equation~(\ref{cont}), from the resolvent (\ref{eq:FRt}) we obtain the ansatz for $f(x)$ in the bulk:
\begin{align}
	f(x)={}&\frac{\sqrt{1-x^2}}{\kappa}+\frac{1}{\pi}\sum_{n,m=0}^\infty\sum_{k=0}^{n+m+1}c_{n,m,k}\kappa^{m+n} \frac{(-1)^n x^{\lambda_k}}{(1-x^2)^{n+1/2}}\notag\\
	&\times \sum_{p=0}^k\frac{k!\lambda_{k-p+1}}{ p!(k-p)!}(i \pi)^{k-p}\ln^p\left(\frac{1-x}{1+x}\right). \label{fR}
\end{align}
Equation (\ref{fR}) is a good ansatz only away from the endpoints of the support of $f(x)$. Indeed, as pointed out by Popov \cite{popov_theory_1977}, higher-order contributions in the perturbative expansion (\ref{fR}) at small $\kappa$ are more and more divergent near $x=\pm 1$. One can easily find the estimate for the perturbative expansion to break down by looking at the consecutive terms in powers of $\kappa$.
They are of the same order when $1-x^2\sim \kappa$, implying the bulk ansatz (\ref{fR}) is only good for $1-x^2\gg \kappa$, i.e., not too close to $x=\pm 1$ \cite{popov_theory_1977}. Such behaviour of $f(x)$ leads to issues when one tries to calculate different moments of $x$, and in particular the zeroth moment that is proportional to the capacitance (\ref{eq:cap}). The formal expression obtained by substituting equation~(\ref{fR}) into the capacitance (\ref{eq:cap}) is divergent. One should thus find a way to treat this problem, since the capacitance is not expected to diverge at any finite $\kappa$. 

A simple solution of the latter problem involves the resolvent (\ref{R}) rather than the function $f(x)$  \cite{volin_quantum_2011,marino_exact_2019}. Let us first expand the resolvent at $x\to \infty$. 
From the definition  (\ref{R}), one expresses it in the form 
\begin{align}
	R(x)=\sum_{n=0}^\infty T_n(\kappa) x^{-n-1}, \label{mom}
\end{align}
where the moments are defined as 
\begin{align}\label{eq:Tn}
	T_n(\kappa)=\int_{-1}^{1}dx x^n f(x).
\end{align}
Since $R(x)=\widetilde R_b(x)+\pi x/\kappa$, by expanding $\widetilde R_b(x)$ of equation~(\ref{Rt}) near $x=\infty$ one obtains the moments $T_n(\kappa)$ as a function of the bulk coefficients $c_{n,m,k}$. In particular, the zeroth moment is given by $T_0(\kappa)=2\pi\mathcal{C}(\kappa)$, where the capacitance is
\begin{align}
	\mathcal{C}(\kappa)=\frac{1}{4\kappa}+\frac{1}{2\pi}\sum_{m=0}^\infty (c_{0,m,0}-2c_{0,m,1})\kappa^m. \label{cap2}
\end{align}
Using $f(x)$ given by equation~(\ref{fR}) makes the integral of $f(x)$ contained in the definition of $T_0$ (\ref{eq:Tn}) divergent due to the divergence of terms with $n>0$ as $|x|\to1$. However, it is very interesting to notice that $T_0(\kappa)$ only contains  the coefficients $c_{0,m,0}$ and $c_{0,m,1}$, i.e., it is determined by the terms with $n=0$ from the ansatz (\ref{fR}). Such truncated ansatz that contains only $n=0$ terms is actually integrable. We were able to explicitly calculate the integral from $-1$ to $1$ of the truncated $f(x)$. Using
\begin{gather}
	\frac{1}{\pi}\int_{-1}^{1} dx \frac{1}{\sqrt{1-x^2}}\ln^p\left(\frac{1-x}{1+x}\right)=(1-\lambda_{p}) (i \pi)^p E_p,\\ 
	\frac{1}{\pi}\int_{-1}^{1} dx \frac{x}{\sqrt{1-x^2}}\ln^p\left(\frac{1-x}{1+x}\right)=2i \lambda_{p} (i \pi)^p p E_{p-1}, 
\end{gather}
where $E_p$ is the Euler number, we performed the summation over $p$. We obtained that the only nonzero contribution is the one arising from $k=0$ or $k=1$, which leads to the right-hand side of equation~(\ref{cap2}) multiplied by $2\pi$.

The preceding discussion implies that the problem of the calculation of the capacitance [see equations~(\ref{eq:cap}) and (\ref{cap2})] simply becomes a determination of the coefficients $c_{0,m,0}$ and $c_{0,m,1}$. As discussed below equation~(\ref{cont}), unlike $c_{0,m,1}$, one cannot obtain the coefficients $c_{0,m,0}$ only from the solution in the bulk. We therefore now consider the problem near the edge.

\subsubsection{Edge regime.}

Let us now focus on the edge regime, which is defined by the limit
\begin{align}
	\kappa\to 0,z\to 1,\qquad t=2\,\frac{z-1}{\kappa}\quad\text{fixed}. \label{edge}
\end{align}
The starting integral equation of the form of equation~(\ref{eq:Love}) in the edge regime at leading order in small $\kappa$ was solved by the Wiener--Hopf method in references~\cite{hutson_circular_1963,popov_theory_1977,pustilnik_fate_2015}. However, here we need more terms of the expansion in $\kappa$, since the resolvent in the edge regime is needed to fix the unknown coefficients $c_{0,m,0}$. They will enable us to find the capacitance (\ref{cap2}) at higher orders in small $\kappa$. In order to find an ansatz, we use the Laplace transform $\hat R(s)$ of $\widetilde R(z=1+\kappa t/2)$ defined by
\begin{align}\label{eq:LT}
	\widetilde R(1+\kappa t/2)=\int_{0}^{\infty} ds e^{-s t}\hat R(s).
\end{align}
This enables us to write the difference equation (\ref{diff}) as an equation that holds for $s<0$ of the form \cite{volin_mass_2010,marino_exact_2019}
\begin{align}
	\sin(s)\left[e^{i s}\hat R(s+i0)+e^{-i s}\hat R(s-i0)\right]=0.\label{diffL}
\end{align}
Equation (\ref{diffL}) imposes constraints on the form that  $\hat R(s)$ can have \cite{volin_mass_2010,volin_quantum_2011}, such as the condition that $\hat R(s)$ must be analytic everywhere except on the negative real axis at each order in small $\kappa$ expansion. The general solution of equation~(\ref{diffL}) is given by \cite{volin_mass_2010,marino_exact_2019}
\begin{align}
	\hat R_e(s)={}&\frac{1}{\sqrt{\kappa} s^{3/2}}\exp\left(\frac{s}{\pi}\ln\frac{\pi e}{s} \right)\Gamma\left(\frac{s}{\pi}+1\right)  \sum_{m=0}^\infty\sum_{n=0}^{\infty}Q_{n,m}\frac{\kappa^{m+n}}{s^n}. \label{Rs}
\end{align}
Here the index $e$ refers to the edge, while the coefficients $Q_{n,m}$ are unknown polynomials of $\ln \kappa$ to be determined using the matching procedure. $\Gamma(x)$ denotes the gamma function. We note that equation~(\ref{diffL}) does not uniquely determine the solution (\ref{Rs}). Namely, each half integer instead of $3/2$ in the first term $1/s^{3/2}$ would nullify equation~(\ref{diffL}). However, the lowest one is fixed after comparing equation~(\ref{Rs}) with the inverse Laplace transform of the leading-order term of equation~(\ref{Rt}) evaluated in the edge regime (\ref{edge}), which is given by
\begin{align}
	\widetilde R_b(z=1+\kappa t/2)=-\frac{\pi}{\sqrt{\kappa}}\sqrt{t}.
\end{align}
Its inverse Laplace transform is $\sqrt{\pi}/2\sqrt{\kappa} s^{3/2}$, which matches the leading-order term $\propto Q_{0,0}$ of equation~(\ref{Rs}) at small $s$. This also fixes $Q_{0,0}=\sqrt{\pi}/{2}$.

\subsubsection{Matching the bulk and the edge solutions.}

Now that we have obtained general expressions for the resolvent in the bulk regime [equation~(\ref{Rt})] and in the edge regime [equation~(\ref{Rs})], let us match them in order to fix all the unknown coefficients $Q_{n,m}$ and $c_{n,m,k}$. In order to proceed, one needs to either perform a Laplace transform of equation~(\ref{Rs}) or an inverse Laplace transform of equation~(\ref{Rt}). We choose to do the latter. The matching procedure therefore becomes equivalent to the problem of solving the equation
\begin{align}
	\hat R_b(s)=\hat R_e(s), \label{RsRt}
\end{align} 
where $\hat R_b(s)$ stands for the inverse Laplace transform of the bulk solution (\ref{Rt}) evaluated in the edge regime (\ref{edge}). In other words [see~equation~(\ref{eq:LT})],
\begin{align}\label{eq:ILTR}
	\hat R_b(s)=\mathscr{L}^{-1}[R_b(1+\kappa t/2)].
\end{align}
Here $\mathscr{L}^{-1}[\ldots]$ denotes the inverse Laplace transform. For a function $F(t)$, it is defined as
\begin{align}\label{eq:ILT}
	\mathscr{L}^{-1}[F(t)]=\int_{\varepsilon-i\infty}^{\varepsilon+i\infty} \frac{dt}{2\pi i} e^{st} F(t).
\end{align}
Here $\varepsilon$ is an arbitrary positive constant chosen in such a way that the contour of integration lies to the right of all singularities of $F(t)$.

Equation (\ref{RsRt}) will be solved order by order at small $\kappa$, and thus it is convenient to perform the inverse Laplace transform~(\ref{eq:ILTR}) on the expansion of $\widetilde R_b(1+\kappa t/2)$ in the limit $\kappa\to 0$.
As a consequence, we need to calculate $\mathscr{L}^{-1}[t^{m-1/2}\ln^n t]$ for $n\geq 0$ and at integer $m$. However, the inverse Laplace transform exists only for $m\leq 0$. To deal with this issue we use the equality $\ln^n A=\lim_{x\to0}\frac{\partial^n}{\partial x^n}e^{x \ln A}$ to obtain an analytic continuation  for $m> 0$. As a result, one has the following analytic continuation under the inverse Laplace transform 
\begin{align}
	\mathscr{L}^{-1}[t^{m-1/2}\ln^n t]=\frac{1}{s^{m+1/2}}\left[\frac{e^{-x\ln s}}{\Gamma(1/2-m-x)}\right]^{(n)}_{x=0}.
\end{align} 
Here and in the following we use the notation 
\begin{align}
	\lim_{x\to 0}\frac{\partial^n}{\partial x^n}A(x)= [A(x)]^{(n)}_{x=0}.
\end{align}

The evaluation of equation~(\ref{eq:ILTR}) is tedious. The main steps can be found in reference~\cite{reichert_analytical_2020}. The final result takes the form
\begin{align}\label{eq:Fbilt}
	\hat R_b(s)={}&-\sqrt{\frac{\pi}{4\kappa s}} \sum_{n=0}^\infty \frac{[(2n)!]^2}{(4^n n!)^3}\frac{2n+1}{2n-1} \frac{\kappa^n}{s^{n+1}}\notag\\
	&+\frac{1}{\sqrt{\kappa s}} \sum_{m=0}^\infty\sum_{n=-m}^\infty\sum_{\ell=0}^{n+m+1}\frac{1}{\ell!} \kappa^m {s}^n\left(\ln\frac{\kappa}{4s}\right)^\ell  V_b(n,m,\ell),
\end{align}
where 
\begin{align}\label{vbbbb}
	V_b(n,m,\ell)={}&\sum_{j=\text{max}(0,-n)}^m\sum_{k=\ell}^{n+m+1}  \frac{(-1)^{j}k!}{4^j j! (k-\ell)!} c_{n+j,m-j,k}\notag\\
	&\times \left[ \frac{\Gamma(n+2j+x+1/2)-2j \lambda_k \Gamma(n+2j+x-1/2)}{\Gamma(n-x+1/2)\Gamma(n+j+x+1/2)}\right]^{(k-\ell)} _{x=0}.
\end{align}
We recall $\lambda_k=[1-(-1)^k]/2$. The expression (\ref{Rs}) we can also bring  to the form of equation~(\ref{eq:Fbilt}):
\begin{align}\label{eq:FeILT}
	\hat R_e(s)={}\frac{1}{\sqrt{\kappa s}}\sum_{n=0}^\infty Q_{n,0}\frac{\kappa^n}{s^{n+1}}+\frac{1}{\sqrt{\kappa s}} \sum_{m=0}^\infty\sum_{n=-m}^\infty\sum_{\ell=0}^{n+m+1}\frac{1}{\ell!}\kappa^m {s}^n \left(\ln\frac{\kappa}{4s}\right)^\ell V_e(n,m,\ell),
\end{align}
where
\begin{gather}\label{veeee}
	V_e(n,m,\ell)=\sum_{j=\text{max}(\ell,n+1)}^{m+n+1} \frac{1}{\pi^{j}(j-\ell)!} Q_{j-n-1,m+n+1-j}  \left[e^{x \ln\frac{4\pi e}{\kappa}}\Gamma(1+x) \right]^{(j-\ell)}_{x=0}.
\end{gather}
Equation (\ref{RsRt}) now becomes equivalent to
\begin{gather}
	V_b(n,m,\ell)=V_e(n,m,\ell), \label{VV}
\end{gather}
provided 
\begin{gather}
	Q_{n,0}=-\frac{\sqrt{\pi}}{2}\frac{[(2n)!]^2}{(4^n n!)^3} \frac{2n+1}{2n-1},\label{QQ}
\end{gather}
which is obtained from the terms that involve the single summation in equations~(\ref{eq:Fbilt}) and (\ref{eq:FeILT}).

It is interesting to note that one can actually determine all the coefficients $c_{n,m,k}$ and $Q_{n,m}$ only from matching  the bulk and edge solutions, i.e., without first finding the coefficients $c_{n,m,k>0}$ using equation~(\ref{diff}) in the bulk, as we discussed earlier. Indeed, in order to find the coefficients $c_{n_1,n_2,n_3}$ or $Q_{n_1,n_2}$, one needs to solve equation~(\ref{VV}) for $n=n_1,m=n_2,\ell=n_3$ and $n=-n_1-1,m=n_1+n_2,\ell=0$, respectively. This leads to a recursive procedure which can be implemented on a computer \cite{marino_exact_2019, volin_mass_2010,volin_quantum_2011}. In Appendix~\ref{ilt} we illustrate the procedure through an example where we find the coefficients needed to obtain the first two corrections of the capacitance.

\subsection{The capacitance at small separations}

Now that all the coefficients $c_{n,m,k}$ can be systematically calculated, we can obtain the capacitance (\ref{cap2}) at $\kappa\ll 1$ to the desired order. We provide here the capacitance with three corrections,
\begin{align}\label{smalllkappa}
	\mathcal{C}(\kappa)={}&\frac{1}{4\kappa}+\frac{\L-1}{4\pi}+\frac{\kappa}{16\pi^2}(\L^2-2)+\frac{ \kappa^2}{64\pi^3}[2\L^2-1-3\zeta(3)]+ \mathcal{O}(\kappa^3),
\end{align}
where $\L=\ln(16\pi/\kappa)$ and $\zeta(n)$ denotes the zeta function. 
Our result is in agreement with the known expressions for the capacitance that were obtained at the linear order in $\kappa$ \cite{tracy_ground_2016,shaw_circulardisk_1970,chew_microstrip_1982}. However, the procedure described in this section could be used to analytically find  an arbitrary number of terms, the only limitation being the computational  time. In reference~\cite{reichert_analytical_2020}, the capacitance is explicitly given to the order $\kappa^7$. 

\subsection{The capacitance at large separations}

\begin{figure}
\centering
\includegraphics[width=0.6\textwidth]{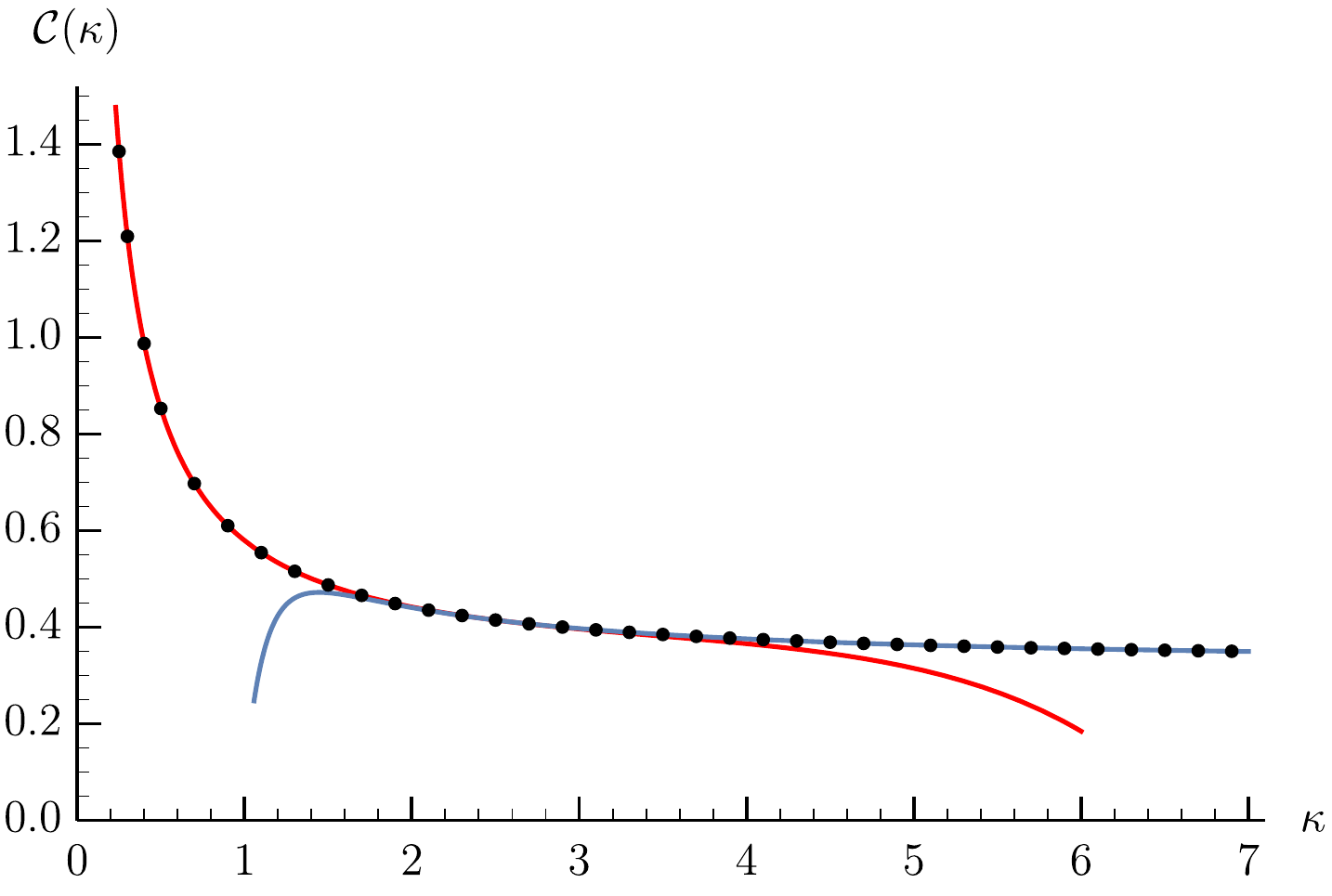}
\caption{\justifying Capacitance $\mathcal C(\kappa)$ as a function of the distance between the plates $\kappa$. The dots represent the exact result obtained by solving numerically equation~(\ref{eq:Love}). The solid red line is the $\kappa\to0$ expansion to $\mathcal{O}(\kappa^6)$ order, while the solid blue line is $\kappa\to\infty$ expansion to $\mathcal{O}(1/\kappa^8)$ order.}
	\label{fig:nx}
\end{figure}

At $\kappa\gg 1$, the Love equation (\ref{eq:LT}) can be analytically solved in a systematic way using the expansion into Legendre polynomials, see section \ref{section:simplesolution}. The capacitance (\ref{eq:cap}) is simply given by
\begin{align}
	\mathcal{C}(\kappa)=2a_0(\kappa),
\end{align} 
where $a_0$ should be obtained from the system of equations (\ref{set}) with the replacement $\lambda\to\kappa$. Once the system is solved at some fixed $M$, the highest-order term in capacitance will be proportional to $1/\kappa^{2M+2}$. For $M=1$ we obtain the result
\begin{align}\label{largekappa}
	\mathcal{C}(\kappa)={}&\frac{1}{\pi}+\frac{2}{\pi^2\kappa}+\frac{4}{\pi^3\kappa^2}-\frac{4(\pi^2-6)}{3\pi^4\kappa^3}-\frac{16(\pi^2-3)}{3\pi^5\kappa^4} +\mathcal{O}(\kappa^{-5}).
\end{align}
In figure~\ref{fig:nx} we show the numerically evaluated capacitance that perfectly matches with the analytical formulae we calculated. 

\subsection{Connection with the Lieb--Liniger model}\label{AppD}

In the context of one-dimensional quantum physics, the Love integral equation (\ref{eq:Love}) is up to a trivial rescaling equivalent to the Lieb integral equation (\ref{eq:LIE}) with the kernel (\ref{eq:kernel}). The dimensionless parameter $\gamma=c/n$ can be connected to the solution of the integral equation (\ref{eq:Love}) as
\begin{align}\label{conn}
	\gamma=2\pi\frac{\kappa}{T_0(\kappa)},
\end{align}
where $T_0(\kappa)$ is defined by equation~(\ref{eq:Tn}). The ground-state function (\ref{eq:e2def}) is given by the expression
\begin{align}\label{eg}
	e_2(\gamma)=4\pi^2\frac{T_2(\kappa)}{T_0(\kappa)^3},
\end{align}
where in the right-hand side one should express $\kappa$ as a function of $\gamma$ using their connection (\ref{conn}). Similarly to $T_0(\kappa)$ [cf.~equation~(\ref{cap2})], the second moment can also be obtained from the resolvent. It takes the form
\begin{align}
	T_2(\kappa)={}&\frac{\pi}{8\kappa}+\sum_{m=0}^\infty \Biggl[\left(\frac{1}{2}c_{0,m,0}-\frac{5}{3}c_{0,m,1}\right)\kappa^m +(4c_{0,m+1,2}+c_{1,m,0}-2c_{1,m,1})\kappa^{m+1}\notag\\ &-8c_{0,m+2,3}\kappa^{m+2} \Biggr]. \label{t2}
\end{align}
Evaluating equation~(\ref{eg}) at $\kappa\ll 1$, corresponding to the regime of weak interactions between bosons $\gamma\ll 1$, one can reproduce equation~(\ref{eq:e2nineterms}). In a similar way one can calculate higher moments $e_4(\gamma)$, $e_6(\gamma)$, etc. We used this in order to obtain the constants  (\ref{eq:a44a66}).

\subsection{A parametric form of the capacitance}

The connection between the Lieb and Love integral equations enables us to express the capacitance directly from the previously obtained results for the system of interacting bosons. Comparing with equation~(\ref{eq:norm}), we obtain that the capacitance (\ref{eq:cap}) can be defined parametrically via $\gamma$ as $\kappa(\gamma)=\gamma\;\! {n}/{Q}$, $\mathcal{C}(\gamma)={n}/{Q}$. Here one should have in mind that $n/Q$ is a function of $\gamma$, see equation~(\ref{eq:Qfin}). We thus arrive at the final result
\begin{align}\label{eq:Cparametric}
	\kappa(\gamma)=\frac{\sqrt\gamma}{2\;\!g(\gamma)},\quad \mathcal{C}(\gamma)= \frac{1}{2\sqrt\gamma\, g(\gamma)},
\end{align}
where the function $g(\gamma)$ is determined by equation~(\ref{eq:gdiff}). Equation (\ref{eq:Cparametric}) is the exact parametric solution for the capacitance at arbitrary separations $\kappa$ enabling one to obtain the capacitance from the ground-state function $e_2(\gamma)$.

The regime of small separations between the plates, $\kappa\ll 1$, corresponds to $\gamma\ll 1$. One should therefore substitute $g(\gamma)$ of equation~(\ref{eq:gsolutionweak}) in the parametric form (\ref{eq:Cparametric}). The obtained result for the capacitance has a significant simplification with respect to the explicit form $\mathcal{C}(\kappa)$, which has a series expansion that begins with the terms given by equation~(\ref{smalllkappa}). In the parametric form there is only one logarithmic term originating from equation~(\ref{eq:gsolutionweak}), unlike in the explicit form $\mathcal{C}(\kappa)$ where the same term proliferates. The function $g(\gamma)$ can be calculated trivially beyond the terms of equation~(\ref{eq:gsolutionweak}) using the result for $e_2(\gamma)$ and the differential equation (\ref{eq:gdiff}). Further corrections will only contain the  power law terms of $\sqrt{\gamma}$, but not any logarithms. We eventually note that in a recent work \cite{bajnok_running_2022},  another parametrisation is found where the expression for the capacitance does not contain logarithms.

\section{Polaron energy spectrum in the Yang--Gaudin Bose gas\label{section:Yang-Gaudin gas}}

When a surrounding medium hosts a distinguishable particle, the properties of that particle are changed due to interactions. The distinguishable particle is transformed into a quasiparticle known as a polaron. Some of its characteristics are the effective mass, the momentum, and the energy. Here we study the polaron within the model of a one-dimensional Bose gas with two internal states. It is described by the previously introduced Hamiltonian (\ref{eq:Horiginal}), provided the wave function obeys the symmetries of a two component, isospin-$\frac{1}{2}$, Bose gas. In this case the model is known under the name Yang--Gaudin \cite{yang_exact_1967,gaudin}.  

Consider the bath of bosons all of the same isospin with one boson of the opposite isospin. This system has three types of elementary excitations. Two of them correspond to the ones of the single-component Lieb--Liniger Bose gas, classified as particlelike type-I and holelike type-II excitations. The third kind of excitations arises due to the presence of an extra boson with the opposite isospin. It is known under the names spin-wave excitation or magnon \cite{fuchs_spin_2005,zvonarev_edge_2009,matveev_spectral_2008} and (iso)spinon \cite{li_exact_2003,robinson_excitations_2017}. This collective excitation we will call a polaron quasiparticle excitation. At low momenta, it has a quadratic dispersion. Since the lowest excitations of the host medium are phonons with linear dispersion, it is energetically favourable for the system to host an excited polaron rather than a phonon. This picture is valid beyond the small-momentum regime as the polaron dispersion lies below the type-II excitation branch. Thus the polaron can be understood as the lowest excitation branch of the system for a given momentum. 

For weak interactions, the polaron excitation energy spectrum can be obtained using the standard perturbation theory. The leading-order contribution arises at the second order. At sufficiently low momentum $p$, it has the form
\begin{align}\label{eq:dispersionperturbationtheory}
	\mathcal{E}(p)=\frac{p^2}{2m}-\frac{mv^2}{K}\left(\frac{\mathrm{arctanh}(p/mv)}{p/mv}-1\right).
\end{align} 
Here $K$ is the Luttinger liquid parameter that is the same as the one for the Lieb--Liniger model and $v=\pi\hbar n/mK$ is the sound velocity. Equation (\ref{eq:dispersionperturbationtheory}) describes an analytic even function of $p$. It has the low-momentum expansion
\begin{align}\label{eq:E(p)polaron}
	\mathcal{E}(p)=\frac{p^2}{2m^*}-\frac{\nu\;\! p^4}{24\hbar^2 n^2m}+\ldots,
\end{align}
where $m^*$ is the effective polaron mass and $\nu$ the dimensionless parameter that controls the quartic term. For the perturbative result (\ref{eq:dispersionperturbationtheory}), they are given by
\begin{align}\label{eq:m*nuperturbationtheory}
	\frac{m}{m^*}=1-\frac{2}{3\pi}\sqrt\gamma,\quad \nu=\frac{24}{5\pi\sqrt\gamma}.
\end{align} 
In this section we will derive the exact results for $m^*$ and $\nu$ and study the polaron energy spectrum in more details. It is based on the published works \cite{ristivojevic_exact_2021,ristivojevic_dispersion_2022,ristivojevic_method_2022}.

\subsection{The Bethe ansatz equations}

The Hamiltonian (\ref{eq:Horiginal}) for the two-component case is solved by the Bethe ansatz \cite{yang_exact_1967,gaudin,li_exact_2003,oelkers_bethe_2006}. Its eigenstates can be classified with respect to the value of the total isospin. In the sector where it has the maximal value $N/2$, the system simplifies to the single-component Lieb--Liniger model. It is characterised by $N$ density quantum numbers, $I_1,I_2,\ldots, I_N$, which define $N$ quasimomenta, $k_1,k_2,\ldots, k_N$. In the case of the total isospin $N/2-1$, which is in our focus, the system acquires an additional spin quantum number $J$ that defines the spin rapidity $\eta$. The Bethe ansatz equations are given by \cite{li_exact_2003,oelkers_bethe_2006}
\begin{subequations}\label{BA}
	\begin{gather}
		e^{i Lk_j}=\frac{k_j-\eta-i c/2}{k_j-\eta+i c/2}\prod_{l=1\atop{l\neq j}}^{N}\frac{k_j-k_l+i c}{k_j-k_l-i c},\quad j=1,2,\ldots,N,\\
		1=\prod_{l=1}^{N}\frac{\eta-k_l-i c/2}{\eta-k_l+i c/2}.
	\end{gather}
\end{subequations}
Here $L$ is the system size and $c$ the interaction strength. After taking the logarithm, the system (\ref{BA}) can be expressed in the form
\begin{subequations}\label{eq:BE}
	\begin{gather}\label{eq:BE1}
		L k_j-\sum_{l=1}^N \theta(k_j-k_l)={2\pi}I_j+\pi-\theta(2k_j-2\eta),\quad j=1,2,\ldots,N,\\\label{eq:BE2}
		2\pi J=-\sum_{l=1}^N \theta(2\eta-2k_l).
	\end{gather} 
\end{subequations}
Here $I_j=n_j-(N+1)/2$, where $n_j$ are integers, $J$ is an integer or odd half-integer depending on whether $N$ is even or odd, and the scattering phase shift is given by equation~(\ref{eq:phaseshift}). The energy $E$ and the momentum $p$ of the system are given by
\begin{align}\label{eq:EP}
	E=\frac{\hbar^2}{2m}\sum_{j=1}^N k_j^2,\quad p=\hbar \sum_{j=1}^N k_j.
\end{align}
Note that the spin rapidity $\eta$ indirectly enters to the energy and the momentum~(\ref{eq:EP}) through the Bethe ansatz equations.

In the special case $\eta\to+\infty$, equation~(\ref{eq:BE1}) becomes independent of the spin quantum number $J$ that enters equation~(\ref{eq:BE2}), and describes the quasimomenta of the single-component system of bosons described by the Lieb--Liniger model. Its ground state is realised for the quantum numbers $I_j$ given by equation~(\ref{eq:quantumnumbers}). The density of quasimomenta (\ref{eq:Fhokj}) in the thermodynamic limit satisfies the integral equation (\ref{eq:LIE}) with the kernel (\ref{eq:kernel}). The density of quasimomenta enables us to calculate the momentum and the energy of the system as a function of the Fermi quasimomentum $Q$. However, if needed, one can express $Q$ in terms of the density $n$ using their connection (\ref{eq:n}).

As follows from equation~(\ref{eq:BE2}), the spin quantum number takes the maximal value $J=N/2$ at $\eta\to+\infty$. Let us study the case of finite $\eta$ where $I_j$ assumes the values given by equation~(\ref{eq:quantumnumbers}). Then equations~(\ref{eq:BE}) describe the excited state of the system that hosts a magnon or in our picture a polaron quasiparticle excitation. The momentum of the system in the excited state coincides with the momentum of the polaron excitation. It can be obtained directly from equations~(\ref{eq:EP}) and (\ref{eq:BE1}), which in the thermodynamic limit gives 
\begin{align}\label{eq:p}
	p(Q,\eta)=\hbar \int_{-Q}^{Q} dk\:\! \rho(k,Q)\left[\pi-\theta(2k-2\eta)\right].
\end{align}
Note that $p$ explicitly depends on the Fermi quasimomentum $Q$ and the spin rapidity $\eta$.

The evaluation of the energy of the system in the excited state is more involved. At finite $\eta$, the quasimomenta become shifted by $\Delta k_j=k_j(\eta)-k_j(\eta\to+\infty)=\mathcal{O}(1/L)$. From equation~(\ref{eq:BE1}) it then follows
\begin{align}\label{eq:Deltak}
	L\Delta k_j={}&\pi-\theta(2k_j-2\eta)+\sum_{l=1}^{N}\theta'(k_j-k_l)(\Delta k_j-\Delta k_l)+\mathcal{O}(1/N).
\end{align}
The formal expression $\rho(k,Q)=\frac{1}{L}\sum_{j=1}^{N}\delta(k-k_j)$ substituted into equation~(\ref{eq:LIE}) leads to
\begin{align}\label{eq:LIEdiscrete}
	1-\frac{1}{L}\sum_{j=1}^N \theta'(k_j-k)=2\pi \rho(k,Q).
\end{align}
After introducing $g(k_j,Q)=L\rho(k_j,Q)\Delta k_j$, equation~(\ref{eq:LIEdiscrete}) enables us to express equation~(\ref{eq:Deltak}) as an integral equation
\begin{subequations}\label{eq:g}
	\begin{gather}
		g(k,Q)+\frac{1}{2\pi}\int_{-Q}^{Q}dq\;\! \theta'(k-q)g(q,Q)=r(k,\eta),\\
		r(k,\eta)=\frac{1}{2}- \frac{\theta(2k-2\eta)}{2\pi}.
	\end{gather}
\end{subequations}
The energy of the system (\ref{eq:EP}) in the thermodynamic limit now becomes $E=N\epsilon_0+\mathcal{E}(Q,\eta)$, where $\epsilon_0$ is given by equation~(\ref{eq:eps0}), while the energy of the polaron excitation corresponding to the momentum (\ref{eq:p}) is given by
\begin{align}
	\mathcal{E}(Q,\eta)=\frac{\hbar^2}{m}\int_{-Q}^{Q} dk\,\! k g(k,Q).
\end{align}
Here $g(k,Q)$ depends on $\eta$ and satisfies equation~(\ref{eq:g}). Using  $\sigma(k,Q)$ that satisfies the integral equation (\ref{eq:sigma}) and the result of Appendix \ref{appendixb} we eventually obtain
\begin{align}\label{eq:eps}
	\mathcal{E}(Q,\eta)=-\frac{1}{2\pi}	\int_{-Q}^{Q} dk\:\! \sigma(k,Q)\theta(2k-2\eta),
\end{align}
where we have used that $\sigma(k,Q)$ is an odd function of $k$. As expected, equation (\ref{eq:eps}) leads to $\mathcal{E}=0$ at $\eta\to+\infty$.

Equations  (\ref{eq:p}) and (\ref{eq:eps}) are exact and determine the parametric form  of the polaron energy spectrum. Our goal is to express it in the explicit form $\mathcal{E}(p)$. Analytically, this is, however, a difficult problem for arbitrary values of the interaction. We can nevertheless study this problem in several special cases.

\subsection{General properties of the polaron energy spectrum}

Let us first study global features of the polaron energy spectrum. The momentum (\ref{eq:p}) and the energy (\ref{eq:eps}) satisfy some general relations that follow from the general properties of $\rho(k,Q)$ and $\sigma(k,Q)$ that are, respectively, even and odd analytic functions of $k$. It then follows that the momentum and energy are analytic functions of $\eta$.  The momentum (\ref{eq:p}) is bounded between $p(Q,\eta\to+\infty)=0$ and $p(Q,\eta\to-\infty)=2\pi\hbar n$. It has the symmetry property around $\pi\hbar n$,
\begin{align}\label{eq:psim}
	p(Q,\eta)-\pi\hbar n=\pi\hbar n-p(Q,-\eta),
\end{align}
which implies $p(Q,0)=\pi\hbar n$. The energy (\ref{eq:eps}) is an even function of $\eta$. It is bounded and reaches the minimum at $\mathcal{E}(Q,\eta\to\pm\infty)=0$, while the maximum occurs at $\eta=0$, which is
\begin{align}
	\mathcal{E}(Q,0)=-\frac{1}{\pi}\int_{0}^{Q}dk \sigma(k,Q)\theta(2k).
\end{align} 
The energy at the maximum thus satisfies the inequality
\begin{align}\label{eq:ineq}
	\mathcal{E}(Q,0)<\int_{0}^{Q}dk \sigma(k,Q).
\end{align}
The quantity on the right-hand side of equation~(\ref{eq:ineq}) formally denotes the energy of the type-II excitation with the momentum $\pi\hbar n$ in the Lieb--Liniger model. Therefore, the energy of the polaron excitation at its maximum is smaller than the energy of the type-II excitation. Having in mind that the latter excitation branch has the smallest energy among all branches at small momenta, it is expected that the same picture holds at any momentum.  This conclusion is indeed correct as verified numerically~\cite{zvonarev_edge_2009}.

The parity of the energy (\ref{eq:eps}) with respect to $\eta$, in conjunction with equation~(\ref{eq:psim}), gives the symmetry property of the dispersion when it is expressed explicitly as a function of the momentum\footnote{\linespread{1}\selectfont Note that from now the notation becomes ambiguous since $\mathcal{E}$ was initially defined as a function of two arguments, see  equation~(\ref{eq:eps}).},
\begin{align}\label{eq:symeps}
	\mathcal{E}(\pi\hbar n+q)=\mathcal{E}(\pi\hbar n-q),\quad 0\le q\le\pi\hbar n. 
\end{align}
Since $\mathcal{E}(p)$ is analytic for our system, the property (\ref{eq:symeps}) shows that odd derivatives of $\mathcal{E}(p)$ at its maximum, $p=\pi\hbar n$, nullify. 

For a given set of quasimomenta $\{k_1,k_2,\ldots,k_N,\eta\}$ that satisfies equations~(\ref{BA}), there is another set that also satisfies equations~(\ref{BA}). It is defined  by the shift
\begin{align}\label{eq:LLtr}
	\tilde k_j={}&k_j+\frac{2\pi}{L}l,\quad j=1,2,\ldots,N,\\
	\quad \tilde\eta={}&\eta+\frac{2\pi}{L}l,
\end{align} 
where $l$ is an integer. The energy and momentum of the new set are
\begin{subequations}
	\begin{gather}
		\tilde E=\frac{\hbar^2}{2m}\sum_{j=1}^N \tilde k_j^2=E+\frac{2\pi l\;\! \hbar n}{Nm}\left(p+{\pi l\;\!\hbar n}\right),\\
		\tilde p=\hbar \sum_{j=1}^N \tilde k_j=p+2\pi l\;\! \hbar n,
	\end{gather}	
\end{subequations}
which correspond to the energy and momentum (\ref{eq:EP}) of the original set. In the thermodynamic limit, the energies of the two configurations are the same, $\tilde E=E$, while the momentum is shifted by an integer multiple of $2\pi\hbar n$. Since the energy of the Lieb--Liniger model does not change \cite{lieb_exact_1963} under the transformation (\ref{eq:LLtr}), we can conclude that the polaron energy spectrum is a periodic function of the momentum,
\begin{align}\label{eq:periodicity}
	\mathcal{E}(p)=\mathcal{E}(p+2\pi l\;\!\hbar n).
\end{align}
Since the spectrum also satisfies the reflection property (\ref{eq:symeps}), it is sufficient to study it in the reduced region $0< p\le \pi\hbar n$. The two properties (\ref{eq:periodicity}) and (\ref{eq:symeps}) will then determine the spectrum at other momenta automatically.

\subsection{Low-energy polaron spectrum at arbitrary interactions}

Equations (\ref{eq:p}) and (\ref{eq:eps}) define the polaron spectrum at an arbitrary momentum and interactions. The low-energy spectrum can be obtained using the formalism developed in section~\ref{section:method}. At $\eta\to+\infty$, the momentum (\ref{eq:p}) and the energy (\ref{eq:eps}) nullify. In order to access small $p$ and $\mathcal{E}$ we expand $\theta(2k-2\eta)$ at $\eta/c\gg 1$. Accounting for the leading- and subleading-order terms, we obtain
\begin{subequations}
	\label{eq:pE}
	\begin{gather}
		p=\frac{\hbar c\:\!n}{\eta}- \frac{\hbar c^3 n}{12\eta^3}+\frac{2\hbar c}{\pi \eta^3} A_{0,2}+\ldots,\\
		\mathcal{E}=\frac{\hbar^2 c}{\pi m \eta^2}A_{1,1}+\frac{6\hbar^2c}{\pi m \eta^4}A_{1,3}-\frac{\hbar^2 c^3}{4\pi m\eta^4}A_{1,1}+\ldots,
	\end{gather}  
\end{subequations}
where $A_{j,l}$ is defined by equation~(\ref{eq:Ajl}). Upon elimination of the spin rapidity $\eta$, the low-momentum spectrum acquires the form of equation~(\ref{eq:E(p)polaron}) with the effective polaron mass
\begin{align}
	\frac{m}{m^*}=\frac{2A_{1,1}}{\pi  c\:\! n^2}
\end{align}
and the quartic coefficient given by
\begin{align}
	\nu=\frac{2A_{1,1}}{\pi c n^2}+\frac{96 A_{0,2}A_{1,1}}{\pi^2 c^3 n^3} -\frac{144 A_{1,3}}{\pi c^3 n^2}.
\end{align}
Further simplification of $m^*$ and $\nu$ follows directly from equations~(\ref{eq:polaron}) and (\ref{eq:e2ldefinition}), yielding
\begin{align}\label{eq:m*}
	\frac{m}{m^*}=-\gamma^2 \frac{d}{d\gamma}\left(\frac{e_2(\gamma)}{\gamma^2}\right)
\end{align}
and 
\begin{align}\label{eq:nu}
	\nu=6\gamma^2 \frac{d}{d\gamma}\biggl(\frac{e_4(\gamma)}{\gamma^4}\biggr)-\left(\gamma^2+{24e_2(\gamma)}\right) \frac{d}{d\gamma}\biggl(\frac{e_2(\gamma)}{\gamma^2}\biggr).
\end{align}
Equations (\ref{eq:m*}) and (\ref{eq:nu}) are the exact results for the coefficients of the polaron energy spectrum (\ref{eq:E(p)polaron}) in the Yang--Gaudin model of the Bose gas.

The series expansion results for $m^*$ and $\nu$ follow straightforwardly from equations~(\ref{eq:e2nineterms}) and (\ref{eq:e2largefinal}). At $\gamma\ll 1$ we obtain
\begin{align}\label{eq:m*gamma<<1}
	\frac{m}{m^*}=1-\frac{2\sqrt\gamma}{3\pi}+\frac{4-3\zeta(3)}{16\pi^3} \gamma^{3/2}+\frac{4-3\zeta(3)}{24\pi^4} \gamma^{2}+\mathcal{O}(\gamma^{5/2}),
\end{align}
and
\begin{align}\label{eq:nugamma<<1}
	\nu=\frac{24}{5\pi\sqrt\gamma}-1+\frac{28}{3\pi^2}+ \left(\frac{2}{3\pi}+\frac{7}{\pi^3}-\frac{45\zeta(3)}{4\pi^3}\right) \sqrt{\gamma}+\mathcal{O}(\gamma^{3/2}),
\end{align}
which is in agreement with the result (\ref{eq:m*nuperturbationtheory}) obtained using the perturbation theory. Equation (\ref{eq:m*gamma<<1}) shows that by decreasing the interaction strength, the polaron mass approaches the bare mass of the original particle. This is expected as it becomes decoupled from the surrounding system. We also note the absence of the term linear in $\gamma$ in equations~(\ref{eq:m*gamma<<1}) and (\ref{eq:nugamma<<1}), which is obvious from the form of the derivatives  in equations~(\ref{eq:m*}) and (\ref{eq:nu}). In the regime of strong interactions, $\gamma\gg 1$, we obtain
\begin{align}\label{eq:m*gamma>>1}
	\frac{m}{m^*}=\frac{2\pi^2}{3\gamma}-\frac{4\pi^2}{\gamma^2}+\frac{16\pi^2}{\gamma^3}+\mathcal{O}(\gamma^{-4})
\end{align}
and
\begin{align}\label{eq:nugamma>>1}
	\nu=\frac{2\pi^2}{3\gamma}-\frac{4\pi^2}{\gamma^2}+\frac{8\pi^2(30+\pi^2) }{15\gamma^3}+\mathcal{O}(\gamma^{-4}).
\end{align}
In this case the polaron motion involves the motion of many surrounding bosons. Therefore, the resulting quasiparticle is heavy due to the surrounding cloud, with its mass diverging in the thermodynamic limit for infinite repulsion, see equation~(\ref{eq:m*gamma>>1}). We note that $\frac{m}{m^*}$ and $\nu$ are the same at the two leading orders. This can be understood from equations~(\ref{eq:m*}) and (\ref{eq:nu}) as $e_2(\gamma)$ and $e_4(\gamma)$ are constants at $\gamma\to\infty$. 

Equations~(\ref{eq:m*}) and (\ref{eq:nu}) show that the quadratic and quartic coefficients of the low-energy spectrum of a polaron in the Yang--Gaudin Bose gas are fully determined by the momenta of the quasimomentum distribution (\ref{eq:e2l}) in the Lieb--Liniger model. The latter statement is correct beyond the first two coefficients. Indeed, the series expansion of $\theta(2k-2\eta)$ in equations~(\ref{eq:p}) and (\ref{eq:eps}) is a power law in $k$ with the positive powers and thus the expressions (\ref{eq:pE}) will depend on $A_{j,l}$ defined by equation~(\ref{eq:Ajl}). They can be transformed to $e_{2l}$ using equations~(\ref{eq:Ajlsol}) and (\ref{eq:e2ldefinition}).

The spectrum (\ref{eq:E(p)polaron}) is quadratic only at momenta smaller than
\begin{align}\label{eq:p*polaron}
	p^*\sim \hbar n \sqrt{\frac{m}{m^* \nu}}.
\end{align}
At $\gamma\ll1$ this gives $p^*\sim\hbar n \gamma^{1/4}$. The value of $p^*$ that is less than $\hbar n$ signals the existence of a qualitatively new behaviour of the polaron spectrum at finite momenta that cannot be described by equation~(\ref{eq:E(p)polaron}). This will be discussed in the following subsection. At  $\gamma\gg 1$ we obtain $p^*\sim\hbar n$ and thus at strong interactions we do not expect to have qualitatively new regimes.

\subsection{Polaron energy spectrum at weak interactions}

Let us find the spectrum in the regime of weak interactions, $\gamma\ll 1$. The solution of equations~(\ref{eq:LIE}) and (\ref{eq:sigma}) at two leading orders are given by equation~(\ref{eq:rhoPopov}) and (\ref{eq:sigmaPopov}), respectively. The Fermi quasimomentum is calculated in equation~(\ref{eq:Qfin}) supplemented by equation~(\ref{eq:gsolutionweak}). Substituting them into equations~(\ref{eq:p}) and (\ref{eq:eps}), for $-Q\le \eta\le Q$ we find
\begin{subequations}
	\label{eq:epspweak}
	\begin{align}
		\label{eq:pweak}
		p={}&\hbar n\left[2\phi-\sin(2\phi)+\sqrt\gamma A(\phi)+\mathcal{O}(\gamma) \right],\\
		\mathcal{E}={}&\frac{4\hbar^2 n^2}{3 m}\sqrt{\gamma}\left[\sin^3\phi+ \frac{3\sqrt\gamma}{4}\left(A(\phi) \cos\phi-\frac{1}{2}\right)+\mathcal{O}(\gamma)\right],
	\end{align}
\end{subequations}
where
\begin{align}
	A(\phi)={}&\cos\phi +\frac{\sin(2\phi)}{2\pi}\left(\ln\frac{32\pi}{\sqrt\gamma}-1\right)
	+\frac{2\sin\phi}{\pi} \ln\tan\left(\frac{\phi}{2}\right).
\end{align}
Here we have introduced the parametrisation $\cos\phi=\eta/Q$, where  $0\le\phi\le \pi$. The terms in brackets in equations~(\ref{eq:epspweak}) proportional to $\sqrt\gamma$ arise from the subleading terms of equations~(\ref{eq:rhoPopov}) and (\ref{eq:sigmaPopov}). They should, therefore, be smaller than the corresponding leading-order ones. At small momenta this occurs at 
\begin{align}\label{eq:p0condition}
	p\gg p_0=\hbar n\sqrt{\gamma},
\end{align}
which is the condition for the applicability of the spectrum (\ref{eq:epspweak}). 

The evaluation of equations~(\ref{eq:p}) and (\ref{eq:eps}) for $\eta\ge Q$ at the leading order in $\gamma\ll 1$ gives
\begin{subequations}
	\label{eq:simple}
	\begin{align}
		\label{eq:pweaksmall}
		p={}&\hbar n\sqrt\gamma\;\! e^{-\phi},\\
		\mathcal{E}={}&\frac{\hbar^2 n^2}{2m}\gamma\;\! e^{-2\phi},
	\end{align}
\end{subequations} 
where $\eta=Q\cosh\phi$ and $\phi>0$. The form of equation~(\ref{eq:pweaksmall}) implies $p<p_0$, which is consistent with the condition (\ref{eq:p0condition}) obtained for the complementary region. Therefore, at momenta smaller than $p_0$, we have obtained a quadratic spectrum of the polaron, $\mathcal{E}(p)=p^2/2m$,  which crosses over into the parametric form given by equations~(\ref{eq:epspweak}) at momenta higher than $p_0$. We note that the quadratic energy spectrum (\ref{eq:simple}) is the first term of the general form of the low-energy spectrum (\ref{eq:E(p)polaron}) where $m^*$ and $\nu$ are given by equations~(\ref{eq:m*}) and (\ref{eq:nu}). From the symmetry property (\ref{eq:symeps}), it follows that the spectrum is also quadratic in the vicinity of $2\pi\hbar n$, which corresponds to $\eta\le -Q$.

\begin{figure}
\centering
\includegraphics[width=0.6\textwidth]{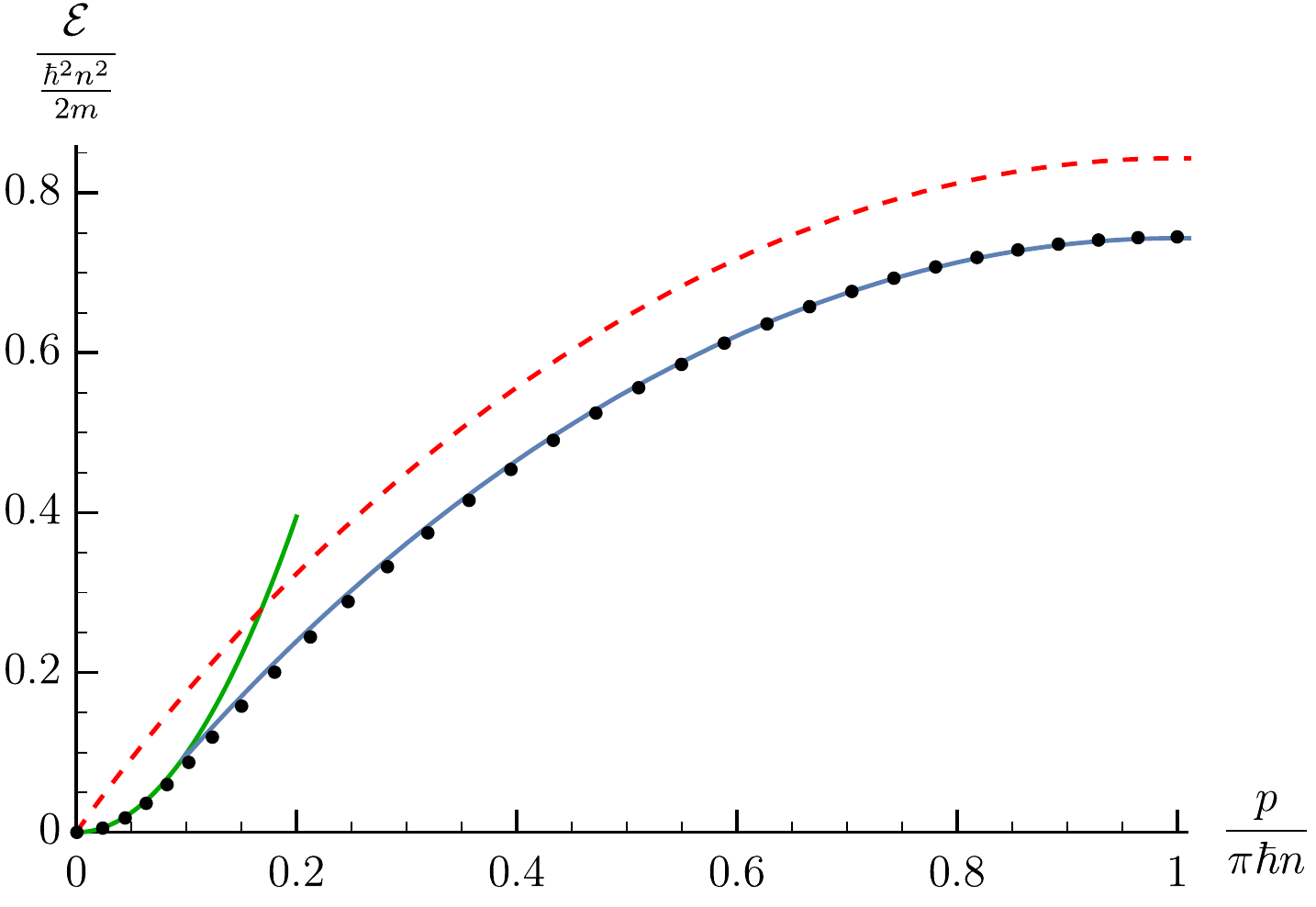}
\caption{\justifying Comparison of the exact result for the polaron dispersion (the dots) and the obtained parametric form (solid curves) given by equations~(\ref{eq:epspweak}) and (\ref{eq:simple}) for $\gamma=0.1$. The dashed curve is the dark soliton dispersion. On the plot is only shown the region of momenta $0\le p\le \pi\hbar n$. At other values of $p$, the dispersion follows from the symmetry (\ref{eq:symeps}) and the periodicity (\ref{eq:periodicity}).} \label{fig1}
\end{figure}

The polaron energy spectrum at weak interactions, $\gamma\ll 1$, and momenta above $p_0$ is given by equations~(\ref{eq:epspweak}). Interestingly, accounting only for the leading-order term (i.e., neglecting the terms proportional to $\sqrt{\gamma}$ in the brackets), equations~(\ref{eq:epspweak}) describe the spectrum of the dark soliton solution  \cite{kulish_comparison_1976,ishikawa_solitons_1980,imambekov_one-dimensional_2012} of the Gross--Pitaevskii equation. It corresponds to type-II excitations in the Lieb--Liniger model \cite{lieb_exact_1963} at momenta higher than $\hbar n\gamma^{3/4}$ \cite{imambekov_one-dimensional_2012,pustilnik_low-energy_2014,ristivojevic_excitation_2014}. However, the polaron excitation energy is always smaller than the energy of type-II excitation with the same momentum, as we have explicitly shown in equation~(\ref{eq:ineq}) at the energy maximum, i.e., at $p=\pi\hbar n$. Equations (\ref{eq:epspweak}) at $\phi=\pi/2$ and thus $A(\phi)=0$ also illustrate this point, leading to 
\begin{align}\label{eq:top}
	\mathcal{E}(\pi\hbar n)=\frac{4\hbar^2 n^2}{3m}\sqrt\gamma\left(1-\frac{3\sqrt\gamma}{8}+\mathcal{O}(\gamma)\right).
\end{align}
The leading-order term in equation~(\ref{eq:top}) is the energy of the dark soliton, while the whole expression (\ref{eq:top}) represents the polaron energy, which is smaller. Notice that the energy of the type-II excitation with the momentum $\pi\hbar n$ does not have the correction proportional to $\gamma$, see equation~(\ref{eq:Ebgweak}), unlike the polaron. In figure~\ref{fig1} we show the exact result obtained numerically for the polaron dispersion and small-$\gamma$ expansion given by equations~(\ref{eq:epspweak}) and (\ref{eq:simple}). The agreement is good even for not particularly small value $\gamma=0.1$, becoming better with decreasing $\gamma$.

\subsection{Polaron energy spectrum at strong interactions}

\begin{figure}
\centering
\includegraphics[width=0.6\textwidth]{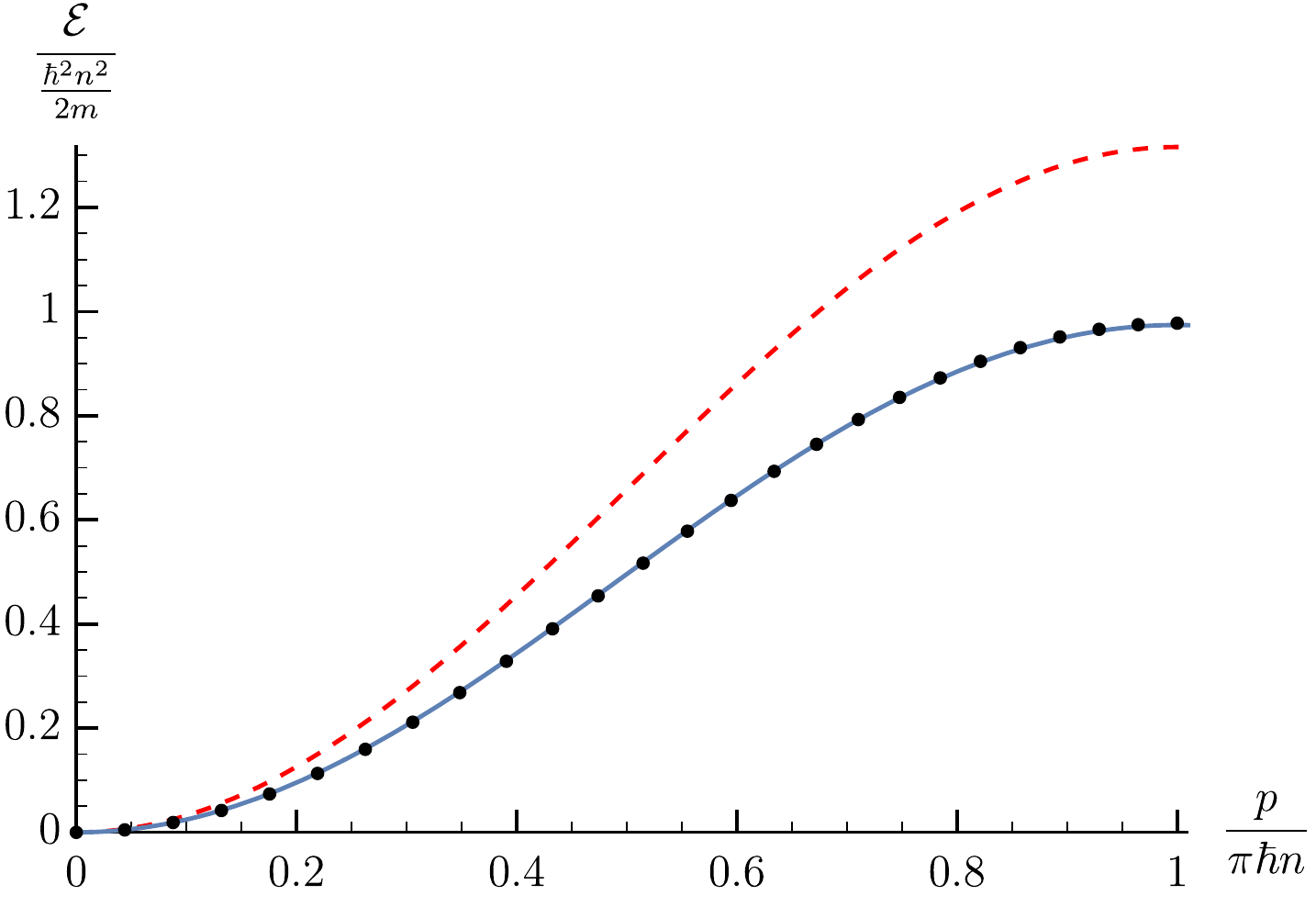}
\caption{\justifying Comparison of the exact result for the polaron dispersion (the dots) and the obtained expression (\ref{eq:displarge}) (solid curve) for $\gamma=20$. The dashed curve represents the dispersion (\ref{eq:displarge}) at the leading order that only accounts for the term proportional to $1/\gamma$.} \label{fig-polarondispersion}
\end{figure}

At strong interactions, $\gamma\gg 1$, the Bethe ansatz equations (\ref{eq:LIE}) and (\ref{eq:sigma}) can be solved using a perturbation theory. Accounting for the first three orders, we find
\begin{gather}\label{eq:rhosignastrong}
	\rho(k,Q)=\frac{1}{2\pi}+ \frac{Q}{\pi^2 c}+\frac{2Q^2}{\pi^3 c^2}+\mathcal{O}(Q^3/c^3),\\ \label{eq:sigmasignastrong}
	\sigma(k,Q)=\frac{\hbar^2 k}{m}+\mathcal{O}(Q^3/c^3).
\end{gather}
This enables us to evaluate equations~(\ref{eq:p}) and (\ref{eq:eps}), giving
\begin{subequations}\label{eq:pEstrong}
	\begin{align}
		p={}&\pi\hbar n\left[1-\frac{2 \arctan\vartheta}{\pi}+\frac{8\pi\vartheta	}{3(1+\vartheta^2)^2 \gamma^2} +\mathcal{O}(\gamma^{-3})\right],\\
		\mathcal{E}={}&\frac{\hbar^2 n^2}{2m}\biggl\{ \frac{8\pi^2}{3(1+\vartheta^2)\gamma}-\frac{16\pi^2}{(1+\vartheta^2) \gamma^2} +\biggl[\frac{64\pi^2}{1+\vartheta^2}+\frac{32\pi^4}{5(1+\vartheta^2)^2} -\frac{128\pi^4}{15(1+\vartheta^2)^3}\biggr]\frac{1}{\gamma^3}\notag\\
		&+\mathcal{O}(\gamma^{-4})\biggr\},
	\end{align}
\end{subequations}
where $\vartheta=2\eta/c$ is kept fixed. Equations (\ref{eq:pEstrong}) give the parametric form of the polaron excitation spectrum at strong interactions, where the parameter $\vartheta$ is a real number. We found that the spectrum (\ref{eq:pEstrong}) can be expressed explicitly in the form
\begin{subequations}\label{eq:displarge}
	\begin{align}\label{eq:ansatz}
		\mathcal{E}(p)=\frac{\hbar^2 n^2}{2m} \sum_{j=1}^{+\infty} C_j(\gamma) \sin^{2j}\left(\frac{p}{2\hbar n}\right).
	\end{align}
	Substituting equations~(\ref{eq:pEstrong}) in expression (\ref{eq:ansatz}) and evaluating it order by order in $1/\gamma$, we find the first three coefficients in the sum. They are
	\begin{align}\label{eq:C1}
		C_1(\gamma)={}&\frac{8\pi^2}{3\gamma}-\frac{16\pi^2}{\gamma^2} +\frac{64\pi^2}{\gamma^3}+\mathcal{O}(\gamma^{-4}),\\ \label{eq:C2}
		C_2(\gamma)={}&-\frac{32\pi^4}{45\gamma^3}+\mathcal{O}(\gamma^{-4}),\\
		C_3(\gamma)={}&-\frac{64\pi^4}{45\gamma^3}+\mathcal{O}(\gamma^{-4}).
	\end{align}
\end{subequations}
We have verified that $C_4,C_5=\mathcal{O}(\gamma^{-5})$. The coefficients $C_j$ therefore decay at least as $\gamma^{-j}$, which makes the series (\ref{eq:ansatz}) rapidly converging. At the maximum, which occurs at $p=\pi\hbar n$, for the polaron energy we thus obtain
\begin{align}
	\mathcal{E}(\pi\hbar n)={}&\frac{4\pi^2\hbar^2 n^2}{3m\;\!\gamma}\biggr[1-\frac{6}{\gamma} +\frac{4(30-\pi^2)}{5\gamma^2}+\mathcal{O}(\gamma^{-3})\biggl].
\end{align}
We notice that accounting for the leading-order term in equation~(\ref{eq:C1}), i.e., at $C_1=8\pi^2/3\gamma$ and thus taking $j=1$, dispersion (\ref{eq:displarge}) reduces to the result obtained in reference~\cite{matveev_spectral_2008}. In figure~\ref{fig-polarondispersion} we compare the exact results for the dispersion with the analytical form (\ref{eq:displarge}). One can observe that even at moderately large $\gamma=20$, the result (\ref{eq:displarge})  taken at the leading order shows significant deviation from the exact one. This occurs due to a relatively large ratio of the subleading and the leading terms in $C_1$. 

Ansatz  (\ref{eq:ansatz}) that remarkably simplifies the parametric dispersion (\ref{eq:pEstrong}) can be understood as a Fourier series of $\mathcal{E}(p)$ on the interval $[0,2\pi\hbar n]$ that satisfies the reflection property (\ref{eq:symeps}) and has even power series starting from $p^2$ around $p=0$. Expanding the form  (\ref{eq:ansatz}) at small $p$, one obtains the dispersion that coincides with equation~(\ref{eq:E(p)polaron}) provided
\begin{gather}\label{eq:C1C2}
	C_1=4\frac{m}{m^*},\quad C_2=\frac{4}{3}\left(\frac{m}{m^*}-\nu\right).
\end{gather}
As a consistency check, we have verified that the substitution of $m/m^*$ given by equation~(\ref{eq:m*gamma>>1}) and $\nu$ calculated in equation~(\ref{eq:nugamma>>1}) into equation~(\ref{eq:C1C2}) leads to the same $C_1$ and $C_2$ as already calculated in equations~(\ref{eq:C1}) and (\ref{eq:C2}).

The ansatz (\ref{eq:ansatz}) is not particularly useful in describing the polaron dispersion in the regime of weak interactions, $\gamma\ll 1$. Comparing the spectrum (\ref{eq:dispersionperturbationtheory}) with equation~(\ref{eq:ansatz}), we obtain
\begin{gather}
	\label{eq:C1C2weak}
	C_1=4+\mathcal{O}(\sqrt\gamma),\quad C_2=-\frac{32}{5\pi\sqrt\gamma}+\mathcal{O}(1),\quad
	C_3=-\frac{128}{7\pi\gamma^{3/2}}+\mathcal{O}(1/\sqrt\gamma).
\end{gather} 
Therefore the series (\ref{eq:ansatz}) will be slowly converging at $p\sim \hbar n$ and $\gamma\ll 1$ and it appears that we need infinitely many terms to accurately describe the dispersion, in striking contrast to very few in the case $\gamma\gg 1$.

A motivated reader can use the procedure presented in this subsection and evaluate further terms of the expansion in $1/\gamma$ in the spectrum (\ref{eq:displarge}). For this one needs the solutions (\ref{eq:rhosignastrong}) and (\ref{eq:sigmasignastrong}) of the Bethe ansatz equations (\ref{eq:LIE}) and (\ref{eq:sigma}) to higher orders, which can be obtained using the systematic method described in section \ref{section:simplesolution}. We notice that the low momentum expansion of the dispersion (\ref{eq:ansatz}) will have a series form with the first two terms given by equation~(\ref{eq:E(p)polaron}). Therefore, the knowledge of the Maclaurin series of the dispersion at strong interactions is sufficient to infer the coefficients $C_j$ of equation~(\ref{eq:ansatz}). Vice versa is trivially correct since the ansatz (\ref{eq:ansatz}) does not have restrictions on $p$.

The obtained results have a direct application. The polaron dispersion denotes the lower spectral edge for zero-temperature correlation functions such as the dynamic structure factor and the spectral function \cite{matveev_spectral_2008,kamenev_dynamics_2009,zvonarev_edge_2009,imambekov_one-dimensional_2012}. The latter are characterised by power-law singularities at the edge. The corresponding exponents are, quite generally, expressed in terms of the sound velocity and the corresponding Luttinger liquid parameter, as well as the derivatives of the dispersion with respect to the momentum and the density  \cite{kamenev_dynamics_2009,imambekov_one-dimensional_2012}. The latter directly follows from the results obtained in this section.

\section{Note on relevant experiments}\label{section:experiments}

First experimental realisations of the Lieb--Liniger model appeared around twenty years ago. Since then, there have been many experimental activities that probed physics of this model. Unlike the Bethe ansatz, the experiments simulate the model only approximately. Nevertheless, the experiments have become one of the driving forces that stimulates current theoretical efforts towards better understanding of integrable systems. In this section we will briefly review some experiments relevant for the topics studied in this review.

\subsection{Initial realisations of the model}

Before the year 2000, it was known how to create Bose--Einstein condensates of atoms in three-dimensional space and the realisation of related degenerate cold Bose gases in one dimension was challenging. Two main experimental techniques used to achieve this after the year 2000 involve optical lattices and atom chips. The former method is based on the dipolar interaction of atoms with the electric field created by a combination of laser beams. Such interaction favors atom confinement along certain directions, leading to a system of parallel tubes filled with atoms. There each tube represents a realisation of a one-dimensional system, provided the effects of the intertube coupling can be controlled, which is the case. Therefore for our purposes it suffices to think in terms of a single tube, which will be our simplification in the following. The technique based on atom-chip setup relies on magnetic dipolar interaction of atoms with external magnetic field. The latter is created by a planar system of current-carrying wires that enable magnetic trapping of atoms to the one-dimensional geometry. Both experimental techniques have their advantages and limitations. For example, in atom-chip setups, a single tube of interacting particles is realised, which is not the case for optical lattices. However, the interactions are typically weak in atom chips and achieved temperatures higher than the ones obtained in optical lattices. On the other hand, in optical lattices the effective one-dimensional interaction strength can be tuned. 

In nature, the atoms are three-dimensional, but in tubes they look one-dimensional. This happens as the atoms are tightly confined radially to zero point motion and free to propagate in the third direction. In this case the effective one-dimensional interaction potential between slow atoms has a contact-form, as in the Lieb--Liniger model \cite{olshanii_atomic_1998}. This picture can be spoiled due to the effect of thermal fluctuations that tend to populate radial energy levels. However, if the temperature is smaller than the radial level spacing, the occupation of excited radial energy levels is small and one practically deals with a one-dimensional system.

First experimental realisations of the Lieb--Liniger model with control over a wide range of interaction strengths that include weak- and strong-coupling regimes were achieved in the optical lattice setups \cite{paredes_tonksgirardeau_2004,kinoshita_observation_2004}. Experimentally determined ground-state energy \cite{kinoshita_observation_2004} was found to be in good agreement with the theoretical result for the Lieb--Liniger model modified to account for the potential of a harmonic trap \cite{dunjko_bosons_2001} present in the experiment.

\subsection{Measurement of local correlation functions}

Local correlation functions were measured in several experiments \cite{tolra_observation_2004,kinoshita_local_2005,haller_three-body_2011}. There the main idea is to relate a local correlation function to a physical process that is responsible for particle losses. Monitoring the number of particles in the system over some time interval, one determines its decay rate, which is proportional to the correlation function in question.

In the setup of reference~\cite{kinoshita_local_2005}, initially prepared system of bosonic atoms confined to one-dimension was illuminated, which caused photoassociation of pairs of atoms into excited two-atom molecules. Such molecules are typically so energetic that escape from the system, which is observed as a loss of atoms. Since the photoassociation process most probably occurs at very small atom separations, much smaller than the mean interparticle distance, the corresponding rate is proportional to $g_2$. By measuring the decay of the atom numbers in time, the authors of reference~\cite{kinoshita_local_2005} measured the local two-body correlation function, which was remarkably close to the theoretical result (\ref{eq:g2}) across a 30-fold range of the coupling $\gamma$.

In reference~\cite{haller_three-body_2011}, the local three-body correlation function $g_3$ was measured. In this setup the particle loss was due to the three-body recombination process. Physically, the collision of three particles leads to the formation of a dimer which typically has sufficient energy to escape from the system together with the remaining particle. The recombination rate is proportional to $g_3$. Experimentally measured $g_3$ was found to be in good agreement with the theoretical result (\ref{eq:g3exact1}) over a wide range of $\gamma$. In a related earlier experiment \cite{tolra_observation_2004}, the correlation function $g_3$ was also measured. The obtained result at a single value of $\gamma$ was consistent with the theory.

\subsection{Experimental probes of the quasimomentum distribution}

The quasimomentum distribution is a central quantity that characterizes the many-body eigenstate (\ref{eq:wave function}). It encodes the information about interactions in the system. Its direct measurement is very complicated. However, it has an important relation with the momentum distribution. When the particles of the integrable system prepared in an external (e.g., a box) potential are allowed to expand along one dimension, the interparticle interactions quickly disappear. The particles become arranged with respect to their momenta, not having possibilities to approach each other. The resulting momentum distribution, which is amenable to experimental probes, at large times approaches the quasimomentum distribution. This idea was used in references~\cite{wilson_observation_2020,dubois_probing_2024} to measure experimentally the quasimomentum distribution. The obtained results were in agreement with theoretical calculations adopted to the specific experimental setups.

The momentum distribution $n(p)$ of a uniform system corresponds to the Fourier transform of its one-body density matrix. At large distances it scales as $\langle \Psi^\dagger(x) \Psi(0)\rangle\propto |x|^{-1/2K}$, leading to $n(p)\propto \left(1/p\right)^{1-1/2K}$ at small momenta. The peak in $n(p)$ is a generic feature of bosonic systems that tend to the Bose--Einstein condensed state with zero momentum. However the effect of interactions leads to a power-law singularity, which is characterised by an interaction-dependent exponent in the one-dimensional case \cite{pitaevskii_bose-einstein_2016}. At finite temperatures, the one-body density matrix decays exponentially beyond the healing length $\xi$. This yields the Lorentzian form of the momentum distribution at small momenta $p\ll \hbar/\xi$. The momentum distribution in the one-dimensional interacting Bose gas was measured in several works \cite{paredes_tonksgirardeau_2004,davis_yang-yang_2012,jacqmin_momentum_2012,fabbri_momentum-resolved_2011,meinert_probing_2015}. Measured data showed agreement with theoretical expectations for the specific experimental setups.

\subsection{Probing the excitation spectrum}

The excitation spectrum of the atomic Lieb--Liniger Bose gas was studied using a two-photon Bragg spectroscopy \cite{fabbri_dynamical_2015,meinert_probing_2015}. In this technique, two lasers that operate at different frequencies are used to illuminate the one-dimensional system. The beams are arranged such that the their wavevector difference is along the system. Its magnitude defines the momentum transfer to the system, while the frequency difference gives the energy transfer. The atoms exposed to the laser beams can absorb a photon from one of the beams and emit it into the other in a stimulated emission. A very important feature of Bragg spectroscopy is the possibility to independently tune the momentum transfer from the energy transfer that is achieved by changing the angle between the two beams. This enables a scan over transferred energies for a given momentum transfer. Within the linear-response theory, the energy absorbed from the lasers is proportional to the dynamic structure factor of the one-dimensional system \cite{pitaevskii_bose-einstein_2016}. By measuring the absorbed energy, the dynamic structure factor is probed experimentally. The latter is defined as the Fourier transform of the time-dependent density-density correlation function, $\langle \Psi^\dagger(x,t) \Psi(x,t) \Psi^\dagger(0,0)\Psi(0,0)\rangle$. At zero temperature, it is characterised by the power-law singularity when the frequency and the wavevector correspond to the elementary Lieb-I excitation \cite{khodas_dynamics_2007}. The experimental results showed consistency with this picture. Accounting for the realistic effects present in the experiments  such as finite temperature, presence of many parallel tubes, the trapping potential, etc., in the Bethe-ansatz calculations, theoretical predictions were in agreement with the measured data \cite{fabbri_dynamical_2015,meinert_probing_2015}.

\subsection{Experimental studies of the thermodynamics}

Predictions arising from the Yang--Yang thermodynamics were also tested in various experiments. In reference~\cite{van_amerongen_yang-yang_2008}, an atom-chip setup was used to measure the density profile of a weakly-repulsive Bose gas. Since the temperature was on the order of the level spacing of the transverse modes, their population was significant. Accounting properly for this effect, measured density profile was in agreement with the one theoretically computed from the Yang--Yang formalism after taking into account the trapping potential. In another experiment performed on atom-chip setup \cite{jacqmin_sub-poissonian_2011}, the statistics of atom number fluctuations was measured, which was in agreement with the prediction based on the Yang--Yang theory. In reference~\cite{vogler_thermodynamics_2013}, an optical lattice was used to create a system of parallel tubes filled with interacting Bose gases. Several interaction strengths were considered, covering both regimes of weak and strong interactions. Measured density distribution over a bundle of tubes contained integrated data. Using an inverse Abel transform, the density profile of a single tube was obtained. Fitted data from the Yang--Yang prediction showed remarkable agreement with the experimental data. 

\subsection{The Luttinger liquid physics with one-dimensional Bose gases}

The Luttinger liquid is a paradigmatic model of one-dimensional interacting quantum systems that includes fermions, spins, and bosons. The model describes the low-energy excitations that are the density waves propagating at a constant velocity. In addition to the sound velocity, the Luttinger liquid parameter is another quantity that fully describes the low-energy properties of the system. Direct observation of the Luttinger liquid physics with Bose gases was achieved in the experiment \cite{yang_quantum_2017}. There, a density dip was created in the centre of the system. Observing the waves triggered by the density perturbation, the sound velocity was measured. The Luttinger liquid parameter was then inferred accounting for their mutual relation due to Galilean invariance. In the same experiment the momentum distribution and the density profile were also measured. From the latter the pressure and the entropy were found, showing remarkable agreement with the Yang--Yang theory.

\section{Some open problems}

This section contains discussions on various subjects relevant for the topics studied previously and a list of some open problems.

\subsection{Overview of various exact relations}

This review illustrates that the existence of exact solution (\ref{eq:wave function}) of the model (\ref{eq:Horiginal}) does not imply direct and easy access to the analytical results of particular physical quantities. We have studied various quantities and their interrelations that reduce the amount of independent information one needs to know in order to access physically relevant information. Here we will give a (nonexhaustive) list of such relations.

In reference~\cite{lieb_exact_1963b}, Lieb evoked the thermodynamic relation (\ref{eq:velocitypressure}) between the sound velocity $v$ and the pressure that in terms of the ground-state energy $E_0=N\epsilon\;\! e_2(\gamma)$ can be expressed as equation~(\ref{eq:soundvelocityexplicit}). Due to Galilean invariance, the Luttinger liquid parameter $K$ and the sound velocity are related as in equation~(\ref{eq:mvK}) \cite{haldane_effective_1981}. On the other hand, $K$ can be also found from the value of the density of quasimomenta at the Fermi quasimomentum, $\rho(Q,Q)$, see equation~(\ref{eq:Kll}) \cite{korepin}. Another exact relation is about the mass of elementary excitations \cite{matveev_effective_2016}, which can be expressed in terms of $K$ as in equation~(\ref{eq:m*vv}).

The above-mentioned are the most well-known relations. Let us list some that involve the moments of the quasimomentum distribution defined by equation~(\ref{eq:e2l}). From equation~(\ref{eq:dndQ}) it follows that the Fermi quasimomentum $Q$ is connected to the ground-state function $e_2(\gamma)$ via the differential relation (\ref{eq:gdiff}). The local correlation functions, see for example equations~(\ref{eq:g2}), (\ref{eq:g3exact1}), and (\ref{eq:g4exact}), are also determined by the moments of the quasimomentum distribution. Finally the consecutive moments of the quasimomentum distribution satisfy the differential relation (\ref{eq:diff-diffeq}) or (\ref{eq:e2l1}). The latter can be understood as follows: the knowledge of a high conserved charge in the ground state as a function of the interaction parameter, $e_{2l}(\gamma)$, suffices to find all lower conserved charges, $e_{2l-2}(\gamma)$, $e_{2l-4}(\gamma)$, \ldots, $e_2(\gamma)$, by simple differentiation. We eventually mention that the short-distance expansion of the one-body density matrix in the Lieb--Liniger model can be expressed in terms of the moments \cite{olshanii_short-distance_2003,olshanii_connection_2017}. The moments (\ref{eq:e2l}) also a determine the dispersion relation of a polaron in the Yang--Gaudin Bose gas, see equations~(\ref{eq:E(p)polaron}), (\ref{eq:m*}), and (\ref{eq:nu}).

Let us consider the derivatives of the quasimomentum distribution at the Fermi quasimomentum, $\rho^{(j,0)}(Q,Q)$, where $j$ is a positive integer. They also participate in some exact relations. The derivatives themselves obey the differential relations (\ref{eq:Fho10}), (\ref{eq:Fho20}), (\ref{eq:Fho30}), etc. Interestingly, one of the latter relations can be integrated leading to the equality (\ref{eq:Fho20solution}). The derivatives $\rho^{(j,0)}(Q,Q)$ determine physical quantities. For example, they enter the spectrum of elementary excitations, see equations~(\ref{eq:spectrumlowp}), (\ref{eq:chivv}), and (\ref{eq:nuvv}). They also determine the low-temperature thermodynamics, see equations~(\ref{eq:f}) and (\ref{eq:C4general}). Having in mind that the quasimomenta in the ground state occupy the states between $-Q$ and $Q$, similarly as the fermions fill the Fermi sea, it is not surprising that the low-energy excitations and the low-temperature thermodynamics are determined by the local properties of the density of quasimomenta near the Fermi quasimomentum.

\subsection{From Lieb--Liniger to other Galilean-invariant models}

Throughout this review we have been focused on the Lieb--Liniger model. However it has many things in common with other integrable models that posses Galilean invariance. The central equation (\ref{eq:LIE}) that describes the density of quasimomenta in the ground state has the same form for other such one-component models, provided model-dependent two-particle scattering phase shift $\theta(k)$ is used. From equation~(\ref{eq:LIE}) we have derived the corresponding partial differential equation  (\ref{eq:PDE}) under a minimal assumption that the phase shift is repeatedly differentiable function, see Appendix \ref{appendixc}. In the case of nonsingular phase shifts, the density of quasimomenta is an analytic function and equation~(\ref{eq:PDE}) remains valid. Therefore, equation~(\ref{eq:PDE}) and some relations that follow from it apply beyond the Lieb--Liniger model. Some well-known examples are the hyperbolic Calogero--Sutherland \cite{sutherland}, and the Yang--Gaudin models \cite{gaudin}. The latter class of models are not single-component, but in the special case of attractive spin-$\frac{1}{2}$ fermions, the ground state consists of paired fermions and the resulting equations for the ground state are similar to that of the Lieb--Liniger case.

The hyperbolic Calogero--Sutherland model is a notable example of Galilean-invariant integrable models with noncontact interactions and a nonsingular phase shift  \cite{sutherland}. It is characterised by the interaction potential
\begin{align}\label{eq:potentialCS}
	V(x)=\frac{\hbar^2}{m} \frac{\lambda(\lambda-1)\varkappa^2}
	{\sinh^2(\varkappa x)}
\end{align}
and the two-particle scattering shift
\begin{align}\label{eq:thetaplus} 
	\theta(k)&=i\ln\left(\frac{\Gamma(1+ik/2\varkappa) \Gamma(\lambda-ik/2\varkappa)}{ \Gamma(1-ik/2\varkappa) 
		\Gamma(\lambda+ik/2\varkappa)}\right).
\end{align}
This complicated function has a simple limiting case. At $\varkappa\to+\infty$, $\lambda\to 0^+$, such that $c=2\varkappa\lambda$ is kept fixed, the phase shift (\ref{eq:thetaplus}) coincides with the one of the Lieb--Liniger model that is given by equation~(\ref{eq:phaseshift}). Therefore, the hyperbolic Calogero--Sutherland model can be understood as an integrable deformation of the Lieb--Liniger one, where the range of interaction is smoothly increased from zero towards a finite value $1/\varkappa$. 

The theoretical developments of the hyperbolic Calogero--Sutherland model are not as advanced as that of the Lieb--Liniger one. For example, we have not found the expression for the ground-state energy in terms of the microscopic parameters of that model. A similar statement also applies to all quantities that have connecting relations with the ground-state energy. This is an open problem that deserves to be studied. Conceptually, it is fundamental to understand the effects of finite interaction range on physical quantities. Moreover, in practical realisations the $\delta$-function interaction is also an idealisation and its broadening might result in new physical insights.

The relation between the two integrable models gives some hints how this study can be done, at least in some limiting cases. Expanding the phase shift (\ref{eq:thetaplus}) around the Lieb--Liniger limit, we will account for the deviations between the two models controlled by the small parameter $n/\varkappa$. The Bethe ansatz procedure can then be performed in a perturbative way. The expanded phase shift contains necessary information to find analytically the ground state and low-momentum excitation spectrum of the hyperbolic Calogero--Sutherland model in this particular limit of small interaction range. In the same regime we can study the thermodynamics. At low temperatures, it reduces to the evaluation of the partial derivatives $\rho^{(j,0)}(Q,Q)$ at $n/\varkappa\ll 1$, since the derived expressions (\ref{eq:f}) and (\ref{eq:C4general}) already apply to the hyperbolic Calogero--Sutherland model. 

Another example of Galilean-invariant models is spin-$\frac{1}{2}$ Fermi gas with contact repulsion, known as the Yang--Gaudin model \cite{gaudin}. In the thermodynamic limit, the ground state has the total spin zero and the Bethe-ansatz integral equation that describes the ground state mathematically has the form of equations~(\ref{eq:LIE}) and (\ref{eq:kernel}) with opposite sign of $c$. This model has been studied extensively \cite{guan_fermi_2013}. However there are some unsolved problems. For example, the problem of capacitance of two coaxial thin plates of the same radius held at the same potential reduces to the integral equation of the Yang--Gaudin model \cite{love_electrostatic_1949,gaudin}. Full solution for the capacitance is an open problem. It can be addressed in details as it was done in section \ref{section:capacitance} for coaxial plates held at opposite potentials. Another possibility is to study the local properties of the quasimomentum distribution near the Fermi quasimomentum. This can be done using the approach similar to that of section \ref{section:derivativesrho}. The thermodynamics of the model is, however, more complicated as the quasimomenta become complex and the resulting equations of the thermodynamic Bethe ansatz are more involved than the ones of the Lieb--Liniger model \cite{takahashi}. The analytical treatment of the low-temperature thermodynamics beyond the trivial limits is an open problem.

\subsection{The thermodynamics in other regimes}

The thermodynamics of the Lieb--Liniger model is determined formally exactly by the thermodynamic Bethe ansatz \cite{yang_thermodynamics_1969}. Nevertheless, extracting simple analytical results from it is a nontrivial task. An approach to achieve this at lowest temperatures has been developed in section \ref{section:thermodynamics}. It enabled us to obtain the free energy and other thermodynamic quantities beyond the leading order in temperature, where the latter was known from conformal field theory. The approach is based on the power-series expansion at low temperatures of the exact equation (\ref{eq:YY}) for the pseudoenergy. In this case the contribution to the integral in equation~(\ref{eq:YY}) from the region with positive pseudoenergy is exponentially small. The expansion can thus be performed around the point where the pseudoenergy nullifies, which in the zero-temperature limit coincides with the Fermi quasimomentum. In this way the thermodynamic quantities expressed in the form of power series of temperature are obtained. The coefficients of the resulting series can be expressed in terms of various partial derivatives of the quasimomentum distribution at the Fermi quasimomentum. The latter objects satisfy various relations, systematically studied in section \ref{section:derivativesrho}. They have enabled us to obtain analytical expressions for the thermodynamic quantities in terms of the interaction parameter. The obtained results apply at lowest temperatures, $\tau=T/\epsilon \ll\gamma$. An open problem is the extension of the results of section \ref{section:thermodynamics} to account for other temperature regimes directly from the thermodynamic Bethe ansatz. For example, there is a particular low-temperature regime, $\tau\ll 1$, at very weak repulsion, $\gamma\ll\tau\ll\sqrt\gamma\ll1$, not described by the developed approach. Another problem is about the derivation of the free energy (\ref{eq:f}) from the quasiparticle picture. In order to obtain the quadratic correction in temperature, it is sufficient to study the low-temperature statistical mechanics of phonons with linear spectrum characterised by the sound velocity. For the quartic temperature correction, the spectrum curvature will matter.

\subsection{Local correlation functions at finite temperatures}

There are at least two different approaches that have lead to the exact results for $N$-body local correlation functions for the Lieb--Liniger model in the thermal equilibrium. One approach is based on the emptiness formation probability in a XXZ chain and its scaling limit that yields the desired correlation function in the Lieb--Liniger case \cite{pozsgay_local_2011}. Another approach uses the conjectured exact result for a vertex operator in  the sinh-Gordon field theory, which after taking the nonrelativistic limit leads to the wanted result in the Lieb--Liniger model \cite{bastianello_sinh-gordon_2018,bastianello_exact_2018}. Formal results of the two approaches have algebraically different structure, but they are a priori equivalent. It is an open question how to prove the equivalence between the two results.

Let us look more closely to the result of reference~\cite{pozsgay_local_2011} for the $N$-body local correlation function defined in equation~(\ref{eq:gj}). More general form of equation~(\ref{eq:gNexact}) that applies in the thermal equilibrium state is given by
\begin{align}\label{eq:gNexactT}
	g_N=\frac{(N!)^2}{(2\pi n)^N}\int_{-\infty}^{\infty} dq_1\ldots dq_N \prod_{1\le l<j\le N} \frac{q_j-q_l}{(q_j-q_l)^2+c^2} \prod_{j=1}^{N}(j-1)!f(q_j)\bar\rho_{j-1}(q_j).
\end{align}
Here the Fermi weight
\begin{align}\label{eq:weight}
	f(q)=\frac{1}{1+e^{\mathcal{\bar E}(q)/T}}
\end{align}
is determined by the solution (\ref{eq:YYoriginal}) of the Yang--Yang equation. In equation~(\ref{eq:gNexactT}), the function $\bar\rho_{j}$ is the solution of the linear integral equation
\begin{align}\label{eq:barrhoj}
	\bar\rho_j(k)+\frac{1}{2\pi}\int_{-\infty}^{\infty}\theta'(k-q)f(q)\bar\rho_j(q) =\frac{k^j}{j!}.
\end{align}
In the limit of zero temperature, $T\to 0$, the Fermi weight (\ref{eq:weight}) becomes one for $-Q<q<Q$ and zero otherwise. Therefore, $\bar\rho_j$ reduces to $\rho_j$, see equation~(\ref{eq:Feq}), and equation~(\ref{eq:gNexactT}) reduces to equation~(\ref{eq:gNexact}). It would be important to understand the deviations of the $N$-body local correlation function (\ref{eq:gNexact}) due to the effect of thermal fluctuations. This may be possible to do analytically at low temperatures by extending the approach of section \ref{section:method} to treat equations of the form (\ref{eq:barrhoj}), eventually evaluating equation~(\ref{eq:gNexactT}). Apart from somewhat trivial case $N=2$, this question for arbitrary interaction strength is addressed numerically \cite{kormos_one-dimensional_2010,pozsgay_local_2011,kormos_exact_2011,bastianello_exact_2018,bastianello_sinh-gordon_2018}. We, however, note the existence of analytical results in the limiting cases of interactions \cite{gangardt_local_2003,nandani_higher-order_2016}.

\subsection{About the family of dimensionless parameters $K_j$}

In section \ref{section:derivativesrho} we have introduced the parameters $K_j$, see equation~(\ref{eq:Kjdefinition}). They represent partial derivatives of the quasimomentum distribution at the Fermi level. We have shown that $K_j$ parameters are not independent but hierarchically ordered, satisfying certain differential equations. We have moreover given a recipe how they can be explicitly solved in terms of the interaction parameter. The parameters are physically important as they enter the final results for the low-temperature thermodynamics, see equation~(\ref{eq:C4}). The same parameters also determine the low-momentum spectrum of elementary excitations, see equations~(\ref{eq:chiKj}) and (\ref{eq:nuKj}). It is expected that they will also enter the temperature corrections of the local correlation functions (\ref{eq:gNexactT}), as can be seen from the following. Accounting for low temperatures, after expanding the expression (\ref{eq:barrhoj}) similarly as it was done in section \ref{section:thermodynamics}, the derivatives of the pseudoenergy at the Fermi quasimomentum will appear. In the thermal equilibrium, these derivatives can be expressed in terms of the derivatives of the quasimomentum distribution, i.e., in terms of $K_j$.

The parameters $K_j$ will have another application. They will appear in the evaluated form of the dynamical correlation function at low temperatures. In the asymptotic regime of large distances $x$ and times $t$ such that the ratio $x/t$ is kept fixed, the latter correlation function is given by \cite{korepin_time_1997,esler_temperature_1998}
\begin{align}\label{eq:dcf}
	\langle \Psi(0,0)\Psi^\dagger (x,t)\rangle\simeq\exp\left(\frac{1}{2\pi}\int_{-\infty}^\infty \frac{dq}{2\pi\rho_t(q)}\left|x-\frac{\mathcal{\bar{E}}'(q)}{2\pi\rho_t(q)}t\right| \ln\left|\frac{1-e^{\mathcal{\bar{E}}(q)/T}}{1+e^{\mathcal{\bar{E}}(q)/T}}\right| 
	\right).
\end{align}
Here $\rho_t(q)$ is the density of vacancies and $\mathcal{\bar{E}}(q)$ is the pseudoenergy. They satisfy equations~(\ref{eq:LIETBA}) and (\ref{eq:YYoriginal}), respectively. Evaluation of equation~(\ref{eq:dcf}) is another example where the techniques developed in this review can be applied.

\subsection{Bose gas density in the box potential}

In the case of zero boundary conditions, the Lieb--Liniger model is solvable by the Bethe ansatz \cite{gaudin_boundary_1971}. Then the wave function nullifies at the two ends. A physically observable effect is the local particle density $n(x)$. It is position dependent and nullifies at the two ends. An interesting open problem is the calculation of the exact form of the density profile $n(x)$. Physical insights for this basic question can be obtained from the following considerations.

In the regime of weak interactions, $\gamma\ll 1$, the system can be described by the Gross--Pitaevskii equation (\ref{eq:GPE}) with the boundary conditions that impose the nullification of the solution at $x=0$ and $x=L$. Such equation is exactly solvable in terms of the Jacobi elliptic functions. Rather than studying that, for our purpose it is sufficient to consider a simplified semi-infinite system, thinking that the infinity of the simplified system corresponds to $L/2$ of the original one. Then the solution is a simple hyperbolic function that interpolates between zero and one at the two ends, see equation~(\ref{eq:tanh}). The latter solution shows that the density of the system reaches the mean density at the distances beyond the healing length $\xi=1/n\sqrt\gamma$.\footnote{\linespread{1}\selectfont This is expected to be the case for a superfluid matter.} This picture that originates from the mean-field treatment is actually not entirely correct as the difference $n(x)-n$ has a long-ranged tail that scales as $x^{-2}$ at $x\gg\xi$ \cite{petkovic_density_2023}. It arises from the effect of quantum fluctuations. 

In the regime of infinite repulsion, $\gamma\to\infty$, an exact result for the  boson density can be obtained using the correspondence with the density of noninteracting fermions in an infinite flat box potential of the length $L$. The final expression for the density profile of $N$ bosons is given by \cite{lacroix-a-chez-toine_non-interacting_2018}
\begin{align}\label{eq:friedel}
	n(x)=\frac{N}{L}+\frac{1}{2L}-\frac{\sin\left((2N+1)\pi x/L\right)}{2L\sin(\pi x/L)}.
\end{align}
One can check that the density satisfies the reflection property $n(x)=n(L-x)$ as it must be the case. In the thermodynamic limit, equation~(\ref{eq:friedel}) becomes $n(x)=n-\sin(2\pi nx)/2\pi x$. Therefore near the edges, the local density approaches the mean density at a length scale of the mean interparticle distance. The local density then shows Friedel oscillations characterised by the same length scale. The oscillation amplitude decays as $1/x$. The density profile of a strongly-interacting Bose gas is thus markedly different from the same profile of a weakly-interacting gas. It would be interesting to understand this picture using the exact solution of the model.

\subsection{Two-component models and nested Bethe ansatz}

With an exception of section \ref{section:Yang-Gaudin gas}, in this review we studied the one-component model of interacting bosons. The corresponding Lieb--Liniger model is described by the Hamiltonian (\ref{eq:Horiginal}) with the assumption of the bosonic symmetry of its wave functions with respect to the permutation of the coordinates of the particles. The same Hamiltonian is also integrable for other symmetries of the wave functions \cite{yang_exact_1967,sutherland_further_1968}. Consider, for example, a system with $N$ quantum particles that can be either bosons or spinless fermions, out of which there are $M$ bosons. In the thermodynamic limit where $N$, $M$, and the system size $L$ tend to infinity proportionally, the system is described by the exact system of equations \cite{lai_ground-state_1971,imambekov_exactly_2006}
\begin{subequations}\label{eq:nested}
	\begin{gather}
		\rho(k)+\frac{1}{\pi}\int_{-B}^{B}dq\;\! \theta'(2k-2q)\sigma(q)=\frac{1}{2\pi},\\
		\sigma(k)+\frac{1}{\pi}\int_{-Q}^{Q}dq\;\! \theta'(2k-2q)\rho(q)=0,
	\end{gather}
\end{subequations}
where $B,Q>0$ and the kernel is given by equation~(\ref{eq:kernel}). The densities and the ground-state energy are then given by
\begin{gather}
	\frac{N}{L}=\int_{-Q}^{Q}dk\rho(k),\quad \frac{M}{L}=\int_{-B}^{B}dk\sigma(k),\\
	E=\frac{\hbar^2 L}{2m} \int_{-Q}^{Q}dkk^2\rho(k).
\end{gather}
Analysis of the two coupled integral equation for the mixture is complicated in the regime of weak interactions, $0<c\ll N/L$. Analytically it is known the ground-state energy at the leading order in $c$ \cite{batchelor_exact_2005},
\begin{align}\label{eq:oelkers}
	\frac{E}{L}=\frac{\hbar^2}{2m}\left(\frac{\pi^2}{3}n_f^3 + c n_b^2+2c n_b n_f \right),
\end{align}
where the $n_b=M/L$ and $n_f=(N-M)/L$. The energy (\ref{eq:oelkers}) consists of three terms. The first denotes the energy of free Fermi gas of the density $n_f$ without bosons. The second term describes the energy weakly-interacting bosons of the density $n_b$ at the leading order in the interaction strength, which corresponds to the approximation $e_2(\gamma)=\gamma$ in the Lieb--Liniger model. The third term describes the interaction between the bosons and fermions and can be understood as a mean-field interaction.  

Equations (\ref{eq:nested}) provide an example of the nested Bethe ansatz integral equations. There the unknown functions are coupled, which is an additional difficulty for the analysis of the already complicated regime of weak interactions. The system (\ref{eq:nested}) can be decoupled, but then the kernel becomes more complicated. On the other hand, the numerical experiment of section \ref{section:numericalexperiment} is well-suited to study the system (\ref{eq:nested}) to an unprecedented accuracy and eventually infer the analytical results at weak interactions. Similarly as in the Lieb--Liniger case, equations~(\ref{eq:nested}) can be systematically analysed in regime of strong interactions. Such study can be motivated by a physically relevant question whether the fermions will experience an effective mutual interaction mediated by the coupling with the subsystem of bosons. Its answer is not contained in the mean-field energy (\ref{eq:oelkers}) as there is no term proportional to $n_f^2$ within that accuracy. Moreover, it would be interesting to extend the formalism of section \ref{section:method}  for the nested systems, equations~(\ref{eq:nested}) being one example.

\subsection{Other problems}

There are, of course, many other problems that require particular attention. Without entering into details, here we mention two of them. 

The first problem would be the use of the Cheon--Shigehara model that is dual to the Lieb--Liniger one \cite{cheon_fermion-boson_1999,khodas_dynamics_2007,imambekov_one-dimensional_2012,granet_duality_2022} in order to calculate the quantities that might be more accessible in one of the models than in the other. A recent study \cite{granet_duality_2022} has made the first steps in this direction. Here we should have in mind that the finite radius of convergence of $1/\gamma$ series at strong interactions of, e.g., the ground-state energy studied in subsection \ref{section:radius} as well as for other quantities should guarantee the convergence of various series expansions for the Cheon--Shigehara model at weak interactions. 

The second problem would be understanding of the structure of the short-range expansion of the one-body density matrix. The first three terms of the expansion were expressed in terms of $e_2(\gamma)$ and $e_2'(\gamma)$ some time ago \cite{olshanii_short-distance_2003}. The fourth-order term depends in addition to $e_4(\gamma)$ and $e_4'(\gamma)$, and can be connected to the local three-body correlation function (\ref{eq:g3exact1}) \cite{olshanii_connection_2017}. Beyond that it is not known. However the obtained structure undoubtedly suggests that higher moments of the quasimomentum distribution will enter at higher orders, which on the other hand, determine higher-order local correlation functions. It is thus expected to have connections between the local correlation functions and the short-distance expansion of the one-body density matrix.

\ack{I would like to acknowledge the discussions and collaborations with G.~Astrakharchik, K.~Matveev, M.~Panfil, A.~Petkovi\'{c}, and B.~Reichert that are encoded in certain parts of the review. In addition, I am grateful to J.-S.~Caux, F.~Essler, and S.~Majumdar for the careful reading of the initial version of this review and helpful suggestions. }

%\funding{\justifying This project has in part received financial support from the CNRS through the MITI interdisciplinary programs through its exploratory research program.}

\appendix

\section{Properties of the integral operator \label{appendixA}}

The integral equation (\ref{eq:Feq}) can be considered as a special case of the equation
\begin{align}\label{eq:IEgeneral}
	\left(\mathcal{I}+\frac{1}{\lambda } \mathcal{K}\right)\rho=f
\end{align}
where $\lambda=1$. In equation~(\ref{eq:IEgeneral}) we have suppressed the variables in the arguments of the functions, introduced the parameter $\lambda$ and the operators of the identity $\mathcal{I}$ as well as the nontrivial part of the integral operator $\mathcal{K}$. The existence of the unique and nontrivial solution $\rho$ crucially depends on the spectral properties of the operator $\mathcal{I}+ \mathcal{K}/\lambda$.

For the special choice of the kernel $\theta'(k)$ given by equation~(\ref{eq:kernel}), equation~(\ref{eq:IEgeneral}) in the homogeneous case $f=0$ reduces to the eigenvalue problem
\begin{align}\label{eq:ev}
	\frac{c}{\pi} \int_{-Q}^{Q} dq \frac{\rho(q,Q)}{c^2+(k-q)^2}=	\lambda \rho(k,Q).
\end{align}
In the limit $c\to 0^+$, under the integral we have a representation of the Dirac $\delta$-function. Therefore, $\lambda=1$ is an eigenvalue at $c\to 0^+$. In the opposite regime $c\gg Q$, the left-hand side of equation~(\ref{eq:ev}) is proportional to $1/c$ for normalizable eigenfunctions that we impose. One thus expects $\lambda\sim 1/c$ and the spectrum that satisfies
\begin{align}\label{eq:spectrum}
	0<\lambda<1.
\end{align}		
Careful treatment of the eigenvalue problem  (\ref{eq:ev}) shows that the spectrum is nondegenerate and obeys $0<\lambda\le 1-2\arctan(c/Q)/\pi$ \cite{baratchart_solution_2019}. Therefore, we can conclude that at finite positive $c$ and at $Q>0$, the spectrum of the eigenvalue problem (\ref{eq:ev}) satisfies the condition (\ref{eq:spectrum}).

This consideration shows that for the special value $\lambda=1$, which is of our interest in the paper, equation~(\ref{eq:ev}) has only a trivial solution $\rho=0$. The Fredholm alternative theorem \cite{book-integralequations} then guarantees that equation~(\ref{eq:IEgeneral}) has a unique solution that can be formally expressed as
\begin{align}\label{eq:solrho}
	\rho=\left(\mathcal{I}+\mathcal{K}\right)^{-1} f.
\end{align}	
Here the inverse of the operator is defined by the infinite power series, which is convergent due to the condition (\ref{eq:spectrum}). However, the convergence is very slow at small $c/Q$ \cite{love_electrostatic_1949}, which makes the analytical treatment of the Lieb--Liniger model at weak interactions generally troublesome. For smooth $f$ as in equation~(\ref{eq:Feq}), the solution of the integral equation will be a differentiable function. From equation~(\ref{eq:solrho}), this can be understood as an infinite sum where each term is differentiable.

\section{Property of the pair of integral equations}\label{appendixb}

Consider a pair of integral equations
\begin{gather}\label{b1}
	\mathcal{F}[\sigma(k,Q)]=g(k),\\
	\label{b2}
	\mathcal{F}[\tau(k,Q)]=h(k),
\end{gather}
where the integral operator $\mathcal{F}$ is defined by equation~(\ref{eq:F}) and $g(k)$ and $h(k)$ are arbitrary functions that satisfy minimal requirements (i) there are unique solutions $\sigma(k,Q)$ and $\tau(k,Q)$ and (ii) the solutions satisfy 
\begin{align}\label{b3}
	\int_{-Q}^{Q}dk \int_{-Q}^{Q}dq \sigma(k,Q)\theta'(k-q)\tau(k,Q)=	\int_{-Q}^{Q}dq \int_{-Q}^{Q}dk \sigma(k,Q)\theta'(k-q)\tau(k,Q),
\end{align}
with $\theta'(k)=\theta'(-k)$. For example, for $g(k)$ and $h(k)$ in the form of polynomials, the assumptions will be satisfied. Then we have the relation
\begin{align}\label{b4}
	\int_{-Q}^{Q}dk \sigma(k,Q)h(k)=\int_{-Q}^{Q}dk \tau(k,Q)g(k).	
\end{align}
Equation (\ref{b4}) can be directly showed by multiplying equations~(\ref{b1}) and (\ref{b2}), respectively, by $\tau(k,Q)$ and  $\sigma(k,Q)$. After performing the integration over $k$ in the interval $-Q<k<Q$, and using the assumption (\ref{b3}) one obtains identical left-hand sides of the two equations. The right-hand sides then give the property (\ref{b4}). Equation (\ref{eq:AjlAlj}) directly follows from the property (\ref{b4}) for the choice $g(k)=k^j/j!$, and $h(k)=k^l/l!$, and thus $\sigma(k,Q)=\rho_j(k,Q)$, $\tau(k,Q)=\rho_l(k,Q)$, see equation~(\ref{eq:Feq}).

\section{The derivation of the partial differential equation \label{appendixc}}

In order to derive equation~(\ref{eq:PDE}), we start from the Lieb integral equation (\ref{eq:LIE}) expressed as $\mathcal{F}[\rho(k,Q)]=1/2\pi$, where the linear integral operator $\mathcal{F}[\rho(k,Q)]$ defined by equation~(\ref{eq:F}). Performing the differentiation of $\mathcal{F}$, we directly obtain 
\begin{subequations}
	\label{eq19}	
	\begin{align}
		\frac{\partial}{\partial Q}\mathcal{F}[\rho(k,Q)] ={}& \mathcal{F}\left[ \frac{\partial\rho(k,Q)}{\partial Q} \right] +\frac{\rho(Q,Q)}{2\pi}[\theta'(k-Q)+\theta'(k+Q)],\\
		\frac{\partial^2}{\partial Q^2}\mathcal{F}[\rho(k,Q)] ={}& \mathcal{F}\left[ \frac{\partial^2\rho(k,Q)}{\partial Q^2} \right] +\frac{\rho^{(0,1)}(Q,Q)}{2\pi}[\theta'(k-Q)+\theta'(k+Q)]\notag\\
		&+\frac{d}{dQ}\left\{\frac{\rho(Q,Q)}{2\pi}[\theta'(k-Q)+\theta'(k+Q)]\right\}.
	\end{align}
\end{subequations}
Here we used the notation $\rho^{(0,1)}(Q,Q)=\partial\rho(k,Q)/\partial Q|_{k=Q}$. Similarly, differentiating $\mathcal{F}$ with respect to $k$, after the partial integrations we obtain
\begin{align}\label{eq20}
	\frac{\partial^2}{\partial k^2}\mathcal{F}[\rho(k,Q)] ={}& \mathcal{F}\left[ \frac{\partial^2\rho(k,Q)}{\partial k^2} \right] -\frac{\rho^{(1,0)}(Q,Q)}{2\pi}[\theta'(k-Q)+\theta'(k+Q)]\notag\\
	&-\frac{\rho(Q,Q)}{2\pi}\frac{d}{dk}[\theta'(k-Q)-\theta'(k+Q)].
\end{align}
Here $\rho^{(1,0)}(Q,Q)=\partial\rho(k,Q)/\partial k|_{k=Q}$ and used the property $\rho^{(1,0)}(Q,Q)=-\rho^{(1,0)}(-Q,Q)$ that follows from equation~(\ref{eq:LIE}). The left-hand sides in equations~(\ref{eq19}) and (\ref{eq20}) are zero since we differentiated $\mathcal{F}[\rho(k,Q)]=1/2\pi$. The linear combination of the right-hand sides then yields
\begin{align}
	\label{eq21}
	\mathcal{F}\left[\frac{\partial^2\rho(k,Q)}{\partial Q^2} - \frac{2}{\rho(Q,Q)}\frac{d\rho(Q,Q)}{dQ} \frac{\partial\rho(k,Q)}{\partial Q}-\frac{\partial^2\rho(k,Q)}{\partial k^2}\right]=0,
\end{align}
since $\mathcal{F}$ is a linear operator. We have used the total derivative $\frac{d\rho(Q,Q)}{dQ}=\rho^{(1,0)}(Q,Q)+\rho^{(0,1)}(Q,Q)$ and the parity of $\theta'(k)$. Since equation~(\ref{eq:LIE}) has a unique solution, see the discussion in Appendix \ref{appendixA}, the Fredholm alternative theorem guarantees that equation~(\ref{eq21}) only has a trivial solution. This is equivalent to equation~(\ref{eq:PDE}), which therefore must be satisfied.

\section{The moment-generating function}

The partial differential equation (\ref{eq:PDE}) can be used to derive an expression for the moment-generating function of the quasimomentum distribution. The latter is defined as
\begin{align}\label{eq:falpha}
	f_\alpha(Q)=\int_{-Q}^{Q} dk \rho(k,Q) \cosh(\alpha k),
\end{align}
where $\alpha$ is a real parameter. The moments of $\rho(k,Q)$, cf.~equation~(\ref{eq:e2l}), can be obtained by differentiating $f_\alpha(Q)$ with respect to $\alpha$ and then taking the limit $\alpha\to 0$. On the other hand, $f_\alpha(Q)$ satisfies the differential equation 
\begin{align}\label{eq:mainfalpha}
	\left(\frac{\partial^2}{\partial Q^2}-\frac{2}{\rho(Q,Q)}\frac{d\rho(Q,Q)}{dQ}\frac{\partial}{\partial Q}\right) f_\alpha(Q)=\alpha^2 f_\alpha(Q),
\end{align}
as it can be shown directly by applying the derivatives to the definition (\ref{eq:falpha}) after making use of equation~(\ref{eq:PDE}). Equation (\ref{eq:mainfalpha}) is an exact result.

Let us discuss the consequences of equation~(\ref {eq:mainfalpha}). Consider the case $\alpha =0$ in equation~(\ref {eq:mainfalpha}). It leads to the relation
\begin{align}\label{eq:dndQ}
	\frac{dn}{dQ}=4\pi\rho(Q,Q)^2,
\end{align}
where the integration constant $4\pi$ was set using the free Fermi gas case. Equation (\ref{eq:dndQ}) is the special case of equation~(\ref{eq:dAdQ}). Another useful relation follows from $n^{2l+1}e_{2l}=(\partial^{2l} f_\alpha(Q)/\partial\alpha^{2l})|_{\alpha=0}$, where $l\ge 1$ is an integer. It leads to
\begin{align}\label{eq:main1234}
	\frac{\partial^2}{\partial n^2}(n^{2l+1}e_{2l})=\frac{l(2l-1)}{8\pi^2\rho(Q,Q)^4} n^{2l-1} e_{2l-2}.	
\end{align}
Upon expressing the derivative according to equation~(\ref{eq:derivatives}), equation~(\ref{eq:main1234}) reduces to equation~(\ref{eq:diff-diffeq}) previously obtained in section \ref{section:method}.

\section{Sommerfeld-like expansion} \label{appendix:sommerfeld}

Consider a class integrals of the form
\begin{align}\label{eq:integralJf}
	J[f]=\int_{-\infty}^{+\infty} dq\;\!\frac{f(q)}{1+e^{\mathcal{E}(q)/T}},
\end{align}
where $\mathcal{E}(q)$ is an even monotonically increasing function for positive $q$ that nullifies at $q=\bar Q>0$, i.e., $\mathcal{E}(\bar Q)=0$ and $\mathcal{E}(0)<0$. Reducing the integral (\ref{eq:integralJf}) over positive $q$ and then splitting it into two integrals with the boundaries of integration $(0,\bar Q)$ and $(\bar Q,+\infty)$ we obtain
\begin{align}\label{eq:integralJf1}
	J[f]=\int_{-\bar Q}^{\bar Q} dq f(q)-\int_{0}^{\bar Q} dq\;\!\frac{f(q)+f(-q)}{1+e^{-\mathcal{E}(q)/T}} + \int_{\bar Q}^{+\infty} dq\;\!\frac{f(q)+f(-q)}{1+e^{\mathcal{E}(q)/T}}.
\end{align}
After the change of variables we end up with
\begin{align}\label{eq:Sommerfeld}
	J[f]={}&T\int_0^{+\infty}\frac{dy}{1+e^y}\biggl[\frac{f(\mathcal{E}^{-1}(Ty)) +f(-\mathcal{E}^{-1}(Ty))}{\mathcal{E}'(\mathcal{E}^{-1}(Ty))}- \frac{f(\mathcal{E}^{-1}(-Ty)) +f(-\mathcal{E}^{-1}(-Ty))}{\mathcal{E}'(\mathcal{E}^{-1}(-Ty))}\biggr]\notag\\
	&+\int_{-\bar Q}^{\bar Q} dq\;\! f(q)+T\int_{|\mathcal{E}(0)|/T}^{+\infty} \frac{dy}{1+e^y}\frac{f(\mathcal{E}^{-1}(-Ty)) +f(-\mathcal{E}^{-1}(-Ty))}{\mathcal{E}'(\mathcal{E}^{-1}(-Ty))}.
\end{align}
If we consider the low-temperature case
\begin{align}\label{eq:conditionT}
	T\ll |\mathcal{E}(0)|,
\end{align}
the first integral in equation~(\ref{eq:Sommerfeld}) can be performed term by term after expanding it in a power series of $T$. In the result one obtains only even powers of $T$. There the integrals of the form 
\begin{align}
	\int_0^{+\infty} dy\;\! \frac{y^j}{1+e^y}=(1-2^{-j}) j! \zeta(j+1)
\end{align}
appear. At low temperatures, the second integral in equation~(\ref{eq:Sommerfeld}) acquires temperature-dependent terms due the dependence of $\bar Q$ on $T$.  Finally, the last integral in equation~(\ref{eq:Sommerfeld}) is exponentially small (unless $f(q)$ has a very special form to cancel the exponential smallness from the denominator, which will not be the case in our applications).

\section{On the low-temperature condition}\label{appendixF}

Here we discuss the condition (\ref{eq:conditionT}). Let us first analyse equation~(\ref{eq:e0}). To achieve that we need to evaluate
\begin{align}
	I(n)=\frac{\lambda}{\pi}\int_{-1}^{1}dy \frac{y^n\sqrt{1-y^2}}{\lambda^2+(x-y)^2}.
\end{align}
This can be done starting from
\begin{align}
	\frac{\lambda}{\pi}\frac{1}{\lambda^2+(x-y)^2}=\frac{1}{\pi}\,\textrm{Im} \frac{1}{y-(x+i \lambda)}
\end{align}
and transforming the integral into the trigonometric one that further can be expressed as
\begin{align}
	I(n)=\frac{1}{2^{n+2}\pi}\,\textrm{Im} \frac{1}{i^n}\oint_{|z|=1} \frac{dz(z^2-1)^n(z^2+1)^2}{z^{n+2}[z^2-2i(x+i \lambda)z-1]}.
\end{align}
For $-1<x<1$ and $0<\lambda\ll 1$, the denominator has poles at $z=0$ and $z=i(x+i\lambda)+\sqrt{1-(x+i\lambda)^2}$ that in the complex plane are inside the circle $|z|=1$. Evaluating the residues we obtain the results that for $n=0$ and $n=2$ are given by
\begin{gather}
	I(0)=\sqrt{1-x^2}-\lambda+\mathcal{O}(\lambda^2),\\
	I(2)=x^2\sqrt{1-x^2}+\frac{\lambda}{2}-3\lambda x^2+\mathcal{O}(\lambda^2).
\end{gather}
Note that the latter two expressions are not accurate in the very near vicinity of $x=\pm 1$, since we assumed $\lambda\ll\sqrt{1-x^2}$.

Equation (\ref{eq:e0}) can now be solved at weak interactions. For $-Q<k<Q$, we obtain
\begin{align}
	\mathcal{E}(k)=-\mu\;\!\frac{\sqrt{Q^2-k^2}}{c}+\frac{\hbar^2}{12mc}(Q^2+2k^2)\sqrt{Q^2-k^2}.
\end{align}
As long as $k\ll Q$, the latter expression does not have restrictions, so we can safely put $k=0$. At $\gamma\ll1$ we can use $\mu=2\epsilon\;\!\gamma$ and $Q=2n\sqrt\gamma$, leading to
\begin{align}\label{eq:e(0)}
	\mathcal{E}(0)=-\frac{8}{3}\epsilon\sqrt\gamma.
\end{align}
Here $\epsilon=\frac{\hbar^2 n^2}{2m}$. At strong interactions we can neglect the integral in the integral equation (\ref{eq:e0}) at the leading order. Using $Q=\pi n$ and $\mu=\hbar^2 Q^2/2m$, we obtain
\begin{align}\label{eq:e(0)strong}
	\mathcal{E}(0)=-\pi^2 \epsilon.
\end{align}
Note that $-\mathcal{E}(0)$ corresponds to the maximal energy of Lieb's type-II excitation. Since $K=\pi/\sqrt\gamma$ at $\gamma\ll 1$ and $K=1$ at $\gamma\gg 1$, the condition (\ref{eq:conditionT}) can be expressed as
\begin{align}\label{eq:conditionsommerfeld}
	T\ll \frac{\epsilon}{K}.
\end{align}
This is the condition for the validity of the Sommerfeld-like expansion (\ref{eq:Sommerfeld}). 

The condition (\ref{eq:conditionsommerfeld}) at weak interactions, $\gamma\ll1$, can further be split into two distinct low-temperature regimes. These are
\begin{align}
	\textrm{(i)}\quad\tau\ll \gamma\quad\textrm{and}\quad \textrm{(ii)}\quad \gamma\ll\tau\ll \sqrt\gamma,
\end{align}
where $\tau=T/\epsilon$. The expansion of, e.g., the free energy in even powers of $\tau$, see equation~(\ref{eq:f}), is only possible in the regime (i). The latter can be concluded using the results for the local two-body correlation function $g^{(2)}=\langle \hat\Psi^\dagger \hat\Psi^\dagger \hat\Psi \hat\Psi\rangle/n^2$ and its relation with the dimensionless free energy per particle $f(\gamma,\tau)=F/N\epsilon$ that is given by \cite{kheruntsyan_pair_2003}
\begin{align}
	g^{(2)}=\left(\frac{\partial f(\gamma,\tau)}{\partial\gamma}\right)_{n,\tau}.
\end{align}
We note the results \cite{kheruntsyan_pair_2003}
\begin{align}
	g^{(2)}=\begin{cases}
		1-\frac{2}{\pi}\sqrt\gamma+\frac{\pi}{24\gamma^{3/2}}\tau^2,\quad &\tau\ll\gamma\ll 1,\\
		1+\frac{1}{2\sqrt\gamma}\tau,\quad &\gamma\ll\tau\ll\sqrt\gamma,
	\end{cases}
\end{align}
which is consistent with equation~(\ref{eq:f}).

\section{The detailed solution of the capacitance to three orders\label{ilt}}

Here we explain the recursive procedure to find the capacitance (\ref{cap2}) to $\mathcal{O}(\kappa)$ order. We thus need to evaluate the coefficients $c_{0,0,0},c_{0,0,1},c_{0,1,0}$ and $c_{0,1,1}$. The first correction to the capacitance is given in terms of $c_{0,0,0}$ and $c_{0,0,1}$. The equation for the coefficient $c_{0,0,0}$ is obtained by solving $V_b(0,0,0)=V_e(0,0,0)$. It leads to
\begin{align}
	c_{0,0,0}=c_{0,0,1}(\ln4+\mathbf{C})+\frac{\ln(4\pi e/ \kappa)-\mathbf{C}}{2},
\end{align}
where $\mathbf{C}\approx 0.5772$ is the Euler--Mascheroni constant. Now we should find $c_{0,0,1}$ which is obtained by solving $V_b(0,0,1)=V_e(0,0,1)$, and trivially leads to
$c_{0,0,1}=1/2$. This fixes $c_{0,0,0}=[1+\ln(16 \pi /\kappa)]/2$. We now find the second correction to the capacitance, which is controlled by $c_{0,1,0}$ and $c_{0,1,1}$. We begin with $c_{0,1,0}$, which requires us to solve $V_b(0,1,0)=V_e(0,1,0)$. It yields
\begin{align}
	c_{0,1,0}={}&\frac{3}{8}c_{1,0,0}+c_{0,1,1}(\ln4+\mathbf{C})+c_{1,0,1}\frac{\ln4+2+\mathbf{C}}{8}+c_{0,1,2}\frac{\pi^2-2(\ln4+\mathbf{C})^2}{2}\notag\\
	&-c_{1,0,2}\frac{8+9\pi^2-2(\ln 64+3\mathbf{C}-2)^2}{48}+Q_{0,1}\frac{\ln(4\pi e /\kappa)-\mathbf{C}}{\sqrt\pi}\notag\\
	&-\frac{3[\ln(4\pi e /\kappa)-\mathbf{C}]^2}{64\pi}-\frac{\pi}{128},
\end{align}
where we used $Q_{1,0}=-3\sqrt{\pi}/32$ according to equation~(\ref{QQ}). Therefore, in order to find $c_{0,1,0}$, one needs six other coefficients: $c_{1,0,0}$, $c_{0,1,1}$, $c_{1,0,1}$, $c_{0,1,2}$, $c_{1,0,2}$, and $Q_{0,1}$. The relation for the first one is obtained by solving $V_b(1,0,0)=V_e(1,0,0)$ and reads
\begin{align}
	c_{1,0,0}=c_{1,0,1}(\ln4+\mathbf{C}-2)-c_{1,0,2}\left[(\ln4+\mathbf{C}-2)^2-\frac{\pi^2-8}{2} \right]+\frac{[\ln(4\pi e /\kappa)-\mathbf{C}]^2}{8\pi}+\frac{\pi}{48}.
\end{align}
To obtain $c_{0,1,1}$, we solve $V_b(0,1,1)=V_e(0,1,1)$ and obtain
\begin{align}
	c_{0,1,1}=-\frac{1}{8}c_{1,0,1}-c_{1,0,2}\frac{3\ln4+3\mathbf{C}-2}{4}+2c_{0,1,2}(\ln4+\mathbf{C})+\frac{Q_{0,1}}{\sqrt\pi}-\frac{3[\ln(4\pi e/\kappa)-\mathbf{C}]}{32\pi}.
\end{align}
We now solve $V_b(1,0,1)=V_e(1,0,1)$ to find
\begin{align}
	c_{1,0,1}=2c_{1,0,2}(\ln4+\mathbf{C}-2)+\frac{\ln(4\pi e /\kappa)-\mathbf{C}}{4\pi}.
\end{align}
We now find the last two coefficients with $\ell=2$ [see equation~(\ref{VV})]. Equation $V_b(0,1,2)=V_e(0,1,2)$ gives
\begin{align}
	c_{0,1,2}=\frac{3}{8}c_{1,0,2}-\frac{3}{64 \pi}.
\end{align}
The coefficient $c_{1,0,2}$ is found from $V_b(1,0,2)=V_e(1,0,2)$ and trivially yields $c_{1,0,2}={1}/{8\pi}$. Finally, the remaining coefficient $Q_{0,1}$ is determined by the equation $V_b(-1,1,0)=V_e(-1,1,0)$ and produces
\begin{align}
	Q_{0,1}=c_{0,0,1}\frac{3\ln4+3\mathbf{C}-4}{16\sqrt\pi}+\frac{c_{0,0,0}}{16\sqrt\pi}+\frac{3[\ln(4\pi e/\kappa)-\mathbf{C}]}{32\sqrt\pi}=\frac{\ln(16\pi/\kappa)}{8\sqrt\pi},
\end{align}
where we used the calculated values for $c_{0,0,0}$ and $c_{0,0,1}$. The above relations among the coefficients fix their values, yielding
\begin{align}
	c_{0,1,0}=\frac{\ln^2(16\pi /\kappa)-2}{8\pi},\qquad c_{0,1,1}=0.
\end{align}
The capacitance (\ref{cap2}) now becomes
\begin{align}
	C(\kappa)=\frac{1}{4\kappa}+\frac{1}{2\pi}(c_{0,0,0}-2c_{0,0,1})+\frac{\kappa}{2\pi}(c_{0,1,0}-2c_{0,1,1})=\frac{1}{4\kappa}+\frac{\L-1}{4\pi}+\frac{\kappa}{16\pi^2}(\L^2-2),
\end{align}
where $\L=\ln(16\pi/\kappa)$. The recursive procedure becomes cumbersome when performed manually. It, however, can be implemented on a computer enabling us to find the capacitance at higher orders in $\kappa$.

\section{List of recurrent symbols}

\begin{tabular}{l l l}
\textbf{Symbol} &
	\textbf{Meaning} & \textbf{Relation to other quantities} \\
	\hline
	$\hbar$ & reduced Planck constant	\\
	$m$ & mass of particles& \\
	$c$ & interaction strength& \\	
	$\theta(k)$ & scattering phase shift & $-2\arctan(k/c)$\\
	$\theta'(k)$ & derivative of the phase shift& $-\frac{2c}{c^2+k^2}$\\
	$N$ & number of particles &\\
	$L$ & system size \\
	$n$ & density of particles & $N/L$ \\
	$\gamma$ & dimensionless parameter of the model & $c/n$\\
	$K$ & Luttinger liquid parameter& $4\pi^2\rho(Q,Q)^2$\\
	$v$ & sound velocity & $\frac{\pi\hbar n}{mK}$\\
	$Q$& Fermi quasimomentum\\ 
	$\lambda$ & dimensionless parameter & $c/Q$\\
	$\rho(k,Q)$ & quasimomentum density distribution &\\
	$\rho(Q,Q)$ & quasimomentum density at the Fermi level& $\frac{\sqrt{K}}{2\pi}$\\
	$\epsilon_0$ &ground-state energy per particle& $\frac{\hbar^2n^2}{2m}e_2(\gamma)$\\
	$E$ &energy& \\
	$E_0$ &ground-state energy& $N\epsilon_0$\\
	$\epsilon$ & natural energy unit & $\frac{\hbar^2 n^2}{2m}$ \\
	$e_2(\gamma)$ & ground-state function & $\epsilon_0/\epsilon$\\
	$e_{2l}(\gamma)$ & dimensionless moments of $\rho(k,Q)$ & $\frac{1}{n^{2l+1}}\int_{-Q}^{Q}dk k^{2l}\rho(k,Q)$ \\
	$\mathcal{F}[f(k,Q)]$ & linear integral operator & $f(k,Q)+\frac{1}{2\pi}\int_{-Q}^{Q}dq\;\! \theta'(k-q)f(q,Q)$\\	
	$\rho^{(j,l)}(k,Q)$ & derivative of $\rho(k,Q)$ & $\frac{\partial^2}{\partial k^j \partial Q^l}\rho(k,Q)$\\
	$\dot\rho(Q,Q)$ & total derivative of $\rho(Q,Q)$ & $\frac{d}{d Q}\rho(Q,Q)$\\
	$K_1$ & dimensionless parameter& $-2\pi n K \frac{\rho^{(1,0)}(Q,Q)}{\rho(Q,Q)}$\\
	$K_2$ & dimensionless parameter& $\frac{K_1^2}{2K}+2\pi^2 n^2 K \frac{\rho^{(2,0)}(Q,Q)}{\rho(Q,Q)}$\\
	$m^{*}$ & effective mass of quasiparticle excitations \\
	$\chi$ & cubic coefficient in the excitation energy\\
	$\nu$ & quartic coefficient in the excitation energy\\
	$T$ & temperature\\
	$F$ & Helmholtz free energy &\\
	$\mu$ & chemical potential \\
	$P$& pressure\\
	$E_{\B}$ & boundary energy\\
$\Psi(x),\Psi^\dagger (x)$ & single-particle Bose field operators
\end{tabular}

\bibliographystyle{iopart-num} 
\bibliography{bibliography-jphysa}

\end{document}